\documentclass{aa}  

\usepackage{graphicx}
\usepackage{subcaption}
\usepackage{txfonts}
\usepackage{hyperref}
\usepackage{longtable}
\usepackage{subcaption}
\usepackage{pdflscape}
\usepackage{rotating}
\usepackage{float} 
\usepackage{orcidlink}
\usepackage{threeparttablex}
\usepackage{array}
\newcolumntype{C}[1]{>{\centering\arraybackslash}p{#1}}
\usepackage{caption}
\usepackage{makecell}
\usepackage{threeparttable}
\usepackage[percent]{overpic}
\usepackage{placeins}

\begin{document} 

   \title{SaNDi-SHoP: Searching for Satellites'N'Disks with \\ a Star-Hopping Program \\} \subtitle{ II. Spectrophotometric analysis and orbital monitoring of directly imaged companions\thanks{Corresponding author: \email{andrea.bernardi@mail.udp.cl}}}
\titlerunning{Spectrophotometric analysis and orbital monitoring of 13 substellar companions}
\authorrunning{Bernardi et al.}
\author{A. Bernardi\inst{1,2}\orcidlink{0009-0001-3428-1029}, A. Zurlo\inst{1,2}\orcidlink{0000-0002-5903-8316}, C. Lazzoni\inst{3,2}\orcidlink{0000-0001-7819-9003}, S. Desidera\inst{3}\orcidlink{0000-0001-8613-2589}, D. Mesa\inst{3}\orcidlink{0000-0001-8467-1933}, S. P\'erez\inst{2,4,5}\orcidlink{0000-0003-2953-755X}, P. H. Nogueira\inst{1,2,6}\orcidlink{0000-0001-8450-3606}, D.~Barbato\inst{3}\orcidlink{0009-0005-9862-2549}, A. Dasgupta\inst{1,2,7}\orcidlink{0009-0009-8115-8910}  }
         
   \institute{\inst{1} Instituto de Estudios Astrof\'isicos, Facultad de Ingenier\'ia y Ciencias, Universidad Diego Portales, Av. Ej\'ercito 441, Santiago, Chile \\
   \inst{2} Millennium Nucleus on Young Exoplanets and their Moons (YEMS), Santiago, Chile \\
   \inst{3} INAF-Osservatorio Astronomico di Padova, Vicolo dell'Osservatorio 5, Padova, Italy, 35122-I \\
   \inst{4} Departamento de Física, Universidad de Santiago de Chile, Av. Victor Jara 3659, Santiago, Chile \\
   \inst{5} Center for Interdisciplinary Research in Astrophysics and Space Science (CIRAS), Universidad de Santiago de Chile \\
\inst{6} Department of Physics and Astronomy, Texas A\&M University, College Station, TX 77843-4242, USA \\
\inst{7} European Southern Observatory, Alonso de Córdova 3107, Casilla19, Santiago, Chile}

   \date{Received 22 December 2025 / Accepted 02 July 2026}

  \abstract
   {Over the past decade, advances in high-contrast imaging instrumentation, coupled with extreme adaptive optics systems, have enabled the discovery of tens of planets and brown dwarfs orbiting at wide separations from their host stars ($a \gtrsim 10$\, au). The existence of companions at these separations challenges current planet-formation paradigms, highlighting the importance of high-contrast imaging as the only technique capable of directly probing this region of planetary systems.}
   {In this paper, we present a survey of thirteen planets and brown dwarfs observed with VLT/SPHERE between June 2023 and July 2025. These data provide updated photometry in the 1.0–1.7\, $\mu$m range and new high-precision astrometry, enabling tighter constraints on their orbital properties.}
   {We used the IRDIS subsystem to acquire dual-band $H2H3$ images ($\lambda_{H2} = 1.593\, \mu$m, $\lambda_{H3} = 1.667\, \mu$m) for all companions in our sample. For the three objects located within the IFS field of view (GQ~Lup~B, PZ~Tel~B, and HD~984~B), we additionally obtained low-resolution ($R \sim 50$) near-infrared (0.96–1.34\, $\mu$m) spectra. We combined our new astrometric measurements with those available in the literature to derive updated orbital solutions. The orbital fitting was performed using the \texttt{orbitize!} Python package. For CT~Cha~b, HIP~78530~B, HIP~64892~B, and RX~J1609.5–2105~b, this work provides the first orbital solutions to date.}
   {We derived new photometry for all objects, which, when compared with field dwarfs in color–magnitude diagrams, indicates spectral types ranging from mid-M to mid-L. For the companions observed with IFS, their spectra are best matched by those of M6–M8.5 field dwarfs. Our updated orbital fits provide tighter constraints for nearly all companions and are consistent with non-circular orbits in all cases, potentially disfavoring core-accretion formation within their circumstellar disks.}
   {}


   \maketitle

\section{Introduction}
\label{sec:intro}

The first exoplanet detection around a solar-type star dates back to 1995 with 51 Pegasi b \citep{Mayor1995}. Since then, more than 6000 exoplanets\footnote{\url{exoplanet.eu}} have been discovered, spanning a wide range of properties. Many challenge classical formation scenarios (such as the core accretion model; \citealt{Pollack1996}), as they are found at tens to hundreds of au and often exceed Jupiter’s mass. While radial velocity (RV) and transit techniques are most sensitive to small separations ($a \lesssim$ 8–10\, au; \citealt{Lagrange2023}), and microlensing relies on rare events, direct imaging (DI) uniquely probes the outer regions of planetary systems and enables the characterization of young planets and their environments, including circumplanetary disks \citep{Zurlo2024}.

Following the first DI detection of a substellar companion \citep{Nakajima1995}, key discoveries include 2MASS~J1207334–393254 b \citep{Chauvin2004} and AB~Pic~b \citep{Chauvin2005a}. Over the past decade, advances in high-contrast imaging with facilities such as VLT/SPHERE \citep{Beuzit2019}, Gemini/GPI \citep{Macintosh2014}, Magellan/MagAO \citep{Males2018, Males2024}, and Subaru/CHARIS \citep{Groff2015} have driven major progress. These instruments enabled discoveries such as the HR 8799 multi-planet system \citep{Marois2008}, AF Lep b \citep{Mesa2023, DeRosa2023, Franson2023b}, the accreting protoplanets in PDS 70 \citep{Keppler2018, Mesa2019, Haffert2019}, and WISPIT 2b \citep{Close2025, vanCapelleveen2025}.

More recently, JWST \citep{Gardner2006} has expanded the accessible parameter space, enabling the detection of lower-mass (less than $\sim$1 M\textsubscript{Jup}) and older (up to ages of several Gyr) companions. Examples include the characterization of the circumplanetary disk around CT~Cha~B \citep{Cugno2025}, the detection of a sub-Jovian planet around TWA 7 \citep{Lagrange2025b}, and the cold planet 14 Her c \citep{BardalezGagliuffi2025}.

Beyond characterization, DI provides precise astrometry, enabling orbital determination from multi-epoch measurements. Orbital parameters are key to constraining formation pathways and system stability. In particular, high eccentricities may indicate dynamical scattering \citep{Bryan2016}, while low eccentricities favor in-situ formation \citep{Bowler2020}. However, wide companions remain challenging due to limited orbital coverage, leading to large uncertainties. The inclusion of RV data improves constraints \citep{Blunt2023}, breaking degeneracies between eccentricity and inclination \citep{FerrerChavez2021} and resolving ambiguities in orbital angles.

A companion’s dynamical mass can be inferred from proper motion measurements of the host star by identifying significant differences (typically signal-to-noise ratio, S/N, $>3$) over long baselines. These proper motion anomalies, or astrometric accelerations, reveal orbiting companions but constrain a degenerate combination of mass and semimajor axis, which can be broken with additional data such as RVs or direct imaging. Catalogs based on \textit{Hipparcos}–\textit{Gaia} comparisons, including \citet{Kervella2022} and \citet{Brandt2021} (\textit{Hipparcos}-\textit{Gaia} Catalog of Accelerations, hereafter HGCA), provide a valuable starting point for targeted surveys and have already enabled discoveries such as the brown dwarf HIP~21152~B \citep{Bonavita2022, Franson2023c}.

Another key parameter for classifying substellar companions is spectral type, directly linked to effective temperature and atmospheric properties, including molecular absorption features (e.g., H$_2$O, CH$_4$, TiO; \citealt{Bailey2014}). The discovery of brown dwarfs and exoplanets extended the classical O–M scheme \citep{Cannon1901}, with substellar objects spanning late M to Y types and temperatures from $\sim2600$\, K to below 400 K. However, temperature alone is not sufficient for spectral classification, as absorption features, along with effects related to spectral, spatial, and temporal resolution, play an important role \citep{Sota2011}. Combined with atmospheric models, this enables estimates of surface gravity and, through evolutionary tracks and isochrones, initial evolutionary masses (e.g., Fig. \ref{cmd}).

In this work, we analyze 13 systems hosting directly imaged planets or brown dwarfs. This paper is the second in a series: \citet{Lazzoni2026} introduced the survey and focused on the search for satellites and circumplanetary disks. Here, we present the photometric and orbital characterization of the companions, refining orbital parameters and, in some cases, providing the first orbital solutions. We also report new SPHERE/IRDIS and SPHERE/IFS photometry.

The paper is organized as follows. Section \ref{sec:obs_red} describes the observations and data reduction. Section \ref{sec:phot_spec} presents the photometric and spectroscopic analysis. Section \ref{sec:orbit_fits} details the orbital fits. Section \ref{sec:results} presents the results, which are discussed in Section \ref{sec:discussion} and summarized in Section \ref{sec:conclusions}.

\section{Observations and data reduction}
\label{sec:obs_red}

The observations presented in this work were performed with the Spectro-Polarimeter High-contrast Exoplanet REsearch (SPHERE) instrument. Data were acquired in IRDIFS mode, combining both near-infrared sub-systems: the InfraRed Dual-band Imager and Spectrograph (IRDIS, \citealt{Dohlen2008}, with a circular field of view, FOV, of $\sim5$\, arcsec radius) and the Integral Field Spectrograph (IFS, \citealt{Claudi2008}, with a $\sim1.73$\, arcsec square FOV), operating simultaneously \citep{Zurlo2014, Mesa2015}. IRDIFS combines IRDIS dual-band imaging \citep{Vigan2010} in $H2H3$ filters ($\lambda_{H2}=1.593\pm0.026\, \mu$m, $\lambda_{H3}=1.667\pm0.027\,\mu$m) with IFS spectroscopy in the $YJ$ range (0.96–1.34 $\mu$m, R$\sim54$).

We observed 13 systems hosting directly imaged planets or brown dwarfs, all with high signal-to-noise ratios (S/N $\gtrsim 50$) and separations $\gtrsim 0.5$\, arcsec to minimize speckle contamination. The targets were selected from those observed within the SpHere INfrared survey for Exoplanets (SHINE; \citealt{Desidera2021}) Guaranteed Time Observations (GTO) program, restricting the sample to systems with companions at sufficiently large separations and with high S/N, as described above.

In this paper, we present astrometry, photometry, and orbital characterization for all companions. For the three targets within the IFS FOV (GQ~Lup~B, PZ~Tel~B, and HD~984~B), we also provide spectra in the $Y$ and $J$ bands and refined spectral types based on comparison with field dwarfs. All relevant properties of the target stars in our survey are reported in Table~\ref{masses}, while Table~\ref{available_data} lists the data considered for the orbital fits for each star, along with their \textit{Gaia} Data Release 3 (DR3; \citealt{GaiaDR3}) ID.
\begin{table}
\centering
\caption{Summary of the SPHERE/IRDIS observations in our survey presented in this work. For each target, we report the date, total field rotation, and the average seeing.}
\begin{tabular}{lcp{2cm}c}
\hline\hline
  Name  &	Obs date     & FOV Rotation      & Seeing \\
        &              &   (deg)   &  (arcsec)  \\
\hline \\
TYC~8047-232-1 & 1 October 2023 & 33.7 & 0.44 \\[0.75ex]
PZ~Tel & 21 June 2025 & 39.3 & 0.56 \\ [0.75ex]
HD~984 & 13 October 2024 & 46.1 & 0.67 \\ [0.75ex]
\hline
\end{tabular}
\label{newobs}
\end{table}

The observations presented by \citet{Lazzoni2026} were obtained between June 2023 and July 2025, covering 12 targets. In addition, eight targets were previously observed within the SHINE GTO program. Observations of AB~Pic were presented by \citet{PalmaBifani2023}; CT~Cha, DH~Tau, and GQ~Lup~B by \citet{Lazzoni2020}; and HIP~64892, HIP~78530, TYC~7084-794-1, and TYC~8047-232-1~B by \citet{Langlois2021}. All targets were observed once, except DH~Tau (October 2024 and January 2025) and TYC~8047-232-1 (October 2023 and August 2024). Two targets, PZ~Tel and HD~984, were observed but not included by \citet{Lazzoni2026}. As discussed therein, such features are most detectable within about half the Hill radius; in these systems, this region is smaller than half the FWHM, making detection unfeasible. A second observation of TYC~8047-232-1 was also excluded, as it was incompatible with the star-hopping technique. Details are given in Table \ref{newobs}. \citet{Lazzoni2026} also report observations of HD~130948, which we exclude due to its unresolved brown dwarf binary (HD 130948 BC), preventing reliable astrometry. All observations used here are listed in Table \ref{astrometry}.

Data reduction followed standard pre-processing steps (dark subtraction, flat-fielding, centering, and wavelength calibration) performed by the High Contrast Data Centre \citep{Maire2016b,Delorne2017,Galicher2018}. Post-processing was carried out with the Vortex Image Processing package (\texttt{VIP}\footnote{\url{vip.readthedocs.io}}; \citealt{GomezGonzalez2017, Christiaens2023}), which was also used to extract astrometry and photometry. Further details on the methodology and error budget are provided in \citet{Lazzoni2026}.
\begin{table*}
\centering
\begin{threeparttable}
\caption{Adopted stellar mass, derived companion mass, parallax, and age for the systems in our sample.}
\label{masses}
\begin{tabular}{lcccccc}
\hline\hline
  Name  &  Adopted stellar mass & Derived companion mass  & Parallax$^a$ &  Age & References \\ 
        &  (M\textsubscript{$\odot$}) & (M\textsubscript{Jup}) & (mas) & (Myr) \\
\hline \\
AB~Pic & $0.90\pm0.10$  & $12.9^{+0.3}_{-1.0}$ & $19.945\pm0.012$ & $28\pm11$ & 1, 2, 3 \\ [0.75ex]
TYC~7084-794-1 & $0.55\pm0.05$ & $33.5^{+2.0}_{-2.1}$ & $44.720\pm0.013$ & $137\pm 17$ & 1, 2, 4, 3\\ [0.75ex]
TYC~8047-232-1 & $0.83\pm0.08$ & $13.1^{+0.2}_{-0.3}$ & $11.543\pm0.012$ & $36\pm8$ & 4, 2 \\ [0.75ex]
CT~Cha & $0.80\pm0.03$ & $9.7^{+1.2}_{-1.1}$ & $5.265\pm0.012$ & $1.41^{+0.38}_{-0.30}$ & 5, 6 \\ [0.75ex]
GQ~Lup & $1.03\pm0.20$ & $29.9^{+1.3}_{-7.3}$ & $6.489\pm0.029$ & $3\pm2$ & 7, 8, 9 \\ [0.75ex]
HIP~78530 & $2.7\pm0.1$ & $19.1^{+1.4}_{-0.6}$ & $7.424\pm0.030$ & $11\pm3$ & 1, 2, 10 \\ [0.75ex]
DH~Tau & $0.41\pm0.04$ & $12.6\pm2.5$ & $7.494\pm0.026$ & $1.4\pm0.1$ & 5 \\ [0.75ex]
HIP~64892 & $2.35\pm0.09$ & $40.7^{+4.0}_{-4.7}$ & $8.360\pm0.048$ & $16\pm2$ & 1, 2, 11\\ [0.75ex]
RXJ1609.5-2105 & $0.7\pm0.1$ & $9.5^{+1.5}_{-1.7}$ & $7.244\pm0.018$ & $11\pm3$ & 1, 12, 13 \\ [0.75ex]
HII~1348 & $1.22\pm0.09$ & $53.3^{+3.7}_{-4.0}$ & $6.979\pm0.031$ & $112\pm5$ & 14, 15 \\ [0.75ex]
PZ~Tel & $1.1\pm0.1$ & $55.5^{+6.5}_{-7.3}$  & $21.162\pm0.022$ & $21\pm4$ & 16, 17, 18, 19, 3 \\ [0.75ex]
HD~984 & $1.20\pm0.06$ & $75^{+22}_{-35}$ & $21.877\pm0.025$ & $115\pm85$ & 1, 20, 21 \\ [0.75ex]
$\eta$~Tel & $2.09\pm0.07$ & $50.2^{+5.5}_{-6.4}$ & $20.603\pm0.010$ & $21\pm4$ & 1, 2, 17, 22 \\ [0.75ex]
\hline
\end{tabular}
\begin{tablenotes}
\footnotesize
\item $^a$ Parallaxes from \textit{Gaia} DR3.
\item References. (1) TESS Input Catalog (TIC, \citealt{Stassun2019}); (2) \citet{Squicciarini2026}; (3) \citet{Gratton2024}; (4) \citet{Baraffe1998}; (5) \citet{Sheehan2019}; (6) \citet{Ginski2024}; (7) \citet{Ginski2014}; (8) \citep{MacGregor2017}; (9) \citet{Alcala2014}; (10) \citet{Lafreniere2011}; (11) \citet{Cheetham2018}; (12) \citet{Ireland2011}; (13) \citet{Pecaut2012}; (14) \citet{Geissler2012}; (15) \citet{Dahm2015}; (16) \citet{DAntona1994}; (17) \citet{Tetzlaff2011}; (18) \citet{Jenkins2012}; (19) \citet{ZunigaFernandez2021}; (20) \citet{Meshkat2015}; (21) \citet{Mints2017}; (22) \citet{Desidera2021}.

\end{tablenotes}
\end{threeparttable}
\end{table*}

\section{Photometric and spectroscopic analysis}
\label{sec:phot_spec}
In this section, we present the photometric and spectroscopic analysis of the substellar companions in our sample. We used IRDIS photometry and, for companions inside the IFS FOV, spectra in the $Y$ and $J$ bands to characterize their spectral properties, deriving contrast and absolute magnitudes, as well as refined spectral type classifications.
The IRDIS and IFS companion fluxes were extracted from the datacubes using the \texttt{VIP} Python package, implementing the negative fake companion (NEGFC) method \citep{Marois2010, Lagrange2010}. All details on this procedure are provided in \citet{Lazzoni2026}. 

The data reduction yielded flux ratios between the companions and their host stars in the $H2$ and $H3$ IRDIS filters, as well as in each spectral channel of IFS. The flux ratios in the $H2$ band were converted into companion masses using the ATMO evolutionary models \citep{Phillips2020}, with the exception of HD~984~B. For this companion, the mass was estimated using the AMES-Dusty atmospheric model \citep{Allard2001}, since the upper age limit of HD~984 (200 Myr, \citealt{Meshkat2015}) lies beyond the range covered by the ATMO models, thus requiring the use of the alternative model AMES-Dusty. All stellar and the derived companion masses, along with system ages and \textit{Gaia} DR3 parallaxes, are reported in Table \ref{masses}.

In order to derive the physical fluxes of the companions, a model of the stellar spectrum is required. For the IRDIS photometry in the $H2$ and $H3$ bands, the stellar flux at zero magnitude was estimated at the corresponding wavelengths using photometric zero-points derived from a Vega spectrum. The stellar magnitudes were then interpolated between the $J$ and $H$ bands to obtain values at the $H2$ and $H3$ filter wavelengths. Finally, the calibrated stellar model was multiplied by the companion-to-star fluxes separately for the $H2$ and $H3$ bands, in order to derive the physical fluxes of the companions. For the IFS contrast spectra, the stellar model spectra were obtained using the VOSA tool (VO Sed Analyzer; \citealt{Bayo2008}), which determines the best-fitting stellar spectrum based on multiple photometric points of the stellar SED. For this purpose, we adopted the BT-NextGen atmospheric models \citep{Allard2012}. These model spectra were then used to derive the flux-calibrated IFS spectra of the companions in the $Y$ and $J$ spectral bands. The details of these results are discussed in Section \ref{subsec:photometry}.

\begin{table*}
\centering
\caption{IRDIS $H$-band contrast and absolute magnitudes, and IFS $Y$- and $J$-band absolute magnitudes for the substellar companions in our sample.}
\begin{tabular}{lccccc}
\hline
\hline
Name & \multicolumn{2}{c}{Contrast (mag)} & \multicolumn{3}{c}{Absolute magnitude (mag)}\\ [0.75ex]
     & $\Delta H2$ & $\Delta H3$ &  $M_Y$ &  $M_J$ &  $M_H$ \\
\hline \\
AB~Pic~b  & $7.82^{+0.03}_{-0.02}$ & $7.64\pm0.02$ & ... & ... & $11.32^{+0.04}_{-0.03}$  \\ [0.75ex]
TYC~7084-794-1~B & $5.82\pm0.01$ & $5.659^{+0.008}_{-0.005}$  & ... & ... & $11.27\pm0.03$ \\ [0.75ex] 
TYC~8047-232-1~B & $7.66^{+0.03}_{-0.02}$ & $7.32^{+0.04}_{-0.06}$ & ... & ... & $11.33^{+0.06}_{-0.07}$ \\ [0.75ex]
CT~Cha~b  & $6.87\pm0.02$ & $6.68\pm0.02$ & ... & ... & $9.33\pm0.05$ \\ [0.75ex]
GQ~Lup~B & $5.961^{+0.006}_{-0.005}$ & $5.88\pm0.01$ & $8.99\pm0.01$ & $8.31\pm0.01$ & $7.68\pm0.04$  \\ [0.75ex]
HIP~78530~B  & $7.80\pm0.01$ & $7.60\pm0.05$ & ... & ... & $9.00\pm0.05$ \\ [0.75ex]
DH~Tau~b  & $5.98\pm0.02$  & $6.12\pm0.01$ & ... & ... & $9.25\pm0.03$ \\ [0.75ex]
HIP~64892~B & $7.204^{+0.009}_{-0.007}$ & $6.98\pm0.02$  & ... & ... & $8.58\pm0.04$ \\ [0.75ex]
RX~J1609.5-2105~b & $8.09\pm0.07$ & $7.86^{+0.05}_{-0.08}$ & ... & ... & $11.40^{+0.07}_{-0.08}$ \\ [0.75ex]
HII~1348~B & $6.10^{+0.04}_{-0.03}$ & $5.98^{+0.04}_{-0.03}$ & ... & ... & $10.09^{+0.05}_{-0.04}$ \\ [0.75ex]
PZ~Tel~B & $5.400^{+0.005}_{-0.003}$ & $5.293^{+0.003}_{-0.005}$ & $9.70\pm0.02$ & $9.08\pm0.01$ & $8.46\pm0.05$ \\ [0.75ex]
HD~984~B &  $6.639\pm0.004$ & $6.467^{+0.004}_{-0.003}$  & $11.02\pm0.01$ & $10.28\pm0.01$ & $9.43\pm0.04$ \\ [0.75ex]
$\eta$~Tel~B & $6.915^{+0.006}_{-0.003}$  & $6.710\pm0.003$  & ... & ... & $8.52\pm0.09$ \\ [0.75ex]

\hline
\end{tabular}
\label{mag_irdis}
\end{table*}

\begin{figure*}[ht!]
 \centering
 \includegraphics[width=0.45\linewidth]{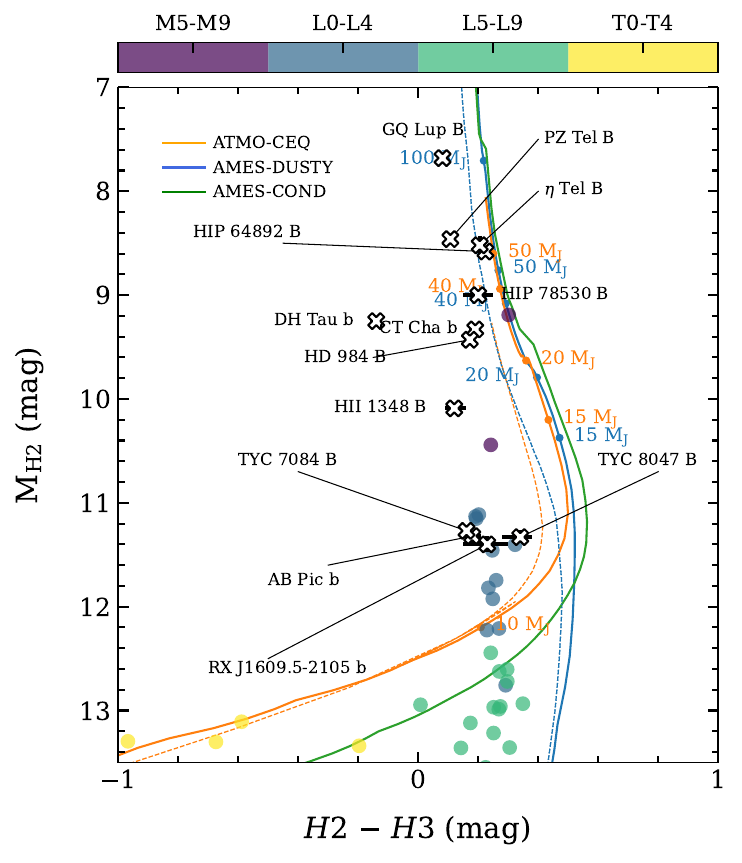}
 \includegraphics[width=0.45\linewidth]{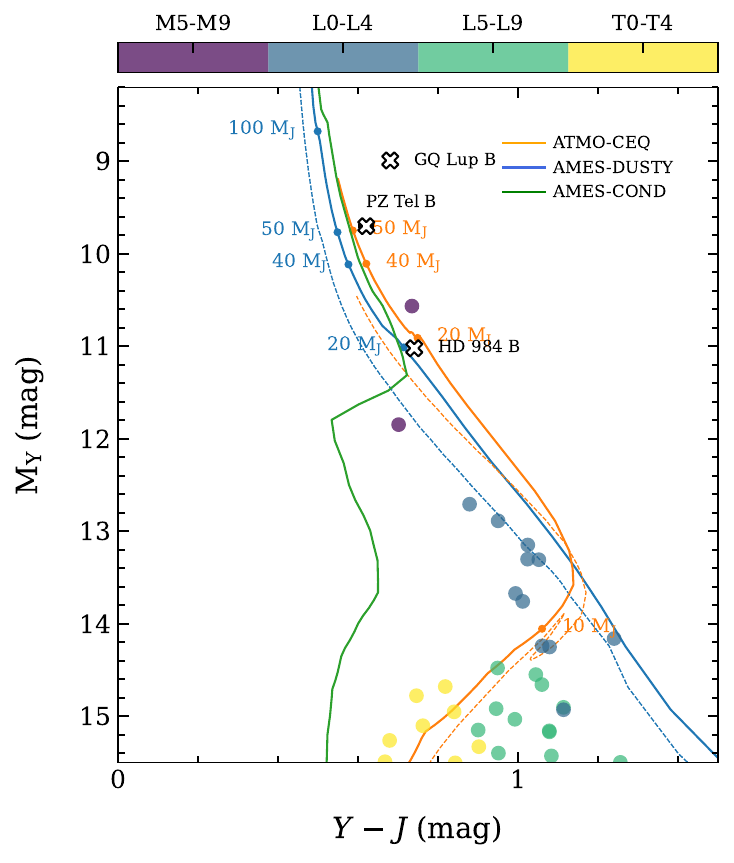}
 \caption{Left: $H2$-$H3$ color magnitude diagram for the objects in our sample. Right: $Y$-$J$ color magnitude diagram for the objects in our sample with IFS observations. ATMO-CEQ, AMES-Cond, and AMES-Dusty evolutionary tracks models at an age of 20 Myr (solid lines) and 100 Myr (dashed lines) are overplotted. In both figures mid-M to late T-type objects from the SpeX Prism Libraries are shown for reference. The color bars above the images refer to the spectral type of the reference objects.}
    \label{cmd}

\end{figure*}

\section{Astrometry and orbital analysis} 
\label{sec:orbit_fits}

From our SPHERE observations, we extracted the astrometric positions of all substellar companions, with average uncertainties of 3.41 mas in separation and 0.12 deg in position angle. The astrometric uncertainties include the centering error of the star behind the coronagraph ($\pm2.5$ mas), the fitting error from the MCMC, and other terms as distortion, pixel scale, true-North correction, and pupil zero-point angle \citep{Maire2021}.

We used our new measurements, together with published astrometry from the literature (all epochs are listed in Table \ref{astrometry}), to fit Keplerian orbits for the 13 systems in our sample. In several cases, no previous orbital fits had been performed, and therefore our work represents the first attempt to characterize the orbits of these companions. Specifically, we were able to derive the first orbital solutions for CT~Cha~b, HIP~78530~B, HIP~64892~B, and RX J1609.5–2105 b.

\begin{table}
\centering
\caption{Priors adopted for the orbital fits.}
\begin{tabular}{lp{4cm}}
\hline
\hline
Parameter  &  Prior \\
\hline \\
Semimajor axis ($a$) & Log-uniform on [$10^{-3}$–$10^4$] au \\ [0.75ex]
Eccentricity ($e$) & Uniform on [0, 1] \\ [0.75ex]
Inclination ($i$) & Uniform in $\sin i$ on [0, 180] deg \\ [0.75ex]
Argument of periastron ($\omega$) & Uniform on [0, 360] deg \\ [0.75ex]
Position angle of nodes ($\Omega$) & Uniform on [0, 360] deg \\ [0.75ex]
Time of periastron passage ($\tau$) & Uniform on [0, 1] \\ [0.75ex]
Total system mass ($M_\mathrm{tot}$) & Gaussian \\ [0.75ex]
Parallax ($\pi$) & Gaussian \\ [0.75ex]
Stellar mass ($M_\star$) & Gaussian \\ [0.75ex]
Companion mass ($M_\mathrm{comp}$) & Log-uniform \\ [0.75ex]
RV offset ($\gamma$) & Uniform on [-5, 5] km s$^{-1}$ \\ [0.75ex]
RV jitter ($\sigma$) & Log-uniform on [10$^{-4}$, 0.05] km s$^{-1}$ \\ [0.75ex]
\hline
\end{tabular}
\label{priors}
\end{table}

Orbit fits were performed using the \texttt{orbitize!} Python package \citep{Blunt2020, Blunt2024}, which employs the \texttt{ptemcee} parallel-tempered MCMC sampler \citep{ForemanMackey2013, Vousden2016}. We adopted standard priors: log-uniform on [$10^{-3}$–$10^4$] au for the semimajor axis ($a$); uniform on [0, 1] for eccentricity ($e$); uniform in $\sin i$ for inclination ($i$); uniform on [$0$, $2\pi$] for $\omega$ and $\Omega$; and uniform on [0, 1] for the time of periastron passage (defined as $\tau = (t_\mathrm{p}-t_\mathrm{ref})/{P}$, with $t_{\mathrm{ref}}=58849$ MJD). Gaussian priors were applied to the total system mass ($M_{\mathrm{tot}}$), based on stellar literature values and photometric companion estimates (Table \ref{masses}), and to the parallax ($\pi$) using \textit{Gaia} DR3 values.

For systems with HGCA data, we included \textit{Hipparcos} \citep{Nielsen2020, VanLeeuwen2007} and \textit{Gaia} DR3 proper motions, adopting a Gaussian prior for the stellar mass ($M_\star$) and a log-uniform prior for the companion mass ($M_{\mathrm{comp}}$). Finally, for systems with available stellar or companion RV measurements, we used a uniform prior on [$-5$, $5$] km s$^{-1}$ for the RV offset ($\gamma$) and a log-uniform prior on [$10^{-4}$–0.05] km s$^{-1}$ for the jitter ($\sigma$). Our priors for the fitted orbital elements are summarized in Table \ref{priors}.

For several companions, a $180$ deg degeneracy in $\omega$ and $\Omega$ arises due to the unknown RV sign. To aid convergence, we restricted $0 \le \omega,\Omega \le 180$ deg and defined the alternative solution as $\omega_1=\omega+\pi$ and $\Omega_1=\Omega+\pi$. This applies to TYC~8047-232-1~B, CT~Cha~b, HIP~78530~B, DH~Tau~b, HIP~64892~B, RX~J1609.5–2105~b, and HII~1348~B.

We drew $10^7$ orbits from the MCMC chains after discarding $8\times10^4$ samples as burn-in (200 steps, 400 walkers), retaining $2.5\times10^4$ steps with a thinning factor of 400. The parallel-tempered setup uses 5600 walkers across 14 temperatures, with the final samples taken from the lowest-temperature ensemble. The \texttt{ptemcee} algorithm periodically proposes swaps between walkers at adjacent temperatures, accepted based on likelihood and temperature differences \citep{Vousden2016}. For two systems (GQ~Lup and PZ~Tel), longer chains were required: 600 walkers, 16 temperatures, 2000 burn-in steps, and $10^5$ recorded steps with thinning of 100.

\begin{table*}
\centering
\caption{Available data, along with \textit{Gaia} DR3 ID, for the objects in our sample}
\begin{tabular}{l l c}
\hline
\hline
Name  &  Available data &	 \textit{Gaia} DR3 ID \\
\hline \\
AB~Pic & Relative astrometry,  HGCA astrometric acceleration & 5495052596695570816 \\ [0.75ex]
TYC~7084-794-1 & Relative astrometry, star-companion RV difference (single epoch) & 2885863400349980288 \\ [0.75ex]
TYC~8047-232-1 & Relative astrometry & 4913138232358905216 \\ [0.75ex]
CT~Cha & Relative astrometry & 5201360671411974912 \\ [0.75ex]
GQ~Lup & Relative astrometry, star-companion RV difference (single epoch) & 6011522757643074304 \\ [0.75ex]
HIP~78530 & Relative astrometry,  HGCA astrometric acceleration & 6243537406369453824 \\ [0.75ex]
DH~Tau & Relative astrometry & 151374202498079872 \\ [0.75ex]
HIP~64892 & Relative astrometry,  HGCA astrometric acceleration & 6088264477373916032 \\ [0.75ex]
RX~J1609.5-2105 & Relative astrometry & 6243841249531772800 \\ [0.75ex]
HII~1348 & Relative astrometry & 66734720809017856 \\ [0.75ex]
PZ~Tel & Relative astrometry, HGCA astrometric acceleration, stellar RV (time series) & 6655168686921108864 \\ [0.75ex]
HD~984 & Relative astrometry, HGCA astrometric acceleration, stellar RV (time series) & 2431157720981843200 \\ [0.75ex]
$\eta$~Tel & Relative astrometry, HGCA astrometric acceleration & 6643602576214115584 \\ [0.75ex]
\hline
\end{tabular}
\label{available_data}
\end{table*}

\section{Results}
\label{sec:results}
\subsection{Photometry and spectroscopy}
\label{subsec:photometry}

We computed $H$-band absolute magnitudes for all objects using $H2$ and $H3$ contrasts and the parallaxes in Table \ref{masses}, constructing a color–magnitude diagram (Fig. \ref{cmd}, left). The two bands were averaged to estimate broadband $H$ magnitudes (Table \ref{mag_irdis}). From IFS spectra, we derived broadband $Y$ ($\lambda_Y = 1.043 \pm 0.070$\, $\mu$m) and $J$ ($\lambda_J = 1.245 \pm 0.120$\, $\mu$m) magnitudes by integrating the spectra, applying Vega zero-points, and using the same parallaxes (Table \ref{mag_irdis}), enabling a second diagram (Fig. \ref{cmd}, right). We also compared IFS spectra of GQ~Lup~B, PZ~Tel~B, and HD~984~B with SpeX Prism Library templates \citep{Burgasser2014}, degraded to R=54. For targets with multiple observations, we performed the analysis on all available epochs and retained the epoch providing the most accurate results. Specifically, we adopted the 1 October 2023 observation for TYC~8047-232-1 and the 2 January 2025 observation for DH~Tau. The color–magnitude diagrams indicate spectral types from mid-M to mid-L (IRDIS) and mid- to late-M (IFS). Several objects lie near AMES and ATMO isochrones, though model differences in $H2$–$H3$ are small and age effects are significant, preventing robust atmospheric constraints from these diagrams alone.

\begin{figure*}
    \centering
    \includegraphics[width=\textwidth]{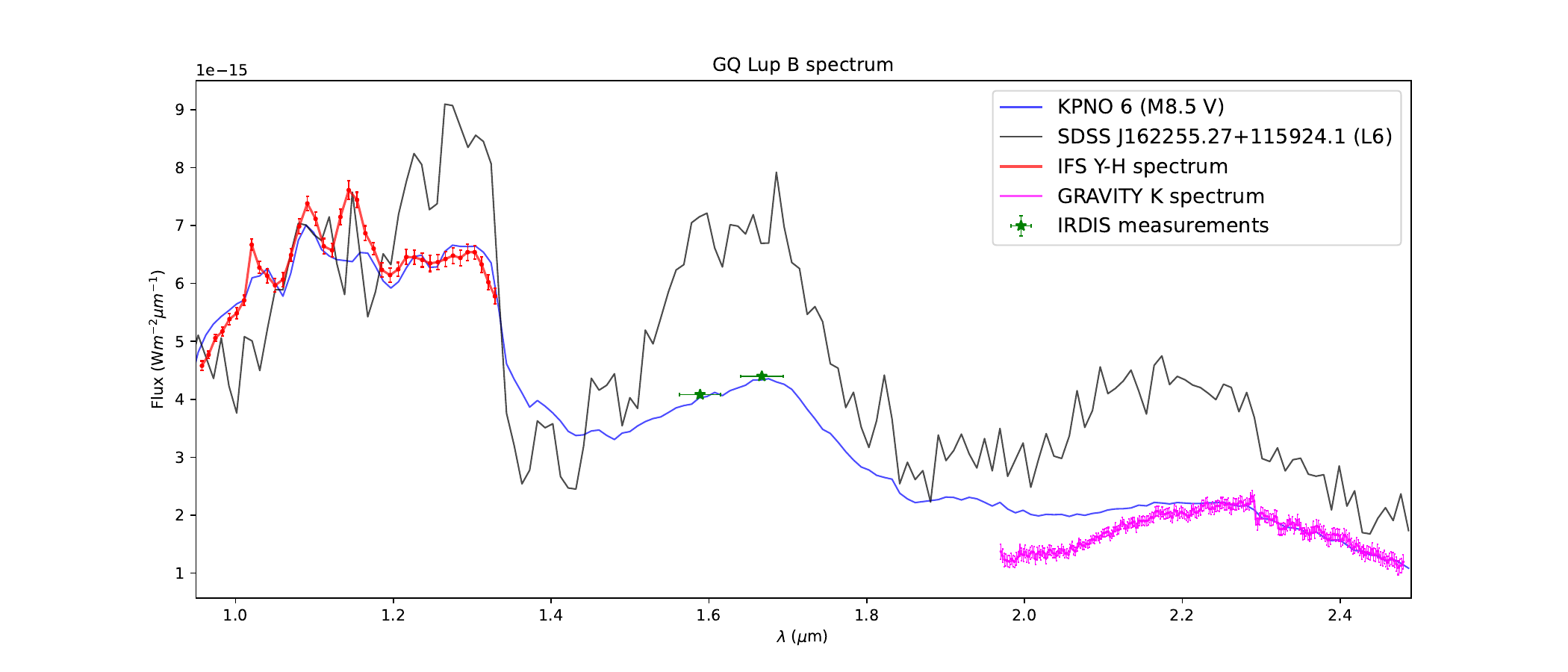}
    \caption{Spectrum of GQ~Lup~B as extracted from IFS and IRDIS observations. A comparison with an M8.5 V field dwarf and an L6 field dwarf is overplotted, together with the GRAVITY $K$-band spectrum \citep{Kammerer2025}.}
    \label{gqlupb_spectrum}
\end{figure*}

Spectral comparisons confirm mid- to late-M types for all three IFS targets. For GQ~Lup~B, the best-fit stellar model ($T_\mathrm{eff}$ = 3600 K, $\log{g}$ = 3) and SpeX templates indicate M8–M8.5, with the best match from KPNO 6 (M8.5\,V; \citealt{Muench2007}). This agrees with \citet{Stolker2021} but not with the L$6\pm0.5$ classification of \citet{Kammerer2025}. To illustrate the discrepancy between our retrieved spectrum and an L6 spectral template, we overplot the spectrum of SDSS J162255.27+115924.1 \citep{Chiu2006}. As shown in Fig. \ref{gqlupb_spectrum}, the L6 template does not reproduce our measurements, with the largest deviations occurring in the $H2$ and $H3$ band fluxes. The GRAVITY $K$-band spectrum from \citet{Kammerer2025} further highlights discrepancies between 2.0 and 2.3 $\mu$m.

\begin{figure*}
    \centering
    \resizebox{\hsize}{!}{\includegraphics{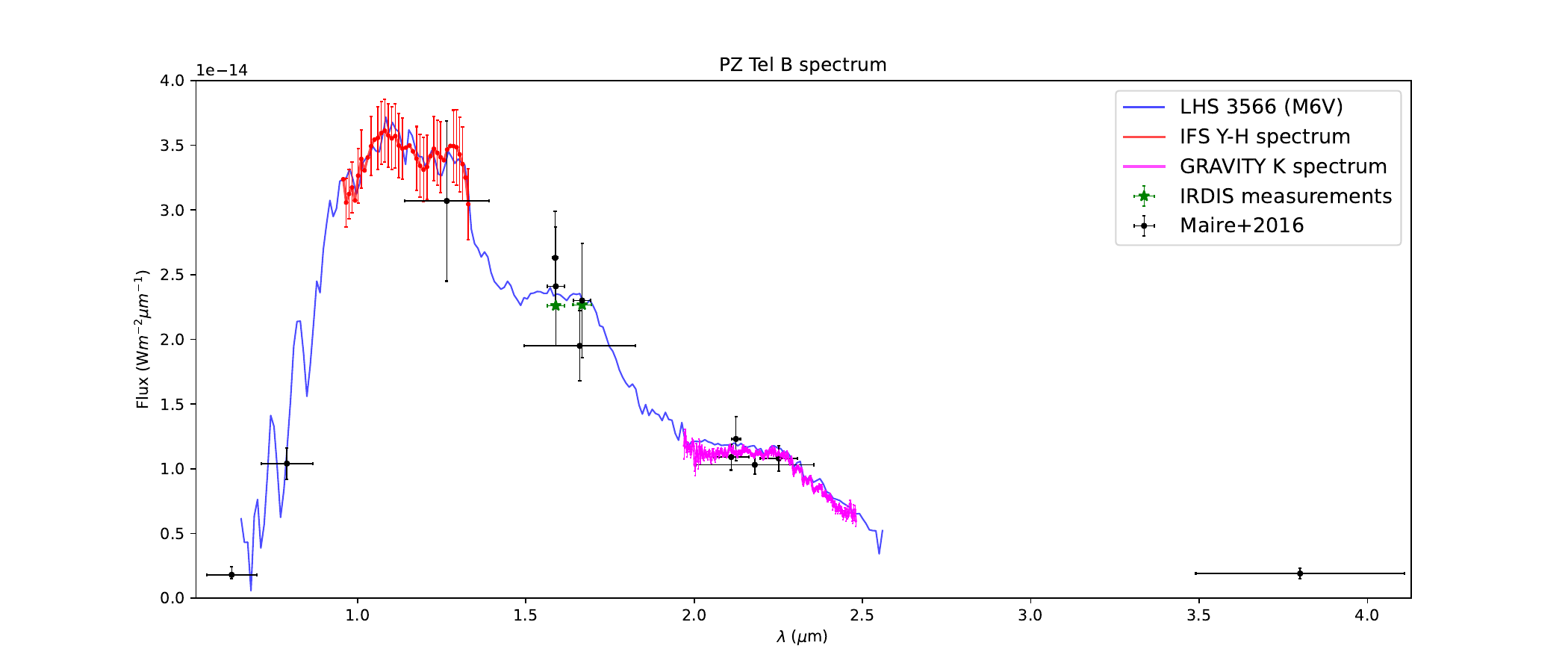}}
    \caption{Spectrum of PZ~Tel~B as extracted from IFS and IRDIS observations. A comparison with an M6\,V field dwarf is overplotted, together with the GRAVITY $K$-band spectrum \citep{Kammerer2025} and the photometry from \citet{Maire2016}.}
    \label{pztelb_spectrum}
\end{figure*}

For PZ~Tel~B, the stellar model ($T_\mathrm{eff}$ = 5200 K, $\log{g}$ = 3.5) and templates indicate M5–M6.5 (Fig. \ref{pztelb_spectrum}), with the best match from LHS 3566 (M6V; \citealt{Burgasser2004}), consistent with \citet{Kammerer2025}. Photometry and spectra agree with \citet{Maire2016}.

\begin{figure*}
    \centering
    \includegraphics[width=\textwidth]{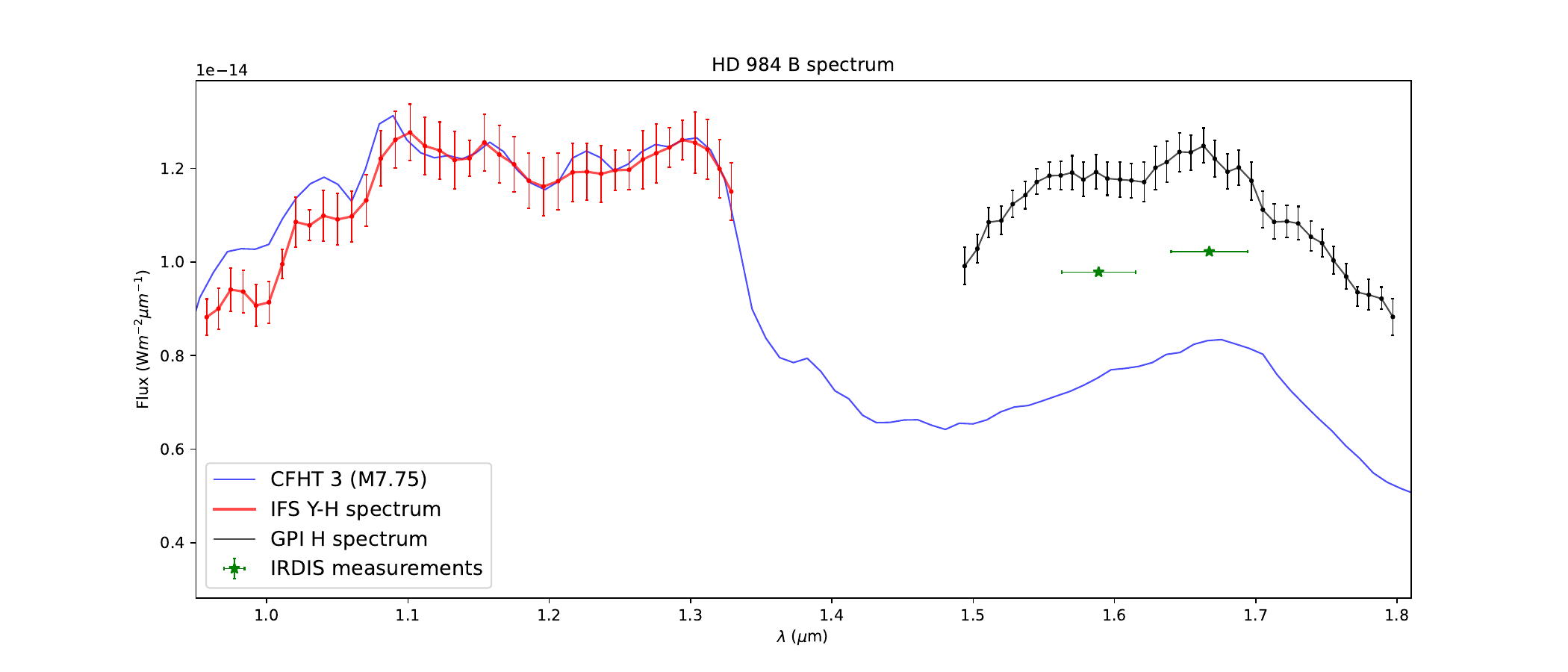}
    \caption{Spectrum of HD~984~B as extracted from IFS and IRDIS observations. The GPI $H$-band spectrum \citep{JohnsonGroh2017} is also included. A comparison with an M7.75 field dwarf is overplotted.}
    \label{hd984b_spectrum}
\end{figure*}

For HD~984~B, the stellar model ($T_\mathrm{eff}$ = 6200 K, $\log{g}$ = 3) and templates indicate M7.75–M8.5 (Fig. \ref{hd984b_spectrum}), with the best match from CFHT 3 (M7.75; \citealt{Muench2007}), consistent with \citet{Franson2022}. However, enhanced $H2$–$H3$ flux suggests a redder spectrum than field dwarfs. Comparison with the GPI $H$-band spectrum \citep{JohnsonGroh2017} supports this behavior. As discussed by \citet{Bonnefoy2015}, who reported a similar behavior for the HR~8799 planets, a possible explanation for this reddening is the presence of additional dust in their atmospheres. In particular, they attributed the observed spectra to an enhanced abundance of photospheric dust compared to field dwarfs. More specifically, they found that a combination of corundum, enstatite, forsterite, and iron grains with different sizes can reproduce the observed spectral reddening. Although a similar explanation could also apply to HD~984~B, we lack sufficient data to draw solid conclusions, and a full atmospheric characterization is beyond the scope of this paper.
\subsection{Orbital fits}
\begin{figure*}
\centering
\begin{overpic}[width=0.6\linewidth]
{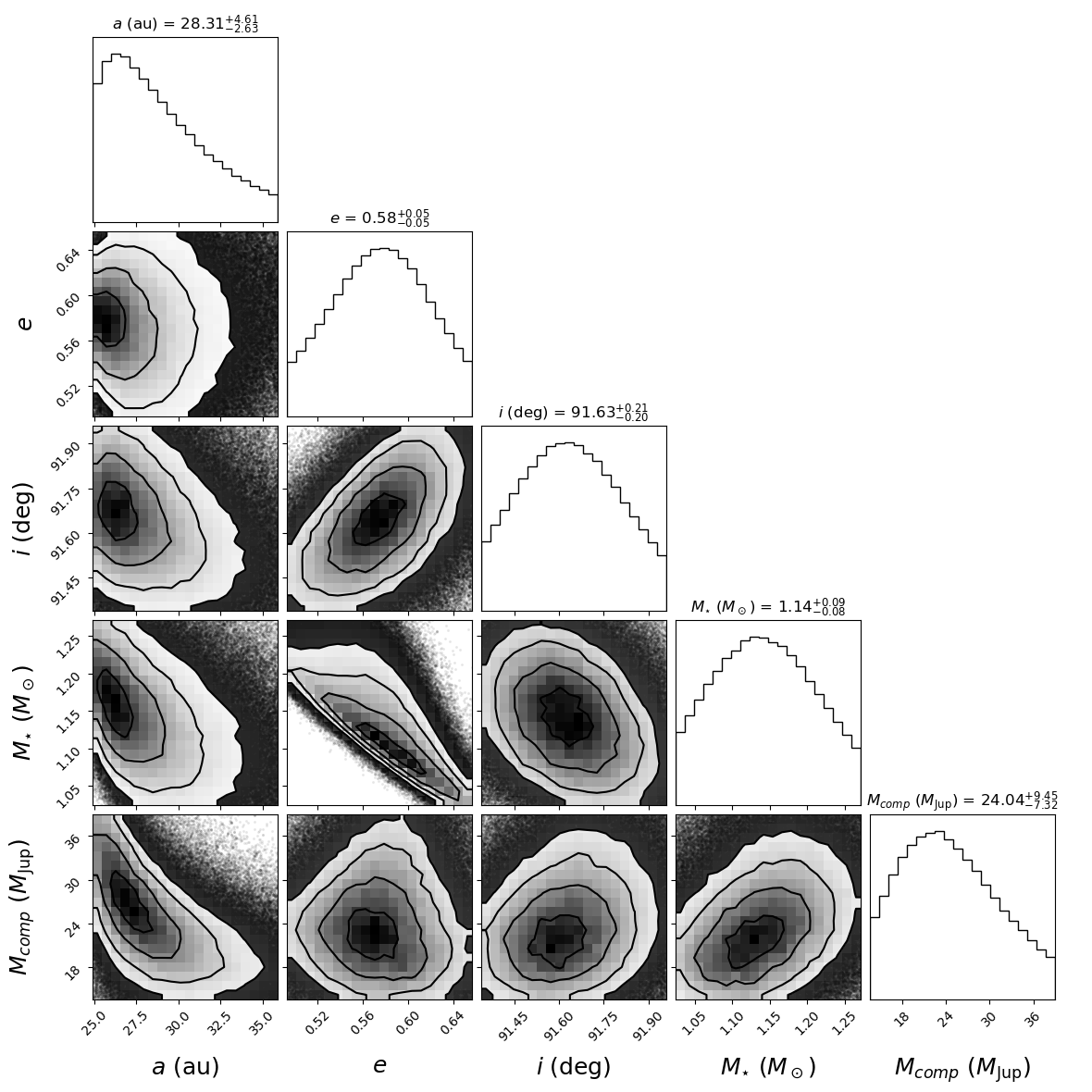}
    \put(61,55){
        \includegraphics[width=0.48\linewidth,trim=5 5 5 5,clip]{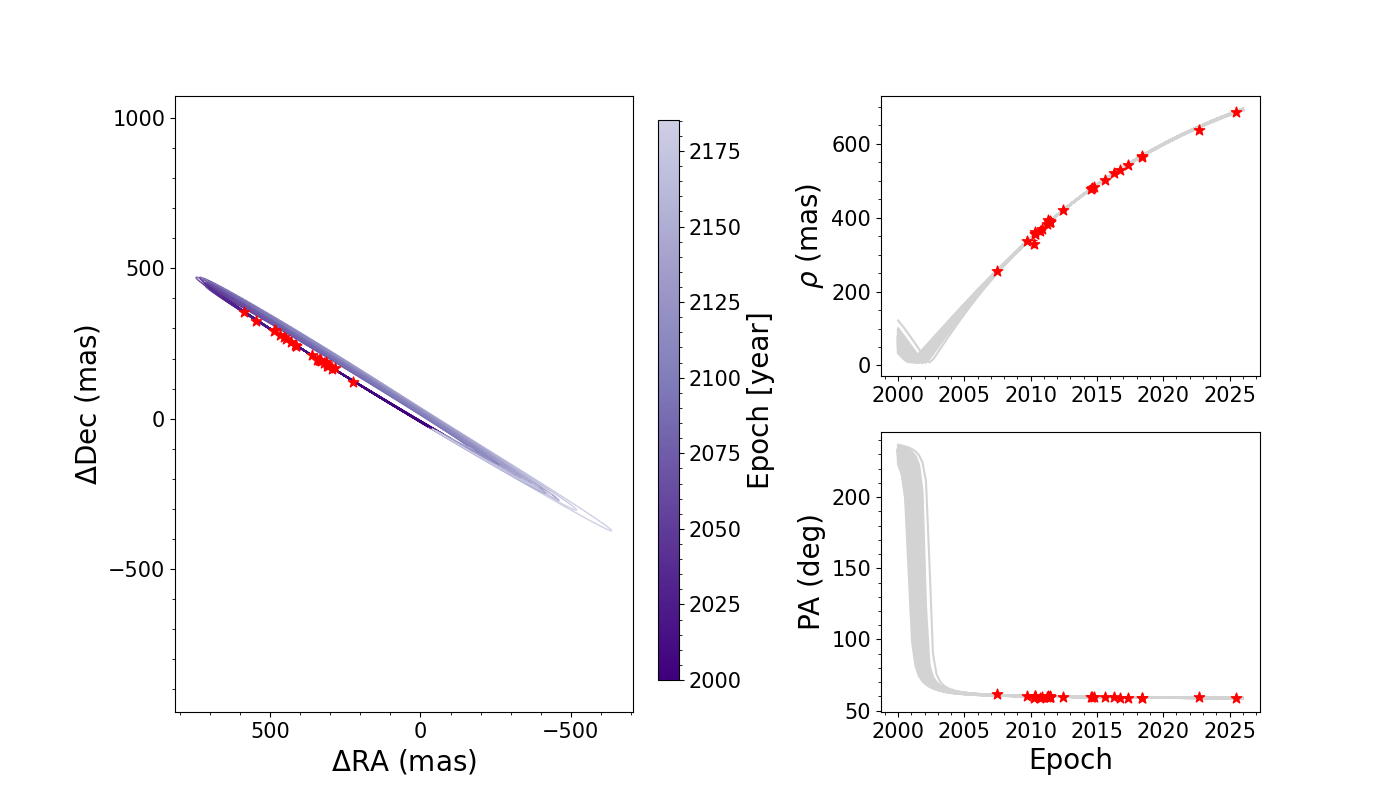}
    }
\end{overpic}
\caption{Corner plot for PZ~Tel~B. Upper right: orbital fit for PZ~Tel~B.}
\label{pztel}
\end{figure*}

In this section, we present the results of the orbital fits for the 13 systems in our sample, which are summarized in Table \ref{fits_results}. Figure \ref{pztel} shows, as an example, 100 orbits randomly drawn from the full set of sampled orbits (left), along with the corner plot of the posterior distributions (right) for PZ~Tel~B. In the Appendix, similar plots for the remaining objects are shown in Fig. \ref{orbit_fits1}, \ref{orbit_fits2}, and \ref{orbit_fits3}, which display 100 randomly drawn orbits for each target. The corresponding corner plots are presented in Fig. \ref{corners1}, \ref{corners2}, and \ref{corners3}.

\subsubsection{AB~Pic}

AB~Pic is a young, K1V-type star \citep{Torres2006}, located at $50.14\pm0.03$\, pc \citep{GaiaDR3}. It is likely a member of the Carina association, with an age of $28\pm11$\, Myr (\citealt{Gratton2024}, obtained considering several literature estimates). Its substellar companion AB~Pic~b was discovered by \citet{Chauvin2005a} using VLT/NACO at $\sim5.5$\, arcsec ($\sim260$\, au), and has a spectral type of L$0.5\pm0.5$ \citep{PalmaBifani2025}.

AB~Pic exhibits a significant proper motion anomaly (e.g., \citealt{Kervella2022, Brandt2021}, SNR $=5.9$), suggesting an additional companion. Although undetected, its properties have been inferred from \textit{Gaia} and \textit{Hipparcos} astrometry, DI limits, and RV data. \citet{Gratton2024} estimated a mass of $8.4\pm4.2$\, M\textsubscript{Jup} at $3.571\pm1.167$\, au, while \citet{Lagrange2025} found ranges of 2.5–9 M\textsubscript{Jup} and 2.5–6 au. This implies the host star is not strictly fixed in astrometry, but we neglected this effect due to the uncertain mass of AB~Pic~c.

We fit seven astrometric measurements over a 20-year baseline, including HGCA absolute astrometry. We adopted a prior of $0.90\pm0.10$\, M\textsubscript{$\odot$} for the stellar mass, based on typical values in the literature (e.g. $0.84\pm0.10$\, M\textsubscript{$\odot$} from the TIC, \citealt{Stassun2019}; $0.92^{+0.10}_{-0.03}$\, M\textsubscript{$\odot$} from \citealt{Squicciarini2026}) We derived a stellar mass of $0.90\pm0.10$\, M\textsubscript{$\odot$}, a companion mass of $10^{+24}_{-8}$\, M\textsubscript{Jup}, a semimajor axis of $237^{+115}_{-55}$\, au, an eccentricity of $0.54^{+0.32}_{-0.28}$\,, and an orbital period of $3800^{+3500}_{-1400}$\, yr.

\subsubsection{TYC 7084-794-1}

TYC~7084-794-1 is a nearby M1V-type star \citep{Torres2006}, member of the AB Doradus moving group with an age of $137\pm17$\, Myr \citep{Gratton2024}, located at $22.361\pm0.006$\, pc \citep{GaiaDR3}. Its substellar companion, TYC~7084-794-1~B, was discovered by \citet{Wahhaj2011} with Gemini/NICI at $\sim3.14$\, arcsec ($67\pm4$\, au). The companion spectral type is estimated to be L$6\pm0.5$ \citep{Kammerer2025} and L$4.5\pm5$ \citep{PalmaBifani2025}.

Our 2023 observation \citep{Lazzoni2026} extends the orbital baseline to 14 years, for a total of six astrometric epochs. We also included a KPIC RV measurement from 12 November 2022, with a companion–star RV difference of $0.71\pm0.10$\, km s$^{-1}$ \citep{Wang2025}. We adopted a prior on the total mass of $0.582\pm0.050$\, M\textsubscript{$\odot$}, based on our estimate of $33.5^{+2.0}_{-2.1}$\, M\textsubscript{Jup} for the mass of the companion, and of $0.55\pm0.05$\, M\textsubscript{$\odot$} for the stellar mass, based on typical values in the literature (e.g. $0.55\pm0.02$\, M\textsubscript{$\odot$} from the TIC, \citealt{Stassun2019}; $0.55\pm0.01$\, M\textsubscript{$\odot$} from \citealt{Squicciarini2026}). We derived a total mass of $0.678^{+0.068}_{-0.042}$\, M\textsubscript{$\odot$}, a semimajor axis of $101^{+15}_{-14}$\, au, an eccentricity of $0.772^{+0.049}_{-0.051}$\,, and an orbital period of $1240^{+330}_{-290}$\, yr.

\subsubsection{TYC 8047-232-1}

TYC~8047-232-1 is a young, K2V-type star \citep{Torres2006}, likely member of the Columba moving group, implying an age of $36\pm8$\, Myr \citep{Gratton2024}. Parallax measurements from \textit{Gaia} DR3 place TYC~8047-232-1 at $86.63\pm0.09$\, pc \citep{GaiaDR3}. The star hosts a substellar companion, TYC~8047-232-1~B, discovered with NTT/SHARP at a separation of $\sim3.2$\, arcsec \citep{Neuhauser2003}, corresponding to a projected semimajor axis of $145.7\pm4.4$\, au. \citet{Chauvin2005b} derived a spectral type of M$9.5\pm1$.

Our 23-year baseline includes 11 astrometric measurements. Our orbit fit was performed by adopting a prior on the total mass of the system of $0.83\pm0.08$\, M\textsubscript{$\odot$}, based on our estimate of $13.1^{+0.2}_{-0.3}$\, M\textsubscript{Jup} for the mass of TYC~8047-232-1~B, and of $0.82\pm0.08$\, M\textsubscript{$\odot$} for the stellar mass, based on typical values in the literature (e.g. $0.8$-$0.9$\, M\textsubscript{$\odot$} from \citealt{Baraffe1998}; $0.79\pm0.9$\, M\textsubscript{$\odot$} from the TIC, \citealt{Stassun2019}; $0.82\pm0.05$\, M\textsubscript{$\odot$} from \citealt{Squicciarini2026}). Finally, we obtained a total dynamical mass of $0.90^{+0.10}_{-0.08}$\, M\textsubscript{$\odot$}, a semimajor axis of $265^{+80}_{-43}$\, au, an eccentricity of $0.82^{+0.10}_{-0.26}$\,, and an orbital period of $4600^{+2600}_{-1200}$\, yr.

\subsubsection{CT~Cha}

CT~Cha is a young K7V-type star \citep{Torres2006} in the Chamaeleon star-forming region, with an age of $1.41^{+0.38}_{-0.30}$\, Myr (\citealt{Sheehan2019}, obtained with \citealt{Feiden2016} non-magnetic models), and located at $190.0\pm0.4$\, pc \citep{GaiaDR3}. Its substellar companion, CT~Cha~b, was discovered by \citet{Schmidt2008} with VLT/NACO at $2.67$\, arcsec. A spectral type of M$9\pm1$ was estimated by \citet{Wu2015}. The companion is accreting, as indicated by Pa$\beta$ \citep{Schmidt2008} and H$\alpha$ \citep{Wu2015} emission, and hosts a circumplanetary disk detected with JWST \citep{Cugno2025}.

We combined our observations with literature data spanning a 17-year baseline. For the 2013 epoch \citep{Wu2015}, multi-band measurements were averaged to derive final astrometry. We performed an orbital fit adopting a prior on the total system mass of $0.81\pm 0.05$\, M\textsubscript{$\odot$}, based on a companion mass of $9.7^{+1.2}_{-1.1}$\, M\textsubscript{Jup}, and a stellar mass of $0.80\pm0.05$\, M\textsubscript{$\odot$}, according to typical values in the literature (e.g. $0.796^{+0.015}_{-0.014}$\, M\textsubscript{$\odot$} from \citealt{Sheehan2019}; $0.9\pm0.2$\, M\textsubscript{$\odot$} from \citealt{Ginski2024}) We derived a total dynamical mass of $0.807^{+0.050}_{-0.051}$\, M\textsubscript{$\odot$}, a semimajor axis of $520^{+380}_{-170}$\, au, an eccentricity of $0.60^{+0.29}_{-0.38}$\,, and an orbital period of $13000^{+18000}_{-6000}$\, yr.

Given the wide separation, only a small orbital arc is covered, leading to large uncertainties. Although a circumplanetary disk could shift the photocenter, we found no wavelength-dependent trends or excess scatter in the residuals, indicating a negligible impact. A more detailed analysis is presented in \citet{Lazzoni2026}.

\subsubsection{GQ~Lup}

GQ~Lup is a young T Tauri K7V-type star \citep{Herbig1977} in the Lupus star-forming region, located at $154.1\pm0.7$\, pc \citep{GaiaDR3}, with an age of $3\pm2$\, Myr (\citealt{Alcala2014}, by comparison with PMS theoretical evolutionary tracks). ALMA and MagAO observations revealed an accretion disk extending to $\sim44$ au and strong H$\alpha$ emission \citep{MacGregor2017, Wu2017}. The system hosts a substellar companion, GQ~Lup~B, discovered with VLT/NACO by \citet{Neuhauser2005} at $\sim740$\, mas ($\sim103$\, au). JWST spectroscopy confirmed warm dust around the companion \citep{Cugno2024}, and its spectral type is L$6\pm0.5$ \citep{Kammerer2025}.

We included 29 astrometric measurements over a 31-year baseline, along with a CRIRES RV measurement from 29 May 2014, yielding a companion–star RV difference of $2.0\pm0.4$\, km s$^{-1}$ \citep{Schwarz2016}. We performed an orbital fit adopting a prior on the total system mass of $1.06\pm0.20$\, M\textsubscript{$\odot$}, based on a companion mass of $29.9^{+1.3}_{-7.3}$\, M\textsubscript{Jup} and a stellar mass of $1.03\pm0.20$\, M\textsubscript{$\odot$}, according to typical values in the literature (e.g. $0.7$\, M\textsubscript{$\odot$} from \citealt{Ginski2014}; $1.03\pm0.05$\, M\textsubscript{$\odot$} from \citealt{MacGregor2017}).

We derived a total mass of $1.19^{+0.39}_{-0.19}$\, M\textsubscript{$\odot$}, a semimajor axis of $90^{+29}_{-10}$\, au, an eccentricity of $0.47^{+0.13}_{-0.16}$\,, and an orbital period of $780^{+510}_{-210}$\, yr. The posterior distributions for $\Omega$ consist of two distint peaks separated by $\sim180$\, deg. We report both the central values, along with their $1\sigma$ confidence interval, in Table \ref{fits_results}. As for CT~Cha, disks around both the host star and companion could in principle affect astrometry. However, we found no wavelength-dependent trends or excess scatter in the residuals, indicating a negligible impact. A more detailed discussion is provided in \citet{Lazzoni2026}.

\subsubsection{HIP~78530}

HIP~78530 is a young B9V-type star \citep{Murphy2020} in the Upper Scorpius association \citep{Gagne2018}, located at $134.7\pm0.5$\, pc \citep{GaiaDR3}, with an age of $11\pm3$\, Myr \citep{Squicciarini2026}. The star shows a marginal HGCA proper motion anomaly (SNR $=0.95$). Its wide substellar companion, HIP~78530~B, was discovered by \citet{Kouwenhoven2005} and later confirmed as comoving by \citet{Lafreniere2011}, at $\sim4.5$\, arcsec ($\sim710$\, au). A spectral type of M$7\pm1$ was estimated by \citet{PalmaBifani2025}.

We included six astrometric epochs over a 17-year baseline, including HGCA absolute astrometry. We considered the 2011 epoch presented by \citet{Bailey2013} as an outlier due to its large uncertainties in both the separation and position angle, which resulted in inflated uncertainties in the posterior distribution. We performed an orbital fit adopting priors of $2.7\pm0.2$\, M\textsubscript{$\odot$} for the stellar mass, based on typical literature values (e.g. $\sim2.5$\, M\textsubscript{$\odot$} from \citealt{Lafreniere2011}; $2.19^{+0.45}_{-0.24}$\, M\textsubscript{$\odot$} from the TIC, \citealt{Stassun2019}; $2.67^{+0.03}_{-0.02}$\, M\textsubscript{$\odot$} from \citealt{Squicciarini2026}). We derived a stellar mass of $2.69\pm0.14$\, M\textsubscript{$\odot$}, a companion mass of $14^{+26}_{-11}$\, M\textsubscript{Jup}, a semimajor axis of $550^{+360}_{-160}$\, au, an eccentricity of $0.65^{+0.19}_{-0.26}$\,, and an orbital period of $7800^{+9300}_{-3200}$\, yr.

\subsubsection{DH~Tau}

DH~Tau is a young M1V-type star \citep{Herbig1977} in the Taurus star-forming region, located at $133.4\pm0.5$ pc \citep{GaiaDR3}, with a mass of $0.41\pm0.04$\, M\textsubscript{$\odot$} and an age of $1.4\pm0.1$\, Myr (\citealt{Sheehan2019}, calculated with \citealt{Feiden2016} non-magnetic models). Its substellar companion, DH~Tau~b, was discovered by \citet{Itoh2005} with Subaru/CIAO at $\sim2.3$\, arcsec ($\sim330$\, au), and has a spectral type of M9–9.5 \citep{Xuan2024}. The companion shows ongoing accretion (H$\alpha$, Pa$\beta$, Br$\gamma$; \citealt{Zhou2014, Bonnefoy2014}) and evidence for a circumplanetary disk from Spitzer/IRAC photometry \citep{Martinez2022}. A candidate $\sim1$\, M\textsubscript{Jup} satellite at $\sim10$\, au has also been reported, though its nature remains uncertain \citep{Lazzoni2020}.

We combined our observations with literature data to obtain 12 astrometric epochs over a 26-year baseline. Several measurements were excluded: early epochs from \citet{Itoh2005} superseded by \citet{Ginski2014}, and two epochs from \citet{Bryan2016} identified as strong outliers ($>15\sigma$ deviations). Moreover, the two epochs from \citet{Bryan2016} were obtained with Keck/NIRC2, whereas our 2015 epoch was collected with SPHERE. As noted in previous studies (e.g., \citealt{Konopacky2016}), the partially transparent NIRC2 coronagraph introduces additional systematic uncertainties; in particular, the 600 mas mask yields astrometric errors of $\sim$4–5 mas \citep{Bowler2018}. Compared to the SPHERE error budget (Section~\ref{sec:orbit_fits}), this implies lower astrometric precision for the NIRC2 data. Furthermore, the \citet{Bryan2016} measurements deviate by more than $15\sigma$ from ours, indicating strong inconsistency. We therefore conclude that these points, rather than our 2015 epoch, should be considered outliers. Our orbital fit, extending the baseline of \citet{Bowler2020} by seven years, adopted a prior on the total system mass of $0.422 \pm 0.040$\, M\textsubscript{$\odot$}, based on a companion mass of $12.6\pm2.5$\, M\textsubscript{Jup}. We derived a total dynamical mass of $0.418^{+0.037}_{-0.039}$\, M\textsubscript{$\odot$}, a semimajor axis of $234^{+102}_{-41}$\, au, an eccentricity of $0.58^{+0.32}_{-0.26}$\,, and an orbital period of $5500^{+4500}_{-1600}$\, yr.

As for CT~Cha, although a circumplanetary disk could affect the photocenter, we found no wavelength-dependent trends or excess scatter in the residuals, indicating a negligible impact. A more detailed discussion is provided in \citet{Lazzoni2026}.

\subsubsection{HIP~64892}
HIP~64892 is a young, B9.5V-type star \citep{Houk1978} located at a distance of $119.6\pm0.7$\, pc \citep{GaiaDR3}, with an estimated age of $16\pm2$\, Myr \citep{Squicciarini2026}. Moreover, this star shows a proper motion difference in the HGCA, even if not significant, with an SNR of 0.78. The star hosts a substellar companion, HIP~64892~B, discovered with VLT/SPHERE and located $\sim1.27$\, arcsec northwest of the primary, corresponding to a projected separation of $\sim159$\, au \citep{Cheetham2018}. Its spectral type was estimated to be M$8\pm1.5$ \citep{Kammerer2025}.

We present a total of five astrometric epochs for the HIP~64892 system, covering a 14-year baseline, along with HGCA absolute astrometry. Our orbit fit was performed by adopting a prior on the stellar mass of $2.35\pm0.09$\, M\textsubscript{$\odot$}, based on typical values in the literature (e.g. $2.35\pm0.09$\, M\textsubscript{$\odot$} from \citealt{Cheetham2018}; $2.74\pm0.37$\, from the TIC, \citealt{Stassun2019}; $2.34^{+0.02}_{-0.01}$\, M\textsubscript{$\odot$} from \citealt{Squicciarini2026}). We obtained a stellar mass of $2.346^{+0.087}_{-0.086}$\, M\textsubscript{$\odot$}, a companion dynamical mass of $12^{+30}_{-10}$\, M\textsubscript{Jup}, a semimajor axis of $134^{+41}_{-33}$\, au and an eccentricity of $0.58^{+0.34}_{-0.37}$\,. The derived orbital period is $1010^{+520}_{-370}$\, yr. 

\subsubsection{RX~J1609.5-2105}

RX J1609.5–2105 is a young M0V-type star \citep{Rizzuto2015}, located at $138\pm0.3$\, pc \citep{GaiaDR3}. It is a member of the Upper Scorpius association, implying an age of $11\pm3$ Myr (\citealt{Pecaut2012}, obtained through \textit{Hipparcos} and 2MASS photometry). Its substellar companion, RX~J1609.5–2105~b, was discovered by \citet{Lafreniere2008} with Gemini/NIRI at $\sim2.2$\, arcsec ($\sim330$\, au), with a spectral type of L$4^{+1}_{-2}$. Its comoving nature was confirmed by \citet{Ginski2014}.

We included 11 astrometric measurements spanning a 19-year baseline, extended by more than 12 years with our new epoch. We performed an orbital fit adopting a prior on the total system mass of $0.71\pm0.20$\, M\textsubscript{$\odot$}, based on a companion mass of $9.5^{+1.5}_{-1.7}$\, M\textsubscript{Jup} and a stellar mass of $0.7\pm0.2$\, M\textsubscript{$\odot$}, based on typical values in the literature (e.g. 0.68-0.77 M\textsubscript{$\odot$} from \citealt{Ireland2011}; $0.62\pm0.08$\, M\textsubscript{$\odot$} from the TIC, \citealt{Stassun2019}). We derived a total dynamical mass of $0.77^{+0.11}_{-0.08}$\, M\textsubscript{$\odot$}, a semimajor axis of $385^{+110}_{-78}$\, au, an eccentricity of $0.21^{+0.14}_{-0.12}$, and an orbital period of $8600^{+4800}_{-2900}$\, yr.

\subsubsection{HII~1348}

HII~1348 is a K5V-type double-lined spectroscopic binary (SB2) with a mass of $1.22\pm0.09$\, M\textsubscript{$\odot$} \citep{Geissler2012}, located at $143.3\pm0.6$\, pc \citep{GaiaDR3}, and with an age of $112\pm5$\, Myr (\citealt{Dahm2015}, through a magnitude estimate of the Pleiades' lithium depletion boundary). Its substellar companion, HII~1348~B, was first identified by \citet{Bouvier1997} at $\sim1.1$\, arcsec and initially classified as a background object, but later confirmed as comoving by \citet{Geissler2012}, who also derived a spectral type of M$8\pm1$.

We included six astrometric epochs spanning a 27-year baseline. We performed an orbital fit adopting a prior on the total system mass of $1.271\pm0.090$\, M\textsubscript{$\odot$}, based on a companion mass of $53.3^{+3.7}_{-4.0}$\, M\textsubscript{Jup}. We also evaluated the impact of the SB2 photocenter motion, deriving a semimajor axis of $a_0 = 0.109 \pm 0.006$\, au \citep{Marcussen2023}, corresponding to $0.760 \pm 0.042$\, mas at the system distance, which is well below our astrometric uncertainties and thus negligible. We derived a total dynamical mass of $1.316^{+0.098}_{-0.089}$\, M\textsubscript{$\odot$}, a semimajor axis of $181^{+49}_{-32}$\, au, an eccentricity of $0.61^{+0.09}_{-0.14}$, and an orbital period of $2100^{+1000}_{-600}$\, yr.

\subsubsection{PZ~Tel}

PZ~Tel is a young G9IV-type star \citep{Torres2006} at $47.25\pm0.05$\, pc \citep{GaiaDR3}. It is a member of the $\beta$ Pictoris moving group, with an age of $21\pm4$\, Myr \citep{Gratton2024}, and shows a significant HGCA proper motion anomaly (SNR $=3.68$). Its brown dwarf companion, PZ~Tel~B, was independently discovered by \citet{Biller2010} and \citet{Mugrauer2010} with Gemini/NICI and VLT/NACO at $\sim350$\, mas ($\sim16.5$\, au), and has a spectral type of M$6\pm0.5$ \citep{Kammerer2025}.

We included 27 astrometric measurements over an 18-year baseline, including HGCA absolute astrometry. Multi-band epochs from \citet{Ginski2014} and \citet{Maire2016} were averaged. We also incorporated 41 HARPS RV measurements (RMS 575 m s$^{-1}$) from the \texttt{HARPS-RVBANK} archive (\citealt{Trifonov2020}), reported in Table \ref{rvs}, inflating their uncertainties by a factor of 10 to account for stellar variability. We performed an orbital fit adopting a prior of $1.1\pm0.1$\, M\textsubscript{$\odot$} for the stellar mass, based on typical values in the literature (e.g. $1.25^{+0.05}_{-0.20}$\, M\textsubscript{$\odot$} from \citealt{DAntona1994}; $1.2\pm0.1$\, M\textsubscript{$\odot$} from \citealt{Tetzlaff2011}; $1.13\pm0.03$ \, M\textsubscript{$\odot$} from \citealt{Jenkins2012}; $1.14$\, M\textsubscript{$\odot$} from \citealt{ZunigaFernandez2021}).

We derived a stellar mass of $1.142^{+0.086}_{-0.081}$\, M\textsubscript{$\odot$}, a companion mass of $24.0^{+9.5}_{-7.3}$\, M\textsubscript{Jup}, a semimajor axis of $28.3^{+4.6}_{-2.6}$\, au, an eccentricity of $0.576^{+0.052}_{-0.054}$, and an orbital period of $140^{+42}_{-24}$\, yr. The posterior distributions for both $\omega$ and $\Omega$ consist of two distint peaks separated by $\sim180$\, deg. We report both the central values, along with their $1\sigma$ confidence interval, in Table \ref{fits_results}.

\subsubsection{HD~984}

HD~984 is a young F7V-type star \citep{Houk1999} at $45.71\pm0.05$\, pc \citep{GaiaDR3}, with an age of $115\pm85$\, Myr (\citealt{Meshkat2015}; obtained through estimates of the coronal X-ray emission and of the stellar rotation). The star exhibits a strong HGCA proper motion anomaly (SNR $=35.2$). Its substellar companion, HD~984~B, was discovered by \citet{Meshkat2015} with VLT/NACO at $\sim190$\, mas ($\sim9.0$\, au), and confirmed as comoving with SINFONI observations. Its spectral type has been estimated as M$6.5\pm1.5$ \citep{Franson2022}.

We included seven astrometric measurements over a 12-year baseline, including HGCA absolute astrometry. The 2015 epoch was averaged over $J$ and $H$ bands. We also incorporated RV data from HARPS (22 epochs; RMS 86.7 m s$^{-1}$; \citealt{Grandjean2020}) and HPF (13 epochs; RMS 48.0 m s$^{-1}$; \citealt{Franson2022}), all reported in Table \ref{rvs}. We performed an orbital fit adopting priors of $1.20\pm0.06$\, M\textsubscript{$\odot$} for the stellar mass, based on typical values in the literature (e.g. $1.2\pm0.06$\, M\textsubscript{$\odot$} from \citealt{Meshkat2015}; $1.15\pm0.06$\, M\textsubscript{$\odot$} from \citealt{Mints2017}; $1.3\pm0.2$\, M\textsubscript{$\odot$} from the TIC, \citealt{Stassun2019}).

We derived a stellar mass of $1.133\pm0.033$\, M\textsubscript{$\odot$}, a companion mass of $68.4^{+2.5}_{-2.7}$\, M\textsubscript{Jup}, a semimajor axis of $22.5^{+2.5}_{-2.3}$\, au, an eccentricity of $0.586\pm0.038$, and an orbital period of $97^{+18}_{-16}$\, yr.

\subsubsection{$\eta$~Tel}

$\eta$~Tel is a young A0V-type star \citep{Houk1975} at $48.5\pm0.2$\, pc \citep{GaiaDR3}. \citep{Desidera2021}. It is a member of the $\beta$ Pictoris moving group (age $21\pm4$\, Myr; \citealt{Gratton2024}) and shows a moderate HGCA proper motion anomaly (SNR $=3.29$). The system hosts a debris disk, first identified from IRAS excess emission \citep{Backman1993} and recently characterized with JWST, revealing thermal dust emission and a broad 20 $\mu$m silicate feature \citep{Chai2024}. Its substellar companion, $\eta$~Tel~B, was discovered by \citet{Lowrance2000} with NICMOS at $\sim4.2$\, arcsec ($\sim200$\, au), with a spectral type of M$7.5$.

We included 17 astrometric measurements over a 27-year baseline, including HGCA absolute astrometry. We performed an orbital fit adopting priors of $2.09\pm0.07$\, M\textsubscript{$\odot$} for the stellar mass, based on typical values in the literature (e.g. $2.2\pm0.1$\, M\textsubscript{$\odot$} from \citealt{Tetzlaff2011}; $2.49\pm0.35$\, M\textsubscript{$\odot$} from the TIC, \citealt{Stassun2019}; $2.09\pm0.03$\, M\textsubscript{$\odot$} from \citealt{Desidera2021}; $2.14^{+0.03}_{-0.02}$ \,M\textsubscript{$\odot$} from \citealt{Squicciarini2026}). We derived a stellar mass of $2.09\pm0.07$\, M\textsubscript{$\odot$}, a companion mass of $11^{+35}_{-8}$\, M\textsubscript{Jup}, a semimajor axis of $148^{+60}_{-34}$\, au, an eccentricity of $0.67^{+0.27}_{-0.41}$, and an orbital period of $1250^{+860}_{-420}$\, yr. The posterior distributions for $\Omega$ consist of two distinct peaks separated by $\sim180$\, deg. We report both the central values, along with their $1\sigma$ confidence interval, in Table \ref{fits_results}.

\section{Discussion}
\label{sec:discussion}
The aim of this study is to refine the spectrophotometric properties and the orbital characterization of the planets and brown dwarf companions in our sample, imaged with VLT/SPHERE, and to relate their orbital properties to potential formation and evolutionary pathways, highlighting possible correlations and trends. Our sample includes 13 substellar companions with dynamical masses between $\sim$7 and 70 M\textsubscript{Jup} and semimajor axes between $\sim$ 25 and 500 au. In Appendix \ref{app:hist}, Fig. \ref{histograms} shows the distributions of the companion masses, semimajor axis, fractional orbital coverage, and eccentricities. Because most of the companions lie in wide separations, the fractional orbital coverage is limited, leading to large uncertainties in the orbital parameters for several systems. Seven targets have less than 1\% of their orbits sampled, while only four show an orbital coverage greater than 3\% (Table \ref{fits_results}).

\subsection{Comparison with past works}
\label{subsec:pastworks}

For most objects, previous studies have already constrained their orbital and physical properties. Here we compare those results with our updated fits (Table \ref{fits_results}), highlighting agreements and differences.
\begin{itemize}
\item AB~Pic: orbital fits based on NACO and SPHERE astrometry yielded $a=190^{+200}_{-50}$\, au, $i=90\pm12$\, deg, and a poorly constrained moderate–high eccentricity \citep{PalmaBifani2023}. Our results agree within $1\sigma$, with smaller uncertainties from the extended baseline. We confirm an almost edge-on configuration ($\sim89$–$104$\, deg), though limited coverage ($\sim0.49$ \%) still leaves parameters like $e$ poorly constrained. The evolutionary mass of 8.3–12.6 M\textsubscript{Jup} \citep{PalmaBifani2025} is consistent with our estimates.
\item TYC 7084-794-1: \citet{Bowler2020} derived $a=241^{+85}_{-130}$\, au and $e=0.94^{+0.033}_{-0.026}$\, from DI astrometry.
Our derived semimajor axis agrees within $1\sigma$ with that reported by \citet{Bowler2020} only at the lower boundary of their uncertainty interval, while we retrieve a lower eccentricity. This discrepancy could be attributed to the inclusion of the differential RV measurement, published after \citet{Bowler2020}. Indeed, an orbital fit performed without the RV constraint yields results consistent with \citet{Bowler2020}.  Evolutionary masses range from $23$–$39$\, M\textsubscript{Jup} \citep{Wahhaj2011, Langlois2021, Kammerer2025}, while \citet{PalmaBifani2025} report $7.3\pm1.1$\, M\textsubscript{Jup}. Our estimate agrees with the higher values but differs by $>6\sigma$ from the latter. Our fits yield higher total system masses, but without independent dynamical constraints we cannot disentangle stellar and companion contributions. Improved astrometry (e.g., VLTI/GRAVITY; \citealt{GRAVITY}) and/or an extended orbital coverage is required.

\item TYC 8047-232-1: \citet{Ginski2013} found $a=196$–880 au and a broad $e$ distribution peaking at 0.997. With a decade-long extended baseline, we obtain tighter constraints. Previous masses (20–40 M\textsubscript{Jup}; \citealt{Neuhauser2003, Chauvin2005b}) are consistent within $\sim1.2\sigma$ with our lower, more precise estimate.

\item CT~Cha: evolutionary masses of $14$–24 M\textsubscript{Jup} \citep{Schmidt2008, Wu2015} place CT~Cha~b at the planet–brown dwarf boundary. Our estimate is lower, consistent with a planetary-mass companion.

\item GQ~Lup: \citet{Venkatesan2025}, combining DI, GRAVITY, and RVs \citep{Schwarz2016}, derived $a=97^{+9}_{-7}$\, au and $e=0.35^{+0.10}_{-0.09}$, consistent within $1\sigma$ with our values. Mass estimates of $\sim30$ and $26.4^{+2.9}_{-3.8}$\, M\textsubscript{Jup} \citep{Stolker2021, Kammerer2025} agree with ours, which provides an independent constraint.

\item HIP~78530: while no prior orbit exists, evolutionary masses of 19–26 and $31.4\pm10.5$\, M\textsubscript{Jup} \citep{Lafreniere2011, PalmaBifani2025} agree with our estimate within $\sim1\sigma$. Adopting the latter mass yields orbital parameters consistent with our nominal fit.

\item DH~Tau: \citet{Bowler2020} found $a=330^{+90}_{-160}$\, au with unconstrained $e$. Our results agree, providing slightly tighter constraints. Literature evolutionary masses (8–22 M\textsubscript{Jup}; \citealt{Itoh2005, Bonnefoy2014, Zhou2014, PalmaBifani2025}) are consistent with ours.

\item HIP~64892: evolutionary masses of $\sim29$–37 and $40.5^{+10.2}_{-11.7}$\, M\textsubscript{Jup} \citep{Cheetham2018, Kammerer2025} are both consistent with our slightly higher estimate within $1\sigma$.

\item RXJ1609.5-2105: previous estimates of $8^{+4}_{-2}$ and $14^{+2}_{-3}$\, M\textsubscript{Jup} \citep{Lafreniere2008, Pecaut2012} placed the companion in the planetary regime. Our value is slightly lower, consistent with the former and within $\sim2.5\sigma$ of the latter.

\item HII~1348: \citet{Weible2025} derived $a=140^{+130}_{-30}$\, au and $e=0.78^{+0.12}_{-0.29}$. Our posteriors are consistent, providing more precise constraints. Their mass estimate (60–63$\pm2$\, M\textsubscript{Jup}; consistent with \citealt{Geissler2012}) agrees with our slightly lower value within $\sim1.3$–$2.6\sigma$.

\begin{figure*}[ht!]
    \centering

    \begin{subfigure}{0.48\linewidth}
        \centering
        \includegraphics[width=\linewidth]{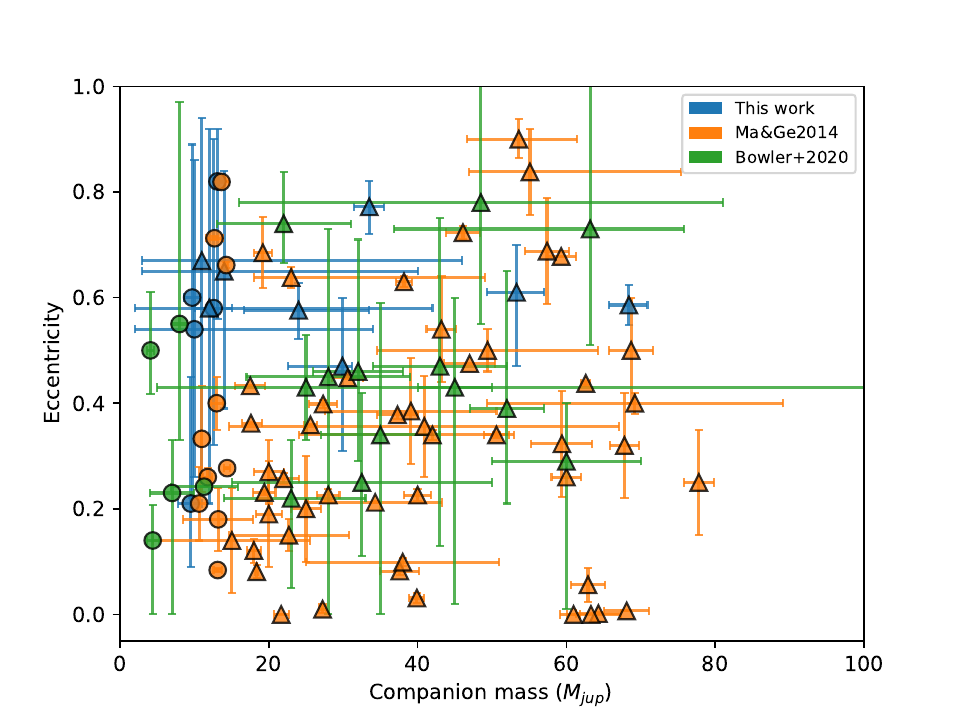}
    \end{subfigure}
    \hfill
    \begin{subfigure}{0.48\linewidth}
        \centering
        \includegraphics[width=\linewidth]{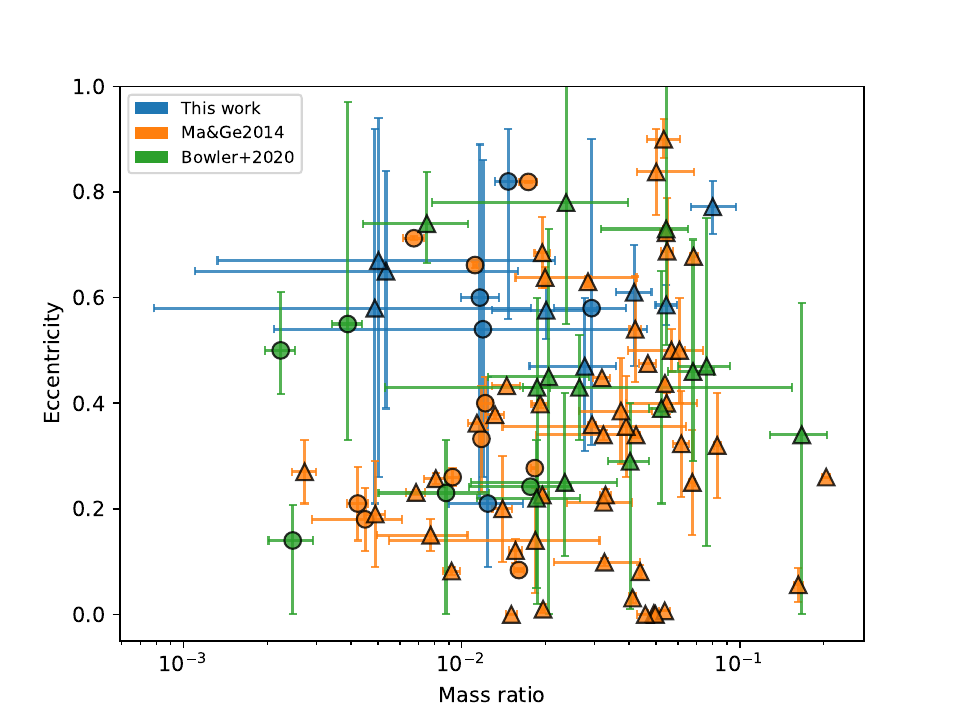}
    \end{subfigure}

    \vspace{0.4cm}

    \begin{subfigure}{0.48\linewidth}
        \centering
        \includegraphics[width=\linewidth]{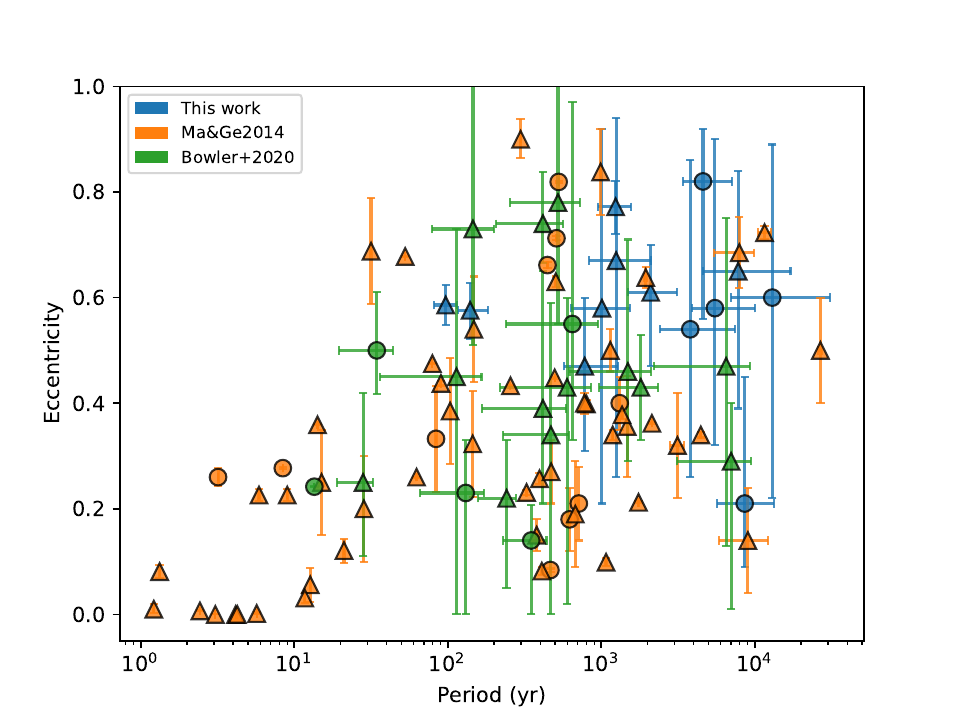}
    \end{subfigure}

    \caption{Top left: eccentricity distribution as a function of the companion mass. Top right: eccentricity distribution as a function of the mass ratio between the companion and the host star. Bottom: eccentricity distribution as a function of the orbital period.} Different colors indicate different surveys, while circles and triangles represent planets and brown dwarfs, respectively. For comparison, objects from \citet{Ma2014} and \citet{Bowler2020} are included.
    
    \label{ecc_plots}

\end{figure*}

\item PZ~Tel: Several studies have constrained the orbit of PZ~Tel~B. The most recent analysis by \citet{Franson2023}, combining astrometry, HGCA acceleration, and HARPS RVs \citep{Trifonov2020}, derived $a = 27^{+25}_{-9}$\, au, $e = 0.52^{+0.08}_{-0.10}$, and $i = 91.73^{+0.36}_{-0.32}$\, deg. Our results are consistent with these and earlier works (e.g., \citealt{Bowler2020, MussoBarcucci2019}), with reduced uncertainties thanks to the extended baseline and broader orbital coverage. We confirm a high eccentricity, in closer agreement with \citet{Franson2023} than with \citet{Bowler2020} or \citet{Beust2016}. The semimajor axis and inclination are consistent with previous estimates, supporting a near edge-on configuration. Mass estimates span $24$–$64$\, M\textsubscript{Jup} \citep{Biller2010, Mugrauer2012, Maire2016, Kammerer2025}, placing the object in the brown dwarf regime. Our derived dynamical mass agrees within $1\sigma$ with \citet{Franson2023}, who retrieved $27^{+25}_{-9}$\, M\textsubscript{Jup}, with reduced uncertainties.

\item HD~984: \citet{Franson2022} combined astrometry, proper motions, and RVs \citep{Grandjean2020, Franson2022}, deriving $a = 28^{+7}_{-4}$\, au, $e = 0.76\pm0.05$, and $i = 120.8^{+1.8}_{-1.6}$\, deg. Our results are consistent with these and earlier studies (e.g., \citealt{Bowler2020}), with reduced uncertainties due to the extended baseline and a significant orbital motion since the last epoch. We find a semimajor axis consistent with \citet{Franson2022} but larger than \citet{Bowler2020}, an inclination consistent with both, and an eccentricity slightly lower than \citet{Franson2022} but higher than \citet{Bowler2020}. These differences reflect improved orbital coverage and the inclusion of astrometric acceleration. The \citet{Franson2022} dynamical mass ($61\pm4$ M\textsubscript{Jup}) agrees with our estimates.

\item $\eta$~Tel: \citet{Chai2024}, combining relative and HGCA astrometry, derived $a = 142^{+18}_{-11}$\, au and $e = 0.50\pm0.10$. Our results are consistent, though still weakly constrained due to the long orbital period and limited coverage. Consequently, the dynamical mass remains poorly determined, with estimates of $42$–$55$\, M\textsubscript{Jup} \citep{Nogueira2024, Chai2024} and $11^{+35}_{-8}$\, M\textsubscript{Jup} from our analysis. As noted by \citet{Chai2024}, the posteriors remain largely prior-driven.
\end{itemize}

\subsection{Eccentricities distributions of brown dwarfs and giant planets}
\label{subsec:ecc}

To investigate differences in eccentricity between giant planets and brown dwarfs, we divided the sample at 15 M\textsubscript{Jup}, close to the deuterium-burning limit \citep{Spiegel2011}, yielding 5 planets and 8 brown dwarfs. Before comparing with other surveys or inferring correlations, we assess the statistical significance of our sample. Unlike larger surveys \citep{Ma2014, Bowler2020}, it comprises only 13 objects, and most eccentricities remain poorly constrained due to limited orbital coverage. Consequently, the observed trends may not reflect intrinsic distributions. However, our sample probes wider separations, enabling tests of whether previously reported trends persist in this regime. The \citet{Bowler2020} sample covers 5–100 au, while \citet{Ma2014} includes periods of $\sim1.2$–9000 days (excluding HIP~78530). In contrast, our closest companion is HD~984~B ($a \sim 22.5$\, au, $P\sim97$\, yr), with ten systems beyond 100 au and four exceeding 250 au.

Figure \ref{ecc_plots} (upper left) shows eccentricity as a function of companion mass for our sample, compared with \citet{Bowler2020} and \citet{Ma2014}. In contrast to these studies, we found no clear distinction between giant planets and brown dwarfs, with both spanning a broad range of eccentricities ($\sim$0.2–0.8). Given that this trend is well established in larger samples, we attribute the difference to our limited sample size, which can bias the reconstructed distribution \citep{Bowler2020}. We also marginally detected a positive correlation between eccentricity and mass ratio (Fig. \ref{ecc_plots}, upper right), as reported by \citet{Bowler2020}, although with lower significance.

One open question from \citet{Bowler2020} is whether eccentricity depends on separation. They found a marginal correlation, more evident for brown dwarfs. Our results (Fig. \ref{ecc_plots}, bottom) are $\bar{e}\approx0.61^{+0.15}_{-0.18}$ for $P<2000$\, yr and $\bar{e}\approx 0.57^{+0.21}_{-0.24}$ at longer periods. These values are consistent within $1\sigma$, preventing confirmation of the trend. This limitation arises from the modest sample size and limited orbital coverage. We therefore conclude that a larger sample and improved orbital coverage are required to better constrain the eccentricity behavior at wide separations.

\begin{figure*}[ht!]
 \centering
 \includegraphics[width=0.45\linewidth]{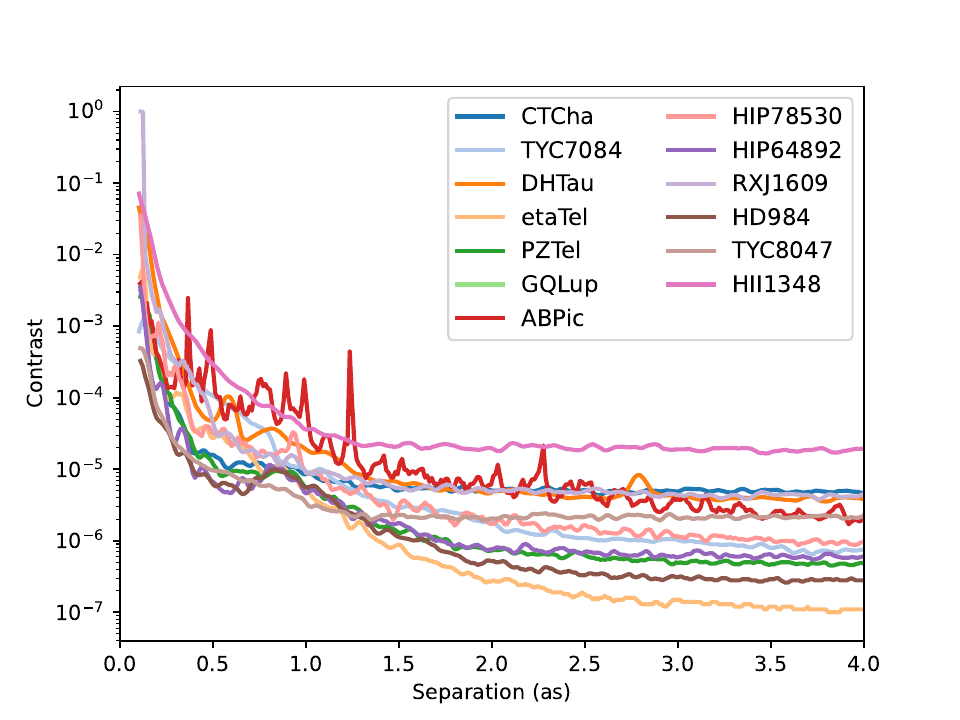}
 \includegraphics[width=0.45\linewidth]{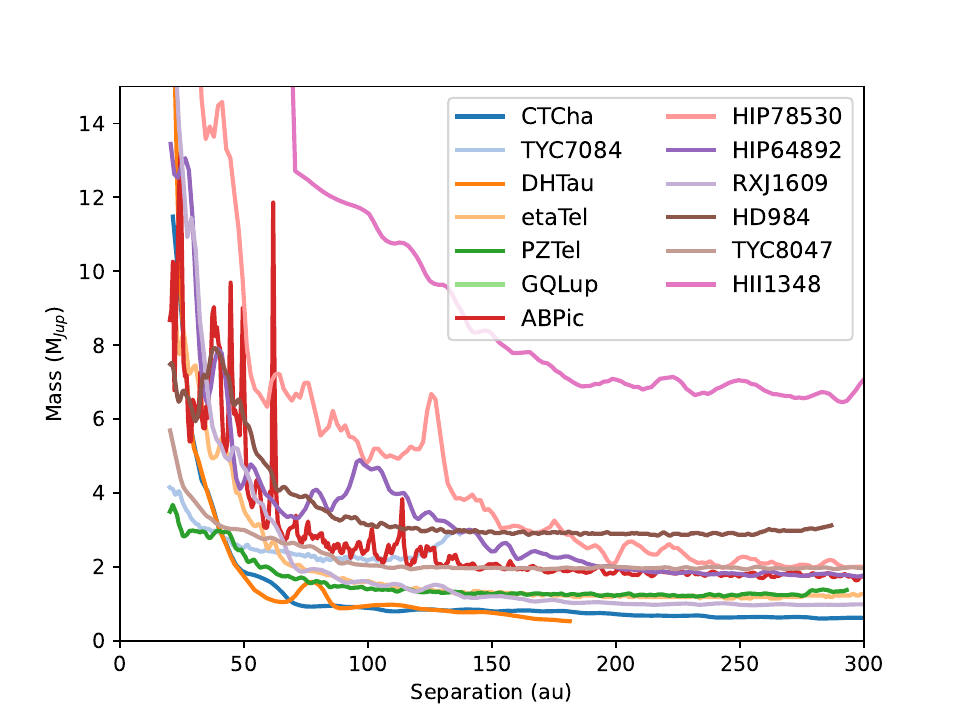}
 \caption{Left: 5$\sigma$ contrast curves in separation versus contrast relative to the host star. Right: corresponding detection limits converted into mass using ATMO evolutionary models.}
    \label{contrasts}

\end{figure*}

\subsection{Possible formation and evolution pathways}
\label{subsec:evol}
The formation pathways of planets and brown dwarfs are closely linked to observable signatures such as their orbital architecture (e.g., \citealt{Boley2009}), including their eccentricity \citep{Marleau2019}. The orbital eccentricity is indeed a key parameter that can provide insights into the formation and subsequent dynamical evolution of these objects. Currently, two main scenarios are thought to describe the formation of planets and stars: core accretion formation within a circumstellar disk \citep{Pollack1996}, typically associated with low eccentricities (up to $\sim0.03$, e.g., \citealt{DAngelo2014,Dong2016,Johansen2017}), and gravitational collapse and fragmentation of a molecular cloud or a disk \citep{Boss1997}, generally resulting in higher orbital eccentricities (up to $\sim0.7$, due to a longer timescale for eccentricity damping, e.g., \citealt{Mayer2004}). Even when accounting for the uncertainties, all our retrieved eccentricities are consistent with non-circular orbits for nearly all substellar companions in our sample. This behavior could be explained by non-core accretion mechanisms, with these objects possibly originating from the fragmentation and collapse of a molecular cloud, representing the low–mass-ratio end of the stellar formation process (e.g., \citealt{Bate2002}).

Even after their formation, additional dynamical processes could have shaped the observed orbital architectures. The presence of substellar companions on highly eccentric orbits may also be explained by gravitational interactions among planets within the same system, leading to planet–planet scattering events (e.g., \citealt{Weidenschilling1996}). These interactions depend strongly on planetary separation and system multiplicity. For two-planet systems, Hill stability requires separations approximately larger than $2\sqrt{3}$ mutual Hill radii under low eccentricity and inclination assumptions \citep{Chambers1996}. For higher multiplicities, no general analytical criterion exists, but numerical integrations show that the time to first close encounter increases approximately exponentially with mutual separation (e.g., \citealt{Chambers1996, Pu2015}). \citet{Veras2009} simulated multi-planet systems and found that two-planet configurations are the most common outcome, with most encounters, collisions, and ejections occurring within $\lesssim100$\, Myr. For an initial spacing of 4.2 mutual Hill radii, the median time to the first encounter is $\sim9\times10^{3}$\, yr and the mean $\sim1$\, Myr. The resulting evolution can produce eccentricities up to $\sim0.8$ and separations of thousands of au.

Efficient planet–planet scattering requires that the circumstellar gas disk has dissipated. Protoplanets embedded in gas experience strong eccentricity damping, often dominating over mutual perturbations and maintaining near-circular orbits \citep{Cresswell2008}. Observations indicate that gas-rich disks typically dissipate within $\sim$3–5 Myr, with few surviving to $\sim10$\, Myr \citep{Fedele2010}. Systems younger than or comparable to this timescale (CT~Cha, GQ~Lup, DH~Tau, RX~J1609.5-2105) may therefore not have undergone prolonged gas-free evolution, limiting scattering efficiency. In contrast, older systems in our sample (up to $\sim200$ Myr) allow sufficient time for dynamical interactions, provided initially compact configurations (a few mutual Hill radii). In such cases, gravitational scattering offers a viable alternative to cloud or disk fragmentation in explaining wide separations and high eccentricities.

These interactions may also involve additional companions: AB~Pic may host another planet, while DH~Tau~b and TYC~8047-232-1~B could host massive satellites \citep{Lazzoni2026}, as well as other undetected bodies. For this purpose, we derived 5$\sigma$ contrast curves for each system in our sample, expressed as contrast versus angular separation from the host star (Fig. \ref{contrasts}, left). The corresponding detection limits are shown in Fig. \ref{contrasts}, right, obtained by converting each contrast curve into mass using the mean ages and distances listed in Table \ref{masses}, together with the ATMO evolutionary models \citep{Phillips2020}. In addition to the already known companions, our observations allow us to rule out the presence of other planets around the stars in our sample at separations up to $\sim$300 au, within a mass range of approximately 1–15 Jupiter masses. We therefore conclude that any gravitational interactions shaping the observed orbits must involve either lower-mass planets or satellites, or more massive objects located at smaller separations, regions that, in our observations, fall beneath the coronagraph.

Additional mechanisms have also been proposed to explain the presence of planets and brown dwarfs at wide separations, including pebble accretion (e.g., \citealt{Johansen2017}), gas-disk migration (e.g., \citealt{Tanaka2002}), and the dynamical recapture of free-floating planets \citep{Perets2012}.

\section{Conclusions}
\label{sec:conclusions}

In this study, we present the photometric and orbital analysis of 13 planets and brown dwarfs from a VLT/SPHERE star-hopping survey. All systems were observed between June 2023 and July 2025, acquiring dual-band IRDIS $H2H3$ images and low-resolution IFS spectra in the $Y$ and $J$ bands. The survey, along with the search for satellites and circumplanetary disks, is presented by \citet{Lazzoni2026}.

Our photometric analysis indicates spectral types from mid-M to mid-L, based on color–magnitude diagrams. For the three objects within the IFS field of view, comparison with empirical templates yields refined estimates: GQ~Lup~B (M8–M8.5), PZ~Tel~B (M5–M6.5), and HD~984~B (M7.75–M8.5). The values for PZ~Tel~B and HD~984~B agree with previous studies, while GQ~Lup~B is consistent with some works (e.g., \citealt{Stolker2021}) but differs from \citet{Kammerer2025}.

The extended astrometric baseline improves orbital constraints for nearly all companions, with significantly reduced uncertainties. However, several companions still have poorly constrained orbital parameters, mainly due to limited orbital coverage (e.g., CT~Cha~b, HIP~78530~B, RX~J1609.5-2105~b). This also affects the derived dynamical masses. With the exception of PZ~Tel~B and HD~984~B, which have the most extended coverage in our sample, the dynamical masses of all other companions remain essentially unconstrained. This highlights the impact of poor orbital coverage and emphasizes the need for additional observations to better constrain the properties of these systems. For four objects (CT~Cha~b, HIP~78530~B, HIP~64892~B, and RX J1609.5–2105 b), we provide the first orbital solutions. All companions exhibit non-circular orbits, with eccentricities ranging from $\sim$ 0.21 to 0.82. Since low-eccentricity orbits are typically associated with core accretion, these results favor alternative formation pathways, such as molecular cloud fragmentation or disk gravitational instability, consistent with the wide separations. We also examined whether dynamical interactions could account for the observed non-zero eccentricities. Planet-planet scattering has been shown to efficiently excite eccentricities up to $\sim0.8$ in initially compact systems, provided that sufficient time has passed for the dissipation of the gas disk. Such interactions may involve additional companions, either ejected during dynamical scattering events or still present in the system, and several targets in our sample are known to host candidate planets or satellites.

We derived 5$\sigma$ contrast curves, converted into mass limits, showing that planets of $\sim$1–15 M\textsubscript{Jup} are excluded out to $\sim$300 au (apart from the detected companions). This suggests that any dynamical interactions must involve either lower-mass planets/satellites or more massive companions at smaller separations, beyond current detection limits.

Future facilities such as the Planetary Camera and Spectrograph (PCS; \citealt{Kasper2021}) on the ELT will provide major improvements, reaching contrasts of $10^{-9}$–$10^{-10}$ and angular resolutions of $\lambda/D \sim 0.008$ arcsec in $H2$, probing sub-au separations in nearby systems. This will enable the detection of much fainter companions and tighter constraints on system architectures. Complementary techniques such as long-baseline interferometry, with instruments like VLTI/GRAVITY and its GRAVITY+ upgrade \citep{GRAVITY+}, offer astrometric precision of $\sim$50 $\mu$as, allowing direct detection of companions at $\sim$1–10 au and opening new parameter space for substellar companions and potential exosatellites.

\begin{acknowledgements}
The authors thank the referee for their comments and suggestions, which have significantly improved the manuscript. We acknowledge support from ANID -- Millennium Science Initiative Program -- Center Code NCN2024\_001. A.B. and A.Z. acknowledge support from Fondecyt Regular grant number 1250249. S.D. and C.L. gratefully acknowledge support from the “Programma di Ricerca Fondamentale INAF 2023” of the Italian National Institute of Astrophysics (INAF Large Grant 2023 “NextSTEPS”). This work has made use of the High Contrast Data Centre, jointly operated by OSUG/IPAG (Grenoble), PYTHEAS/LAM/CeSAM (Marseille), OCA/Lagrange (Nice), Observatoire de Paris/LESIA (Paris), and Observatoire de Lyon/CRAL, and supported by a grant from Labex OSUG@2020 (Investissements d’avenir – ANR10 LABX56). S.P. acknowledges support from FONDECYT 1231663 and FIUF137139-USACH. This paper makes use of the observations collected at the European Southern Observatory under ESO programs 111.24UH.001, 112.25QU.001, 113.26GK.001, 114.276W.001, 095.C-0298(H), 198.C-0209(F), 1100.C-0481(K), 096.C-0241(C), 096.C-0241(A), 097.C-0865(C), 095.C-0298(A), 198.C-0209(D), 096.C-0241(E), 096.C-0241(G), 096.C-0241(B), and 095.C-0298(D)

\end{acknowledgements}
\bibliographystyle{aa}
\bibliography{bibliography}
\begin{appendix}

\begin{figure}[h!]
\section{Orbital fits}
\centering

\begin{subfigure}{0.5\textwidth}
  \centering
  \includegraphics[width=\linewidth]{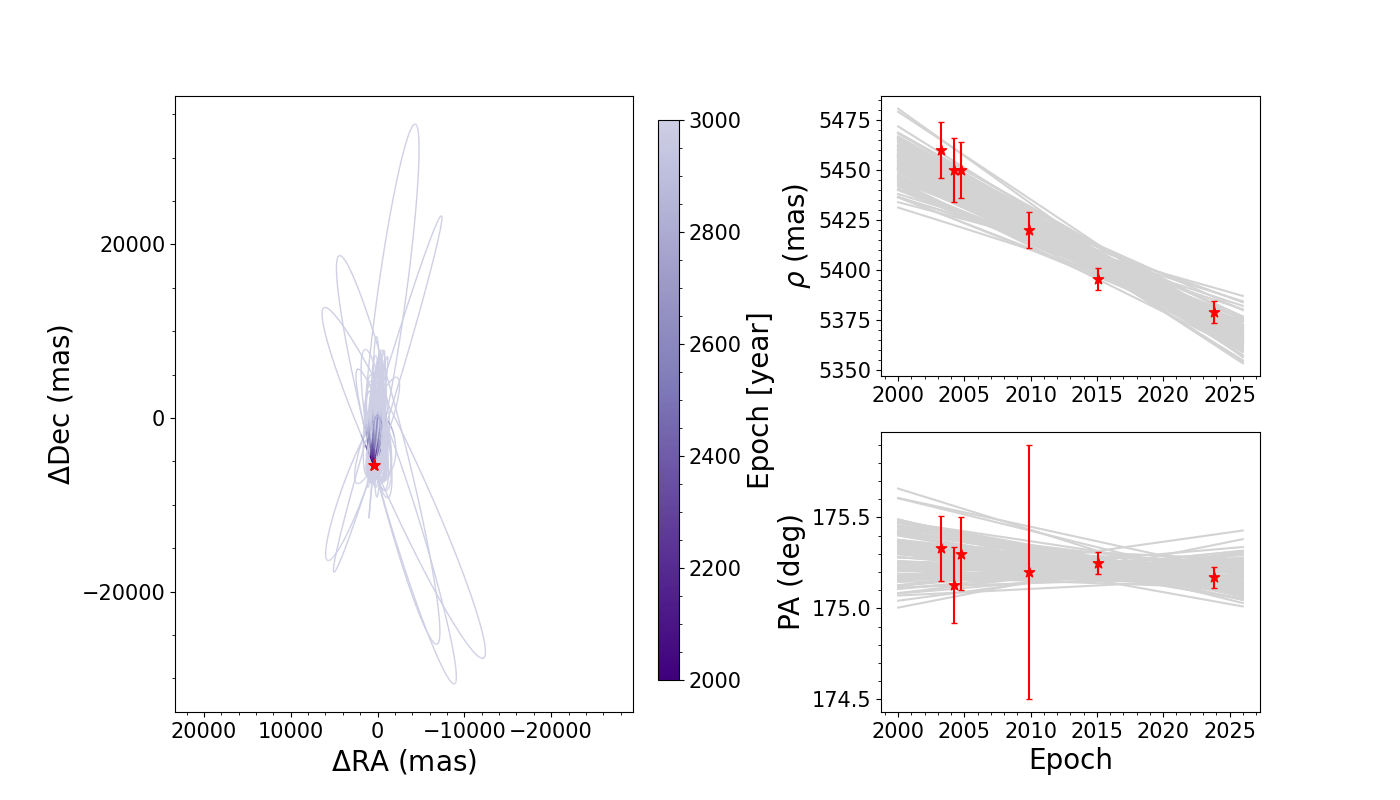}
  \caption{AB~Pic~b}
\end{subfigure}
\begin{subfigure}{0.50\textwidth}
  \centering
  \includegraphics[width=\linewidth]{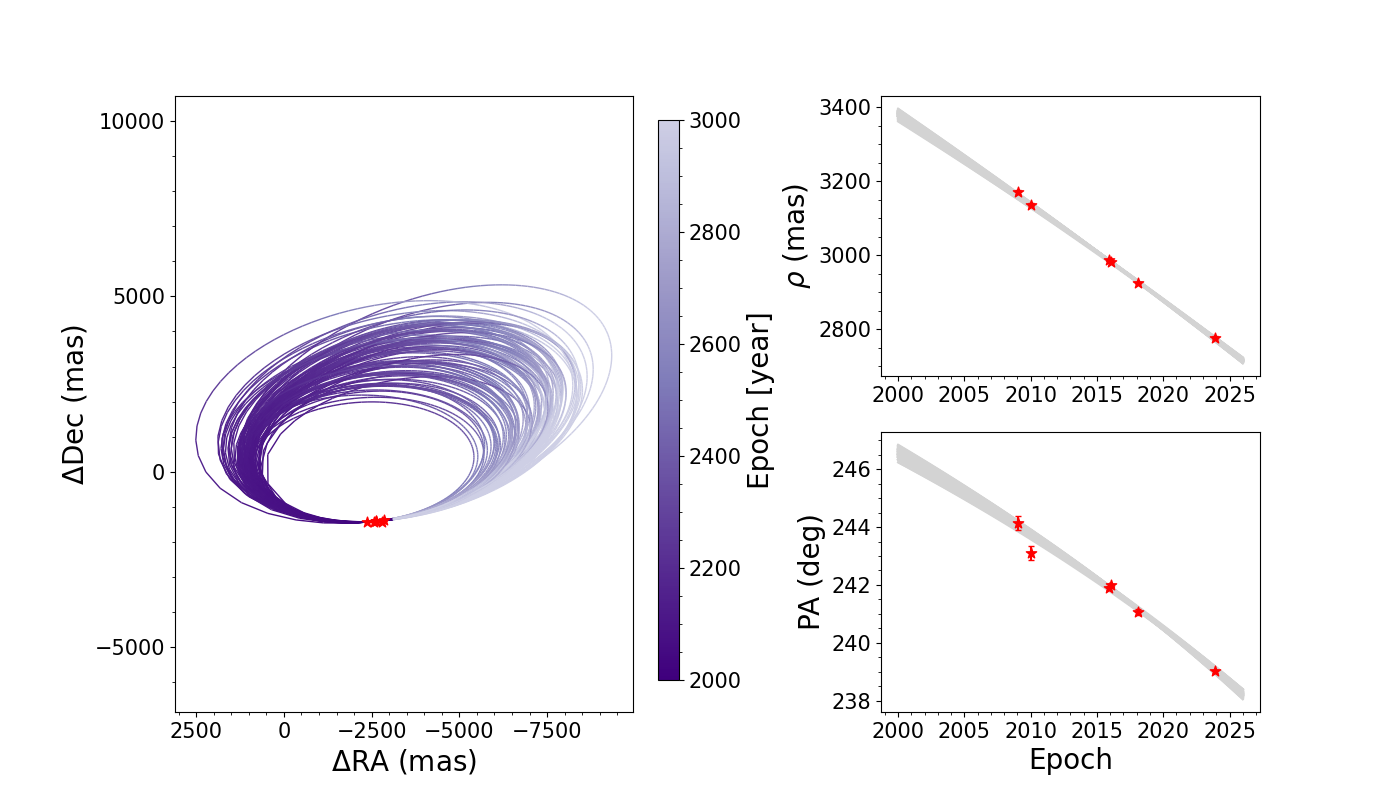}
  \caption{TYC~7084-794-1~B}
\end{subfigure}

\begin{subfigure}{0.50\textwidth}
  \centering
  \includegraphics[width=\linewidth]{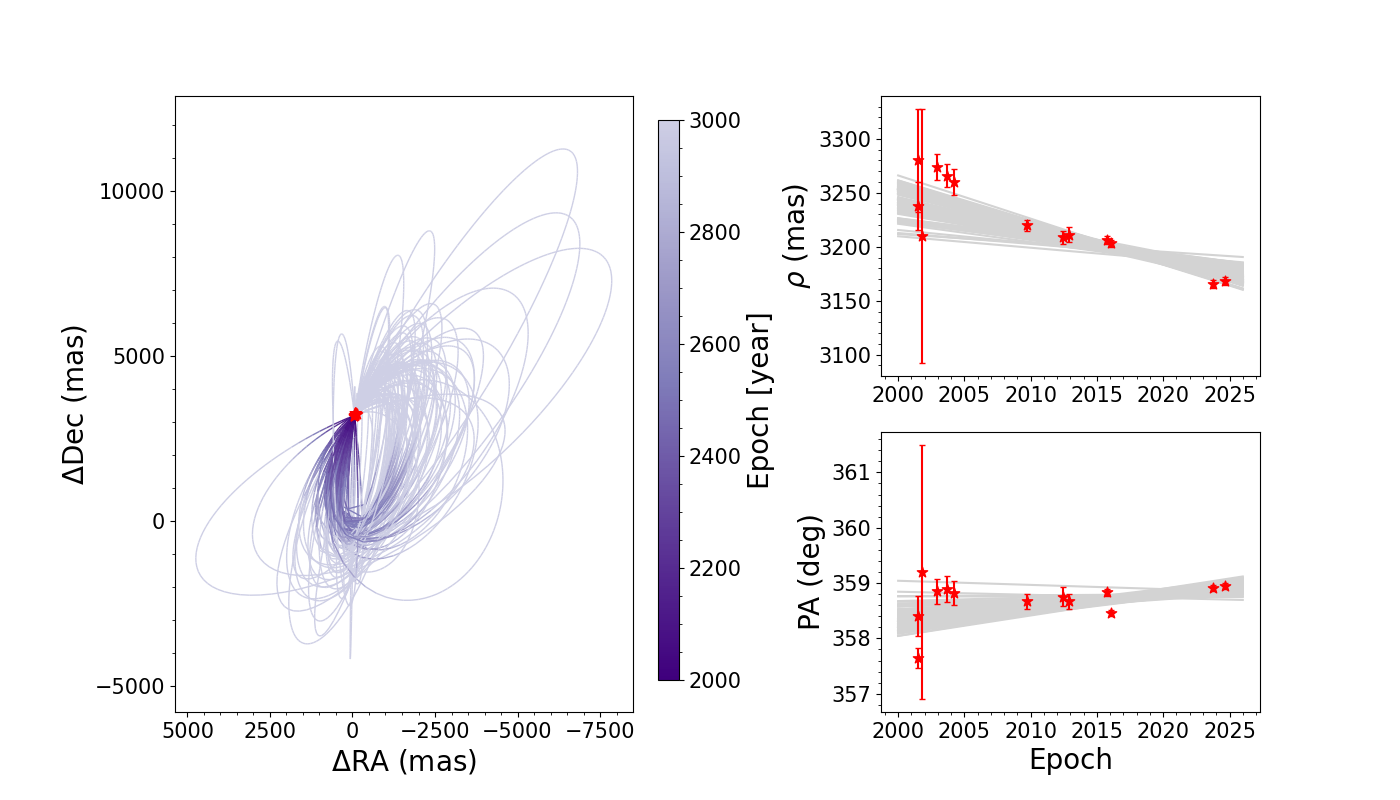}
  \caption{TYC~8047-232-1~B}
\end{subfigure}
\begin{subfigure}{0.50\textwidth}
  \centering
  \includegraphics[width=\linewidth]{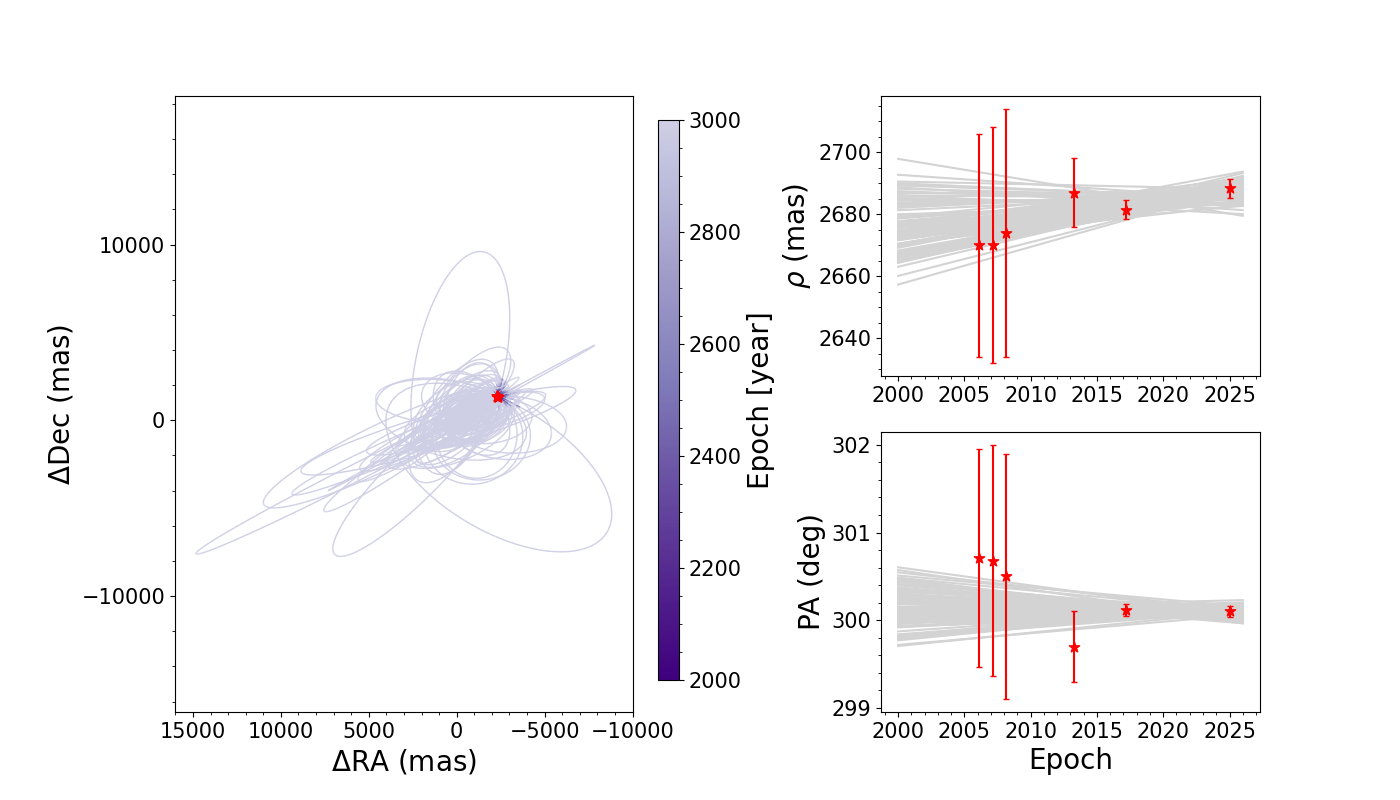}
  \caption{CT~Cha~b}
\end{subfigure}

\caption{Orbital fits for AB~Pic~b (a), TYC~7084-794-1~B (b), TYC~8047-232-1~B (c), and CT~Cha~b (d).}
\label{orbit_fits1}
\end{figure}

\begin{figure}
\centering

\begin{subfigure}{0.50\textwidth}
  \centering
  \includegraphics[width=\linewidth]{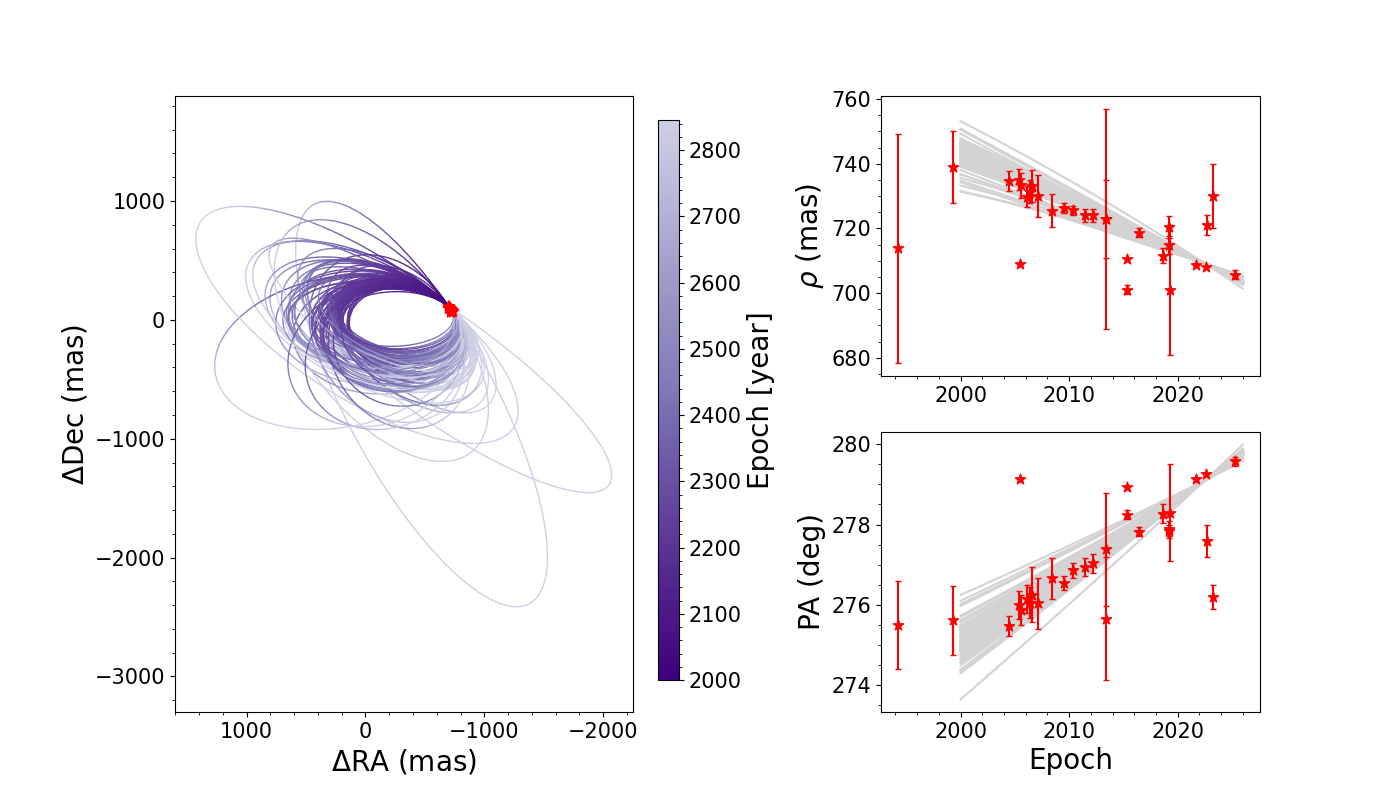}
  \caption{GQ~Lup~B}
\end{subfigure}
\begin{subfigure}{0.50\textwidth}
  \centering
  \includegraphics[width=\linewidth]{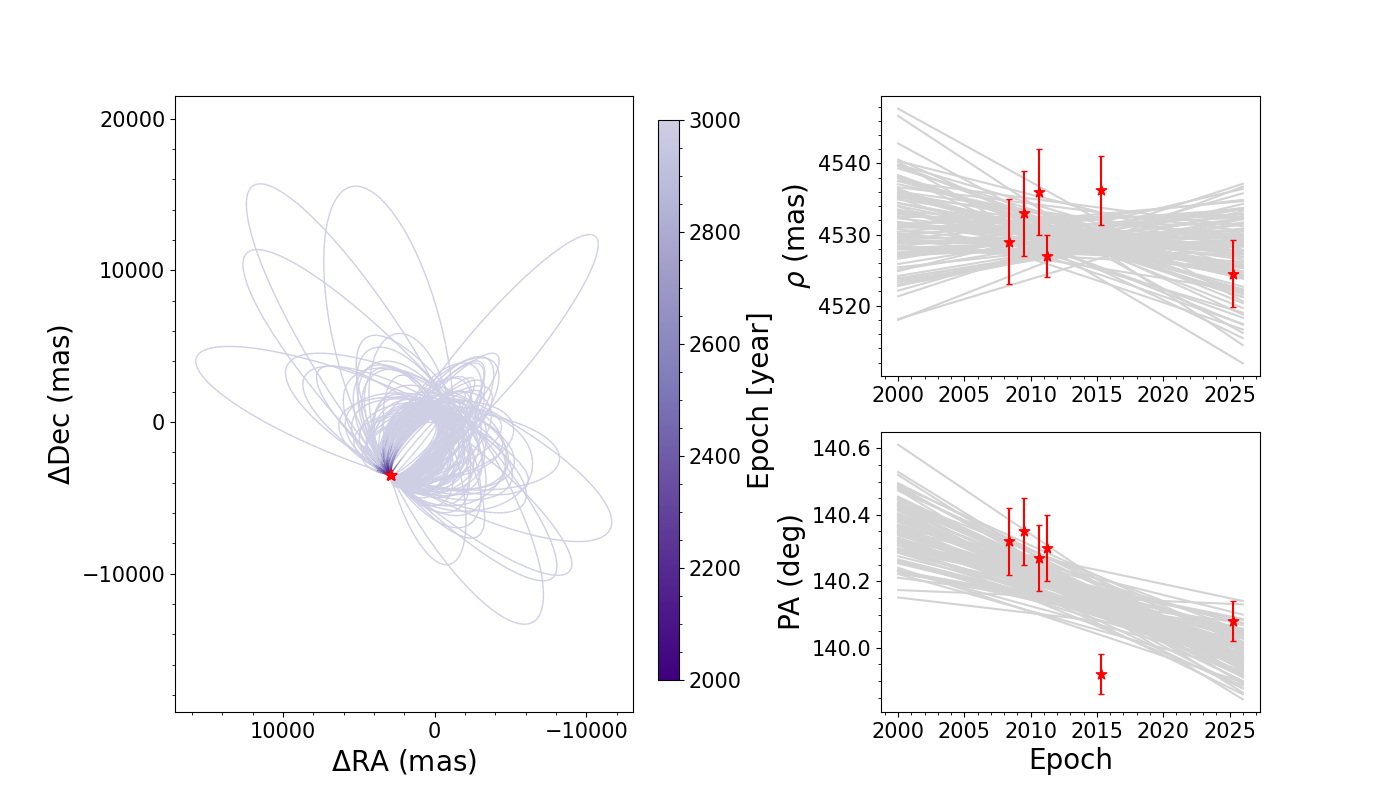}
  \caption{HIP~78530~B}
\end{subfigure}

\begin{subfigure}{0.50\textwidth}
  \centering
  \includegraphics[width=\linewidth]{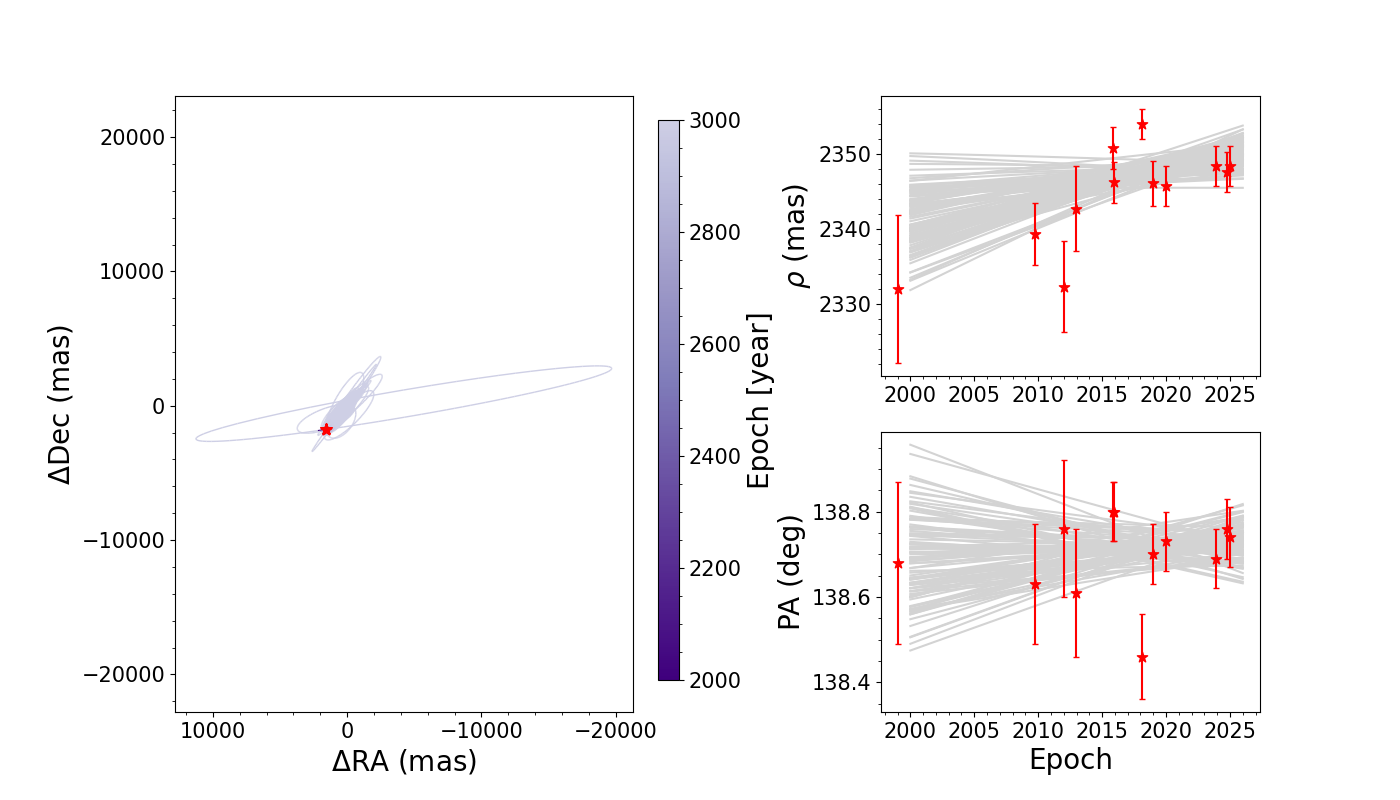}
  \caption{DH~Tau~b}
\end{subfigure}
\begin{subfigure}{0.50\textwidth}
  \centering
  \includegraphics[width=\linewidth]{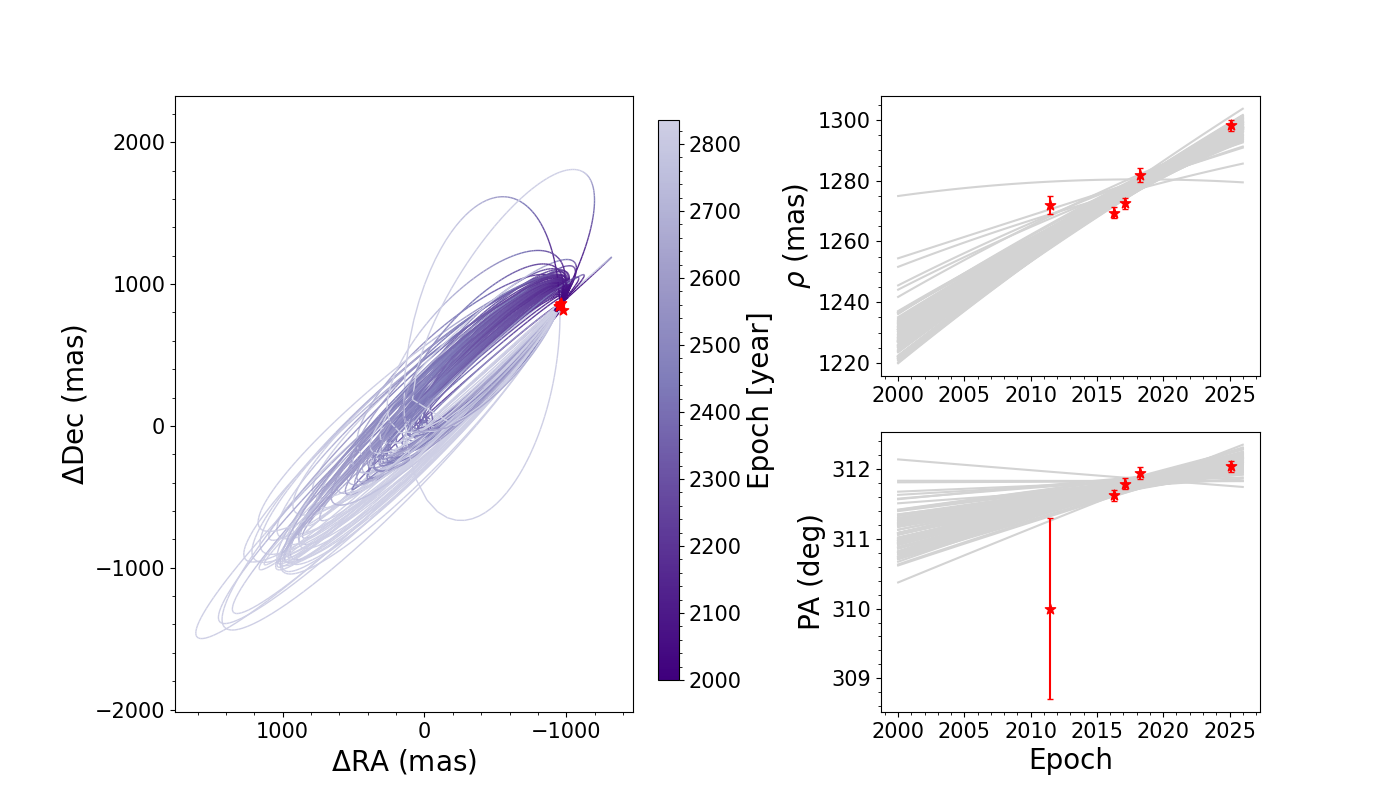}
  \caption{HIP~64892~B}
\end{subfigure}

\caption{Orbital fits for GQ~Lup~B (a), HIP~78530~B (b), DH~Tau~b (c), and HIP~64892~B (d).}
\label{orbit_fits2}
\end{figure}

\begin{figure}
\centering

\begin{subfigure}{0.50\textwidth}
  \centering
  \includegraphics[width=\linewidth]{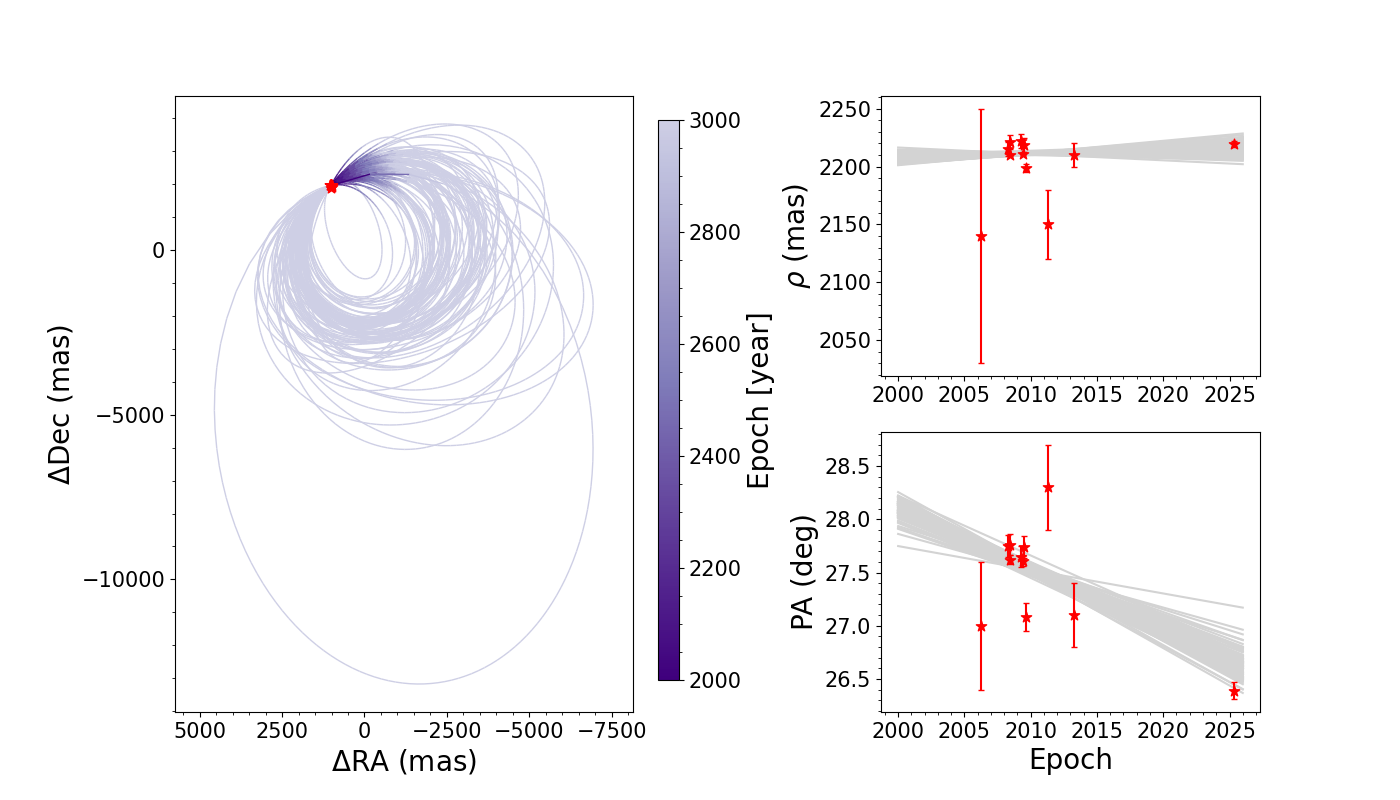}
  \caption{RX~J1609.5-2105~b}
\end{subfigure}
\begin{subfigure}{0.50\textwidth}
  \centering
  \includegraphics[width=\linewidth]{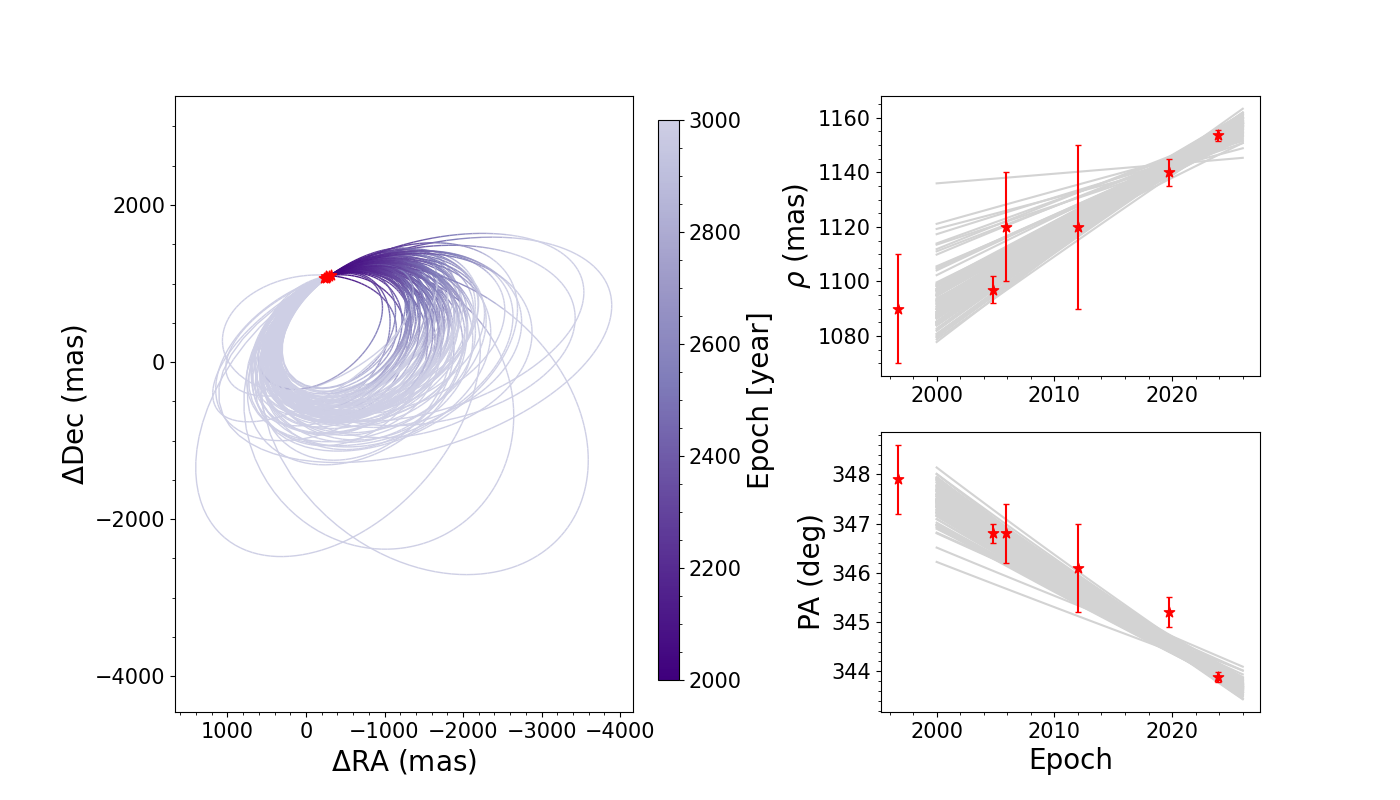}
  \caption{HII~1348~B}
\end{subfigure}

\begin{subfigure}{0.50\textwidth}
  \centering
  \includegraphics[width=\linewidth]{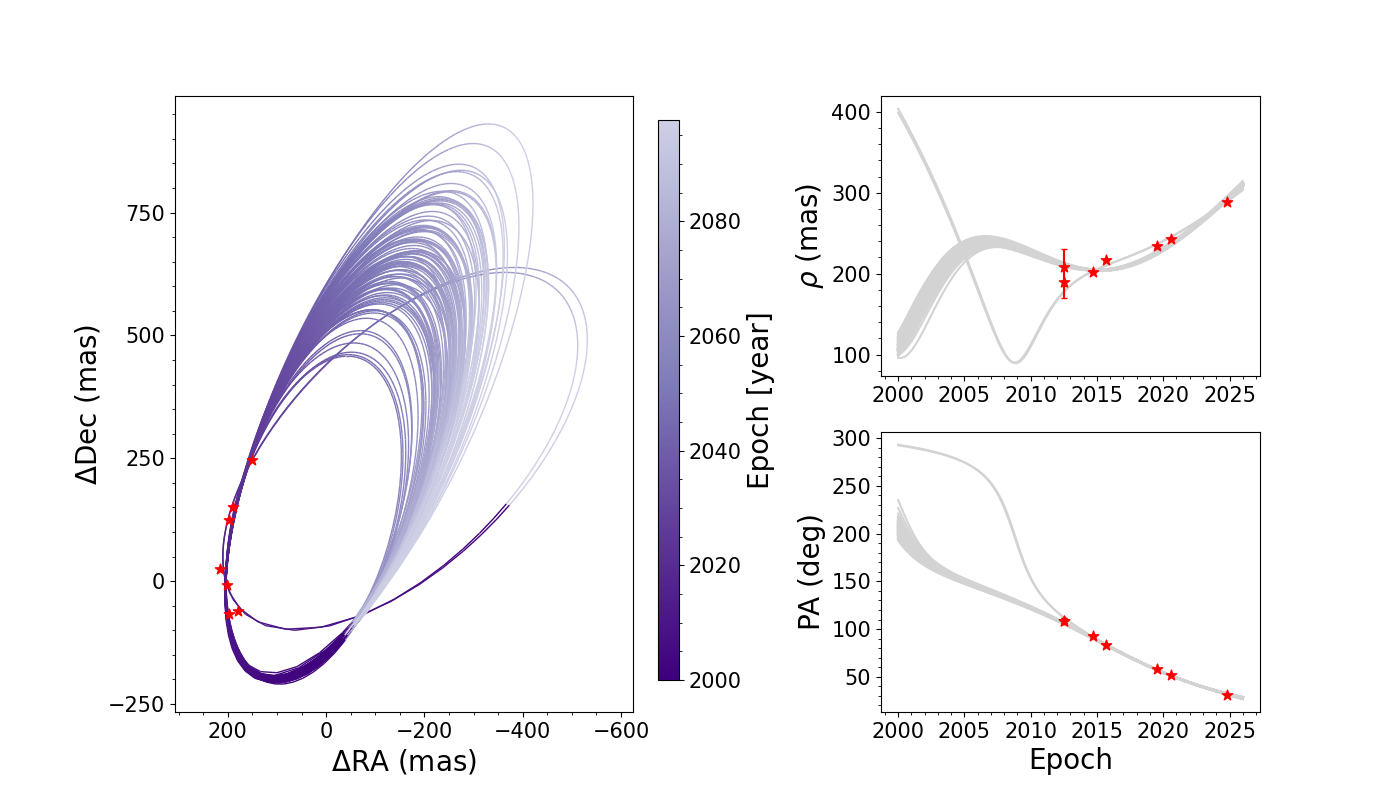}
  \caption{HD~984~B}
\end{subfigure}
\begin{subfigure}{0.50\textwidth}
  \centering
  \includegraphics[width=\linewidth]{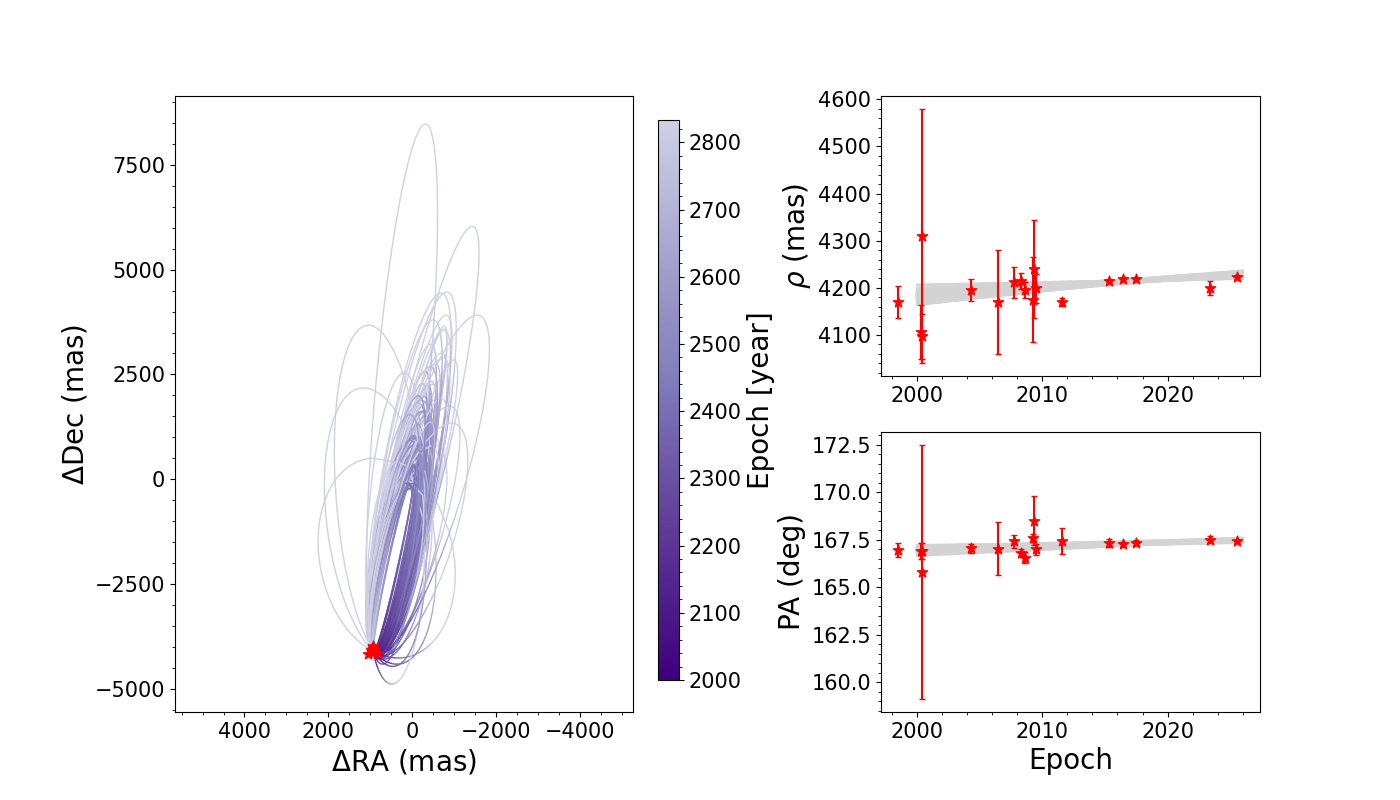}
  \caption{$\eta$~Tel~B}
\end{subfigure}

\caption{Orbital fits for RX~J1609.5-2105~b (a), HII~1348~B (b), HD~984~B (c), and $\eta$~Tel~B (d).}
\label{orbit_fits3}
\end{figure}
\FloatBarrier

\section{Orbital parameters}
\renewcommand{\arraystretch}{1.5}
\begin{sidewaystable*}
\centering
\caption{Orbital parameter solutions for the 13 substellar companions of our survey.}
\resizebox{\textheight}{!}{
\begin{tabular}{lcccccccccccccc}
\hline
\hline
Name & $a$ & $e$ & $i$ & $\omega$ & $\Omega$ & $\tau$ & $\pi$ & $\gamma$ & $\sigma$ & $M_\mathrm{tot}$ & $M_\star$ & $M_\mathrm{comp}$ & $P$ & \makecell{Orbital \\ coverage} \\ [0.75ex]
     & (au) & & (deg) & (deg) & (deg) & & (mas) & (km s$^{-1}$) & (km s$^{-1}$) & (M\textsubscript{$\odot$}) & (M\textsubscript{$\odot$}) & (M\textsubscript{Jup}) & (yr) & (\%) \\
\hline \\
AB~Pic~b & $237^{+115}_{-55}$ & $0.54^{+0.32}_{-0.28}$  & $95.2^{+8.3}_{-6.5}$ & $199^{+34}_{-39}$ & $177.6^{+8.1}_{-3.9}$ & $0.36^{+0.14}_{-0.11}$ & $19.945\pm0.012$ & ... & ... & ...& $0.90\pm0.10$ & $10^{+24}_{-8}$ & $3800^{+3600}_{-1400}$ & $0.54^{+0.32}_{-0.26}$   \\ [0.75ex]
TYC~7084-794-1~B & $101^{+15}_{-14}$ & $0.772^{+0.049}_{-0.051}$ & $141^{+11}_{-7}$ & $142.2^{+7.6}_{-6.6}$ & $259.6^{+3.4}_{-7.4}$ & $0.066^{+0.022}_{-0.012}$ & $44.722\pm0.013$ & ... & ... &  $0.678^{+0.068}_{-0.042}$ & ... & ... & $1240^{+330}_{-290}$ & $1.18^{+0.36}_{-0.25}$  \\ [0.75ex] 
TYC~8047-232-1~B & $265^{+80}_{-43}$ & $0.82^{+0.10}_{-0.26}$ & $52^{+21}_{-33}$ & ($39$ or $219$)$^{+96}_{-30}$ & ($135$ or $315$)$^{+29}_{-121}$ & $0.17^{+0.12}_{-0.05}$ & $11.543\pm0.012$ & ... & ... & $0.90^{+0.10}_{-0.08}$ & ... & ... & $4600^{+2500}_{-1200}$ & $0.50\pm0.18$ \\ [0.75ex]
CT~Cha~b & $520^{+380}_{-170}$ & $0.60^{+0.29}_{-0.38}$ & $89^{+32}_{-34}$ & ($57$ or $237$)$^{+55}_{-39}$ & ($119$ or $299$)$^{+14}_{-36}$ & $0.75^{+0.15}_{-0.20}$ & $5.265\pm0.011$ & ... & ... & $0.807^{+0.050}_{-0.051}$ & ... & ... & $13000^{+18000}_{-6000}$ & $0.15^{+0.12}_{-0.09}$   \\ [0.75ex]
GQ~Lup~B & $90^{+29}_{-10}$ & $0.47^{+0.13}_{-0.16}$ & $50^{+7}_{-14}$ & $198^{+40}_{-58}$ & $88^{+26}_{-31}$ or $256^{+12}_{-16}$ & $0.36^{+0.07}_{-0.12}$ & $6.494^{+0.031}_{-0.026}$ & ... & ... & $1.19^{+0.39}_{-0.19}$ & ... & ... & $780^{+510}_{-210}$ & $4.0^{+1.5}_{-1.6}$  \\ [0.75ex]
HIP~78530~B & $550^{+360}_{-160}$ & $0.65^{+0.19}_{-0.26}$ & $122^{+19}_{-14}$ & ($95$ or $275$)$^{+49}_{-28}$ & ($82$ or $262$)$^{+54}_{-57}$ & $0.41^{+0.12}_{-0.15}$ & $7.433\pm0.028$ & ... & ... & ... & $2.69\pm0.14$ & $14^{+26}_{-11}$ & $7800^{+9300}_{-3200}$ & $0.22^{+0.15}_{-0.12}$   \\ [0.75ex]
DH~Tau~b & $234^{+102}_{-41}$ & $0.58^{+0.32}_{-0.26}$ & $88^{+11}_{-13}$ & ($124$ or $304$)$^{+32}_{-34}$ & ($138$ or $318$)$^{+8.4}_{-21}$ & $0.55^{+0.13}_{-0.14}$ & $7.493\pm0.023$ & ... & ... & $0.418^{+0.037}_{-0.039}$ & ... & ... & $5500^{+4500}_{-1600}$ & $0.47^{+0.19}_{-0.21}$  \\ [0.75ex]
HIP~64892~B & $134^{+41}_{-33}$ & $0.58^{+0.34}_{-0.37}$ & $79^{+4}_{-14}$ & ($50$ or $230$)$^{+45}_{-27}$ & ($131$ or $311$)$^{+3}_{-10}$ & $0.76^{+0.10}_{-0.08}$ & $8.358\pm0.048$ & ... & ... & ... & $2.346^{+0.087}_{-0.086}$ & $12^{+30}_{-10}$ & $1010^{+520}_{-370}$ & $1.36^{+0.78}_{-0.46}$ \\ [0.75ex]
RX~J1609.5-2105~b & $385^{+110}_{-78}$ & $0.21^{+0.14}_{-0.12}$ & $150^{+13}_{-12}$ & ($58$ or $238$)$^{+52}_{-33}$ & ($135$ or $315$)$^{+27}_{-86}$ & $0.86^{+0.09}_{-0.62}$ & $7.245^{+0.016}_{-0.015}$ & ... & ... & $0.77^{+0.11}_{-0.08}$ & ... & ... & $8600^{+4800}_{-2900}$ & $0.22^{+0.11}_{-0.08}$ \\ [0.75ex]
HII~1348~B & $181^{+49}_{-32}$ & $0.61^{+0.09}_{-0.14}$ & $
149^{+15}_{-17}$ & ($39$ or $219$)$^{+27}_{-23}$ & ($141$ or $321$)$^{+18}_{-16}$ & $0.878^{+0.043}_{-0.049}$ & $6.981\pm0.030$ & ... & ... & $1.316^{+0.098}_{-0.089}$ & ... & ... & $2100^{+1000}_{-600}$ & $1.30^{+0.51}_{-0.42}$  \\ [0.75ex]
PZ~Tel~B &  $28.3^{+4.6}_{-2.6}$ & $0.576^{+0.052}_{-0.054}$ & $91.63^{+0.21}_{-0.20}$ & ($58.91$ or $238.78$) $^{+0.12}_{-0.13}$ & $38^{+14}_{-10}$ or $234^{+13}_{-11}$ & $0.836^{+0.042}_{-0.037}$ & $21.162\pm0.022$ & $-0.01^{+0.12}_{-0.02}$ & $0.04999^{+0.00001}_{-0.00002}$ & ... & $1.142^{+0.086}_{-0.081}$ & $24.0^{+9.5}_{-7.3}$ & $140^{+42}_{-24}$ & $12.9^{+2.7}_{-3.0}$  \\ [0.75ex]
HD~984~B & $22.5^{+2.5}_{-2.3}$ & $0.586\pm0.038$ & $115.0^{+1.1}_{-0.9}$ & $320.0^{+3.7}_{-8.5}$ & $153.7^{+4.3}_{-1.8}$ & $0.815^{+0.032}_{-0.037}$ & $21.871\pm0.025$ & $0.252^{+0.028}_{-0.054}$ & $0.04968^{+0.00024}_{-0.00052}$ & ... & $1.133\pm0.033$ & $68.4^{+2.5}_{-2.7}$ & $97^{+18}_{-16}$ & $12.6^{+2.5}_{-2.0}$ \\ [0.75ex]
$\eta$~Tel~B & $148^{+60}_{-34}$ & $0.67^{+0.27}_{-0.41}$ & $81^{+4}_{-14}$ & $177\pm72$ & $168.1^{+6.1}_{-4.2}$ or $347.7^{+5.0}_{-6.5}$ & ${0.55^{+0.14}_{-0.15}}$  & $20.601^{+0.098}_{-0.094}$ & ... & ... & ... & $2.09\pm0.07$ & $11^{+35}_{-8}$ & $1250^{+860}_{-420}$ & $2.2^{+1.1}_{-0.9}$ \\ [0.75ex]

\hline
\end{tabular}
}
\label{fits_results}
\end{sidewaystable*}

\begin{figure*}[h!]
    \section{Histograms}
    \label{app:hist}
    \centering

    \begin{subfigure}[t]{0.24\textwidth}
        \includegraphics[width=\linewidth]{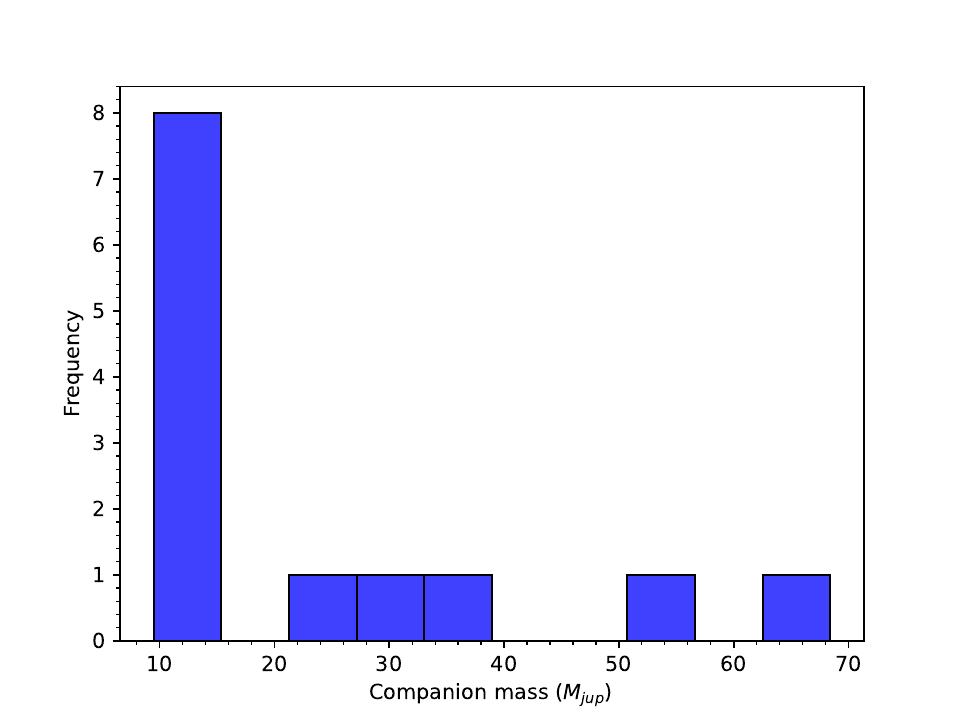}
        \caption{Companion mass}
    \end{subfigure}
    \hfill
    \begin{subfigure}[t]{0.24\textwidth}
        \includegraphics[width=\linewidth]{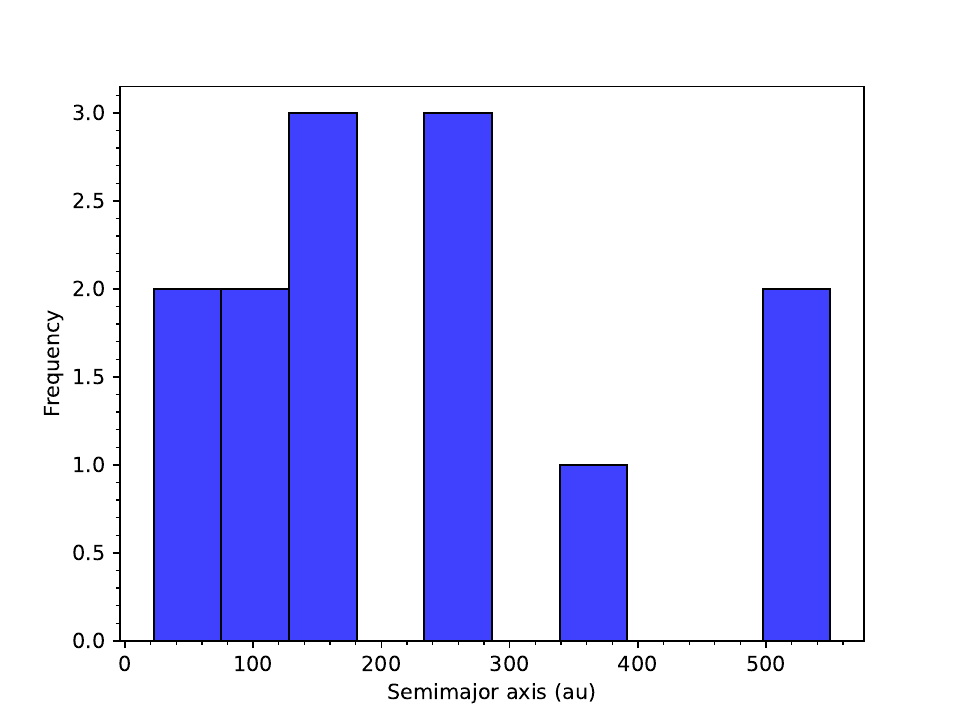}
        \caption{Semimajor axis}
    \end{subfigure}
    \hfill
    \begin{subfigure}[t]{0.24\textwidth}
        \includegraphics[width=\linewidth]{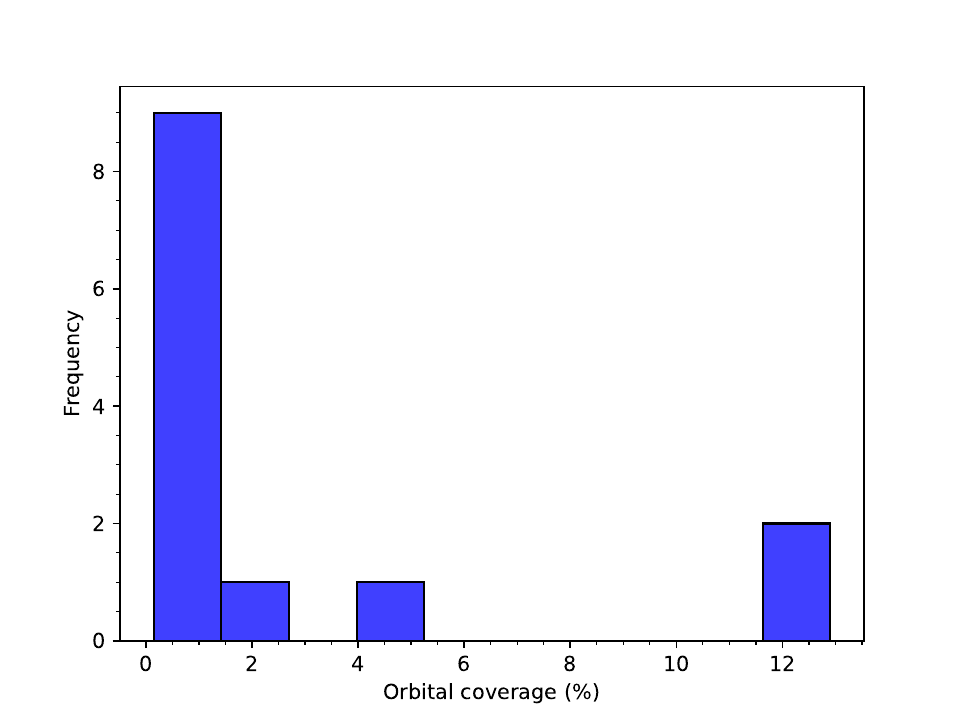}
        \caption{Fractional orbital coverage}
    \end{subfigure}
    \hfill
    \begin{subfigure}[t]{0.24\textwidth}
        \includegraphics[width=\linewidth]{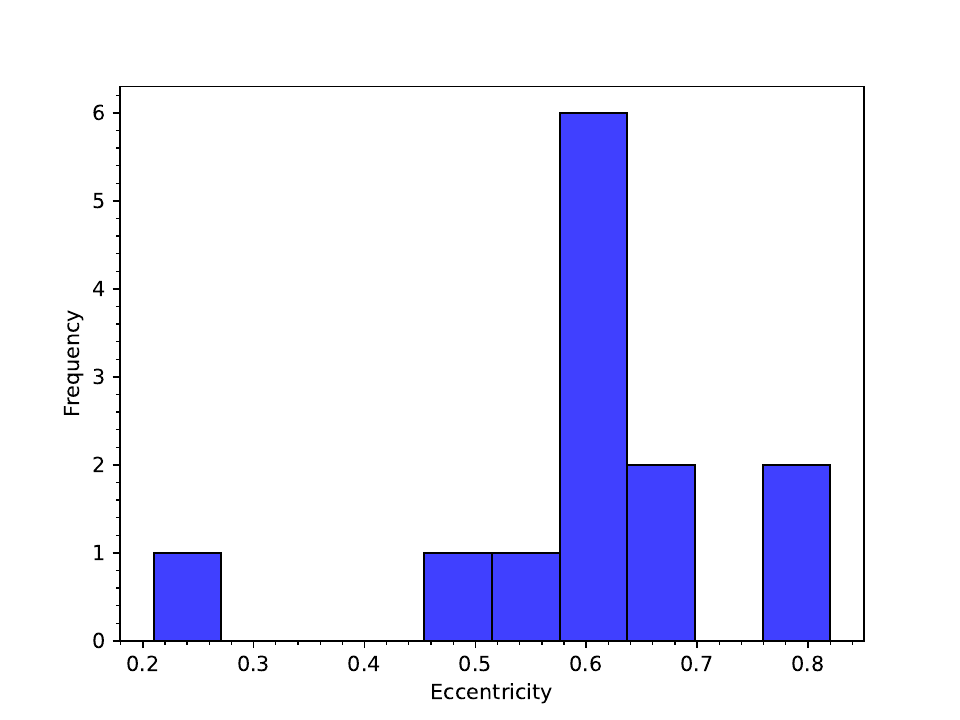}
        \caption{Eccentricities}
    \end{subfigure}

    \caption{Histograms for the companion masses, semimajor axis, fractional orbital coverage, and eccentricities for the 13 objects in our sample.}
    \label{histograms}
\end{figure*}

\begin{figure*}[h!]
\section{Corner plots}
\centering

\begin{subfigure}{0.44\textwidth}
  \centering
  \includegraphics[width=\linewidth]{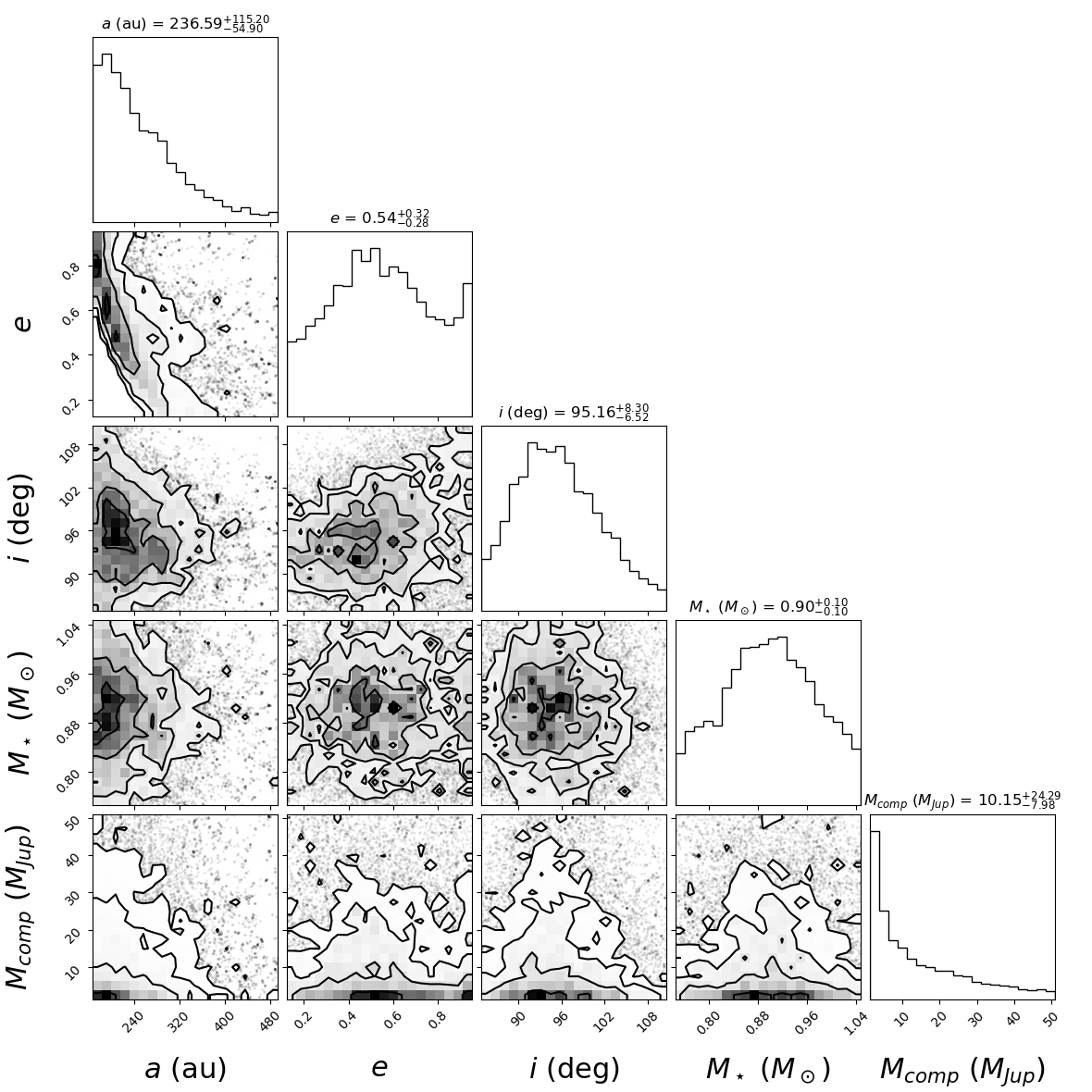}
  \caption{AB~Pic~b}
\end{subfigure}
\begin{subfigure}{0.44\textwidth}
  \centering
  \includegraphics[width=\linewidth]{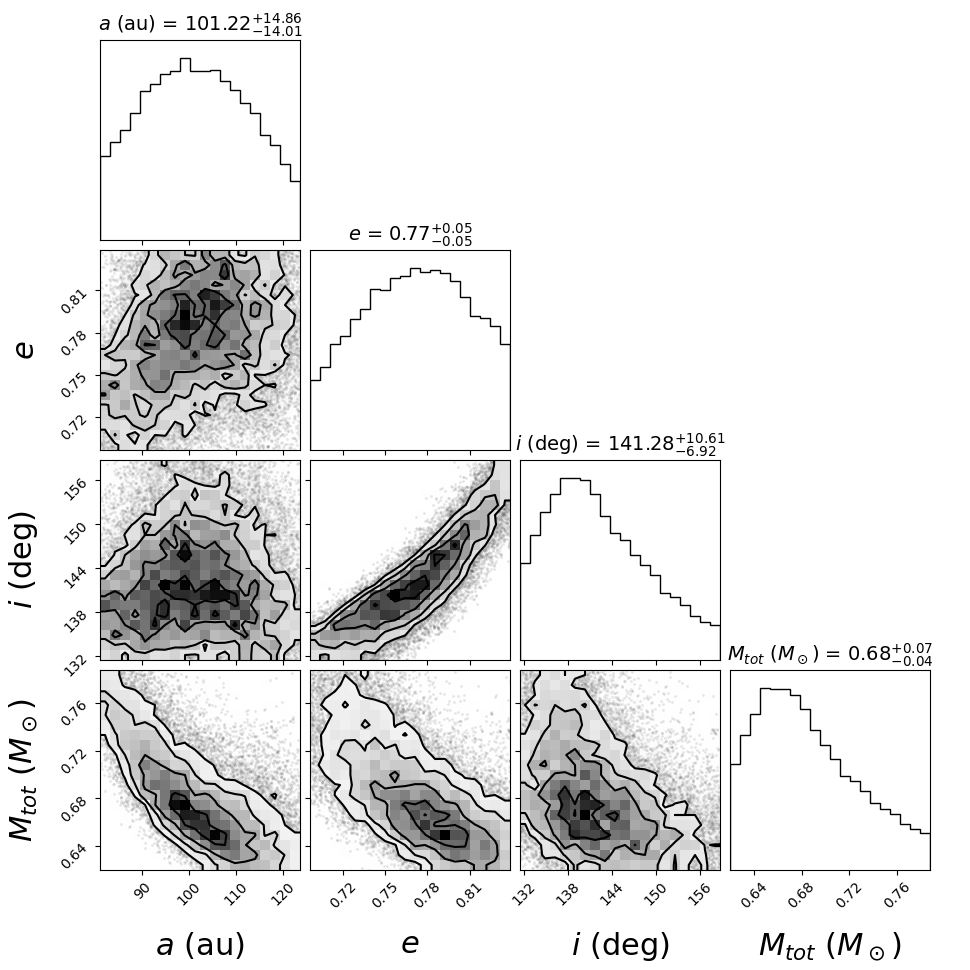}
  \caption{TYC~7084-794-1~B}
\end{subfigure}

\begin{subfigure}{0.44\textwidth}
  \centering
  \includegraphics[width=\linewidth]{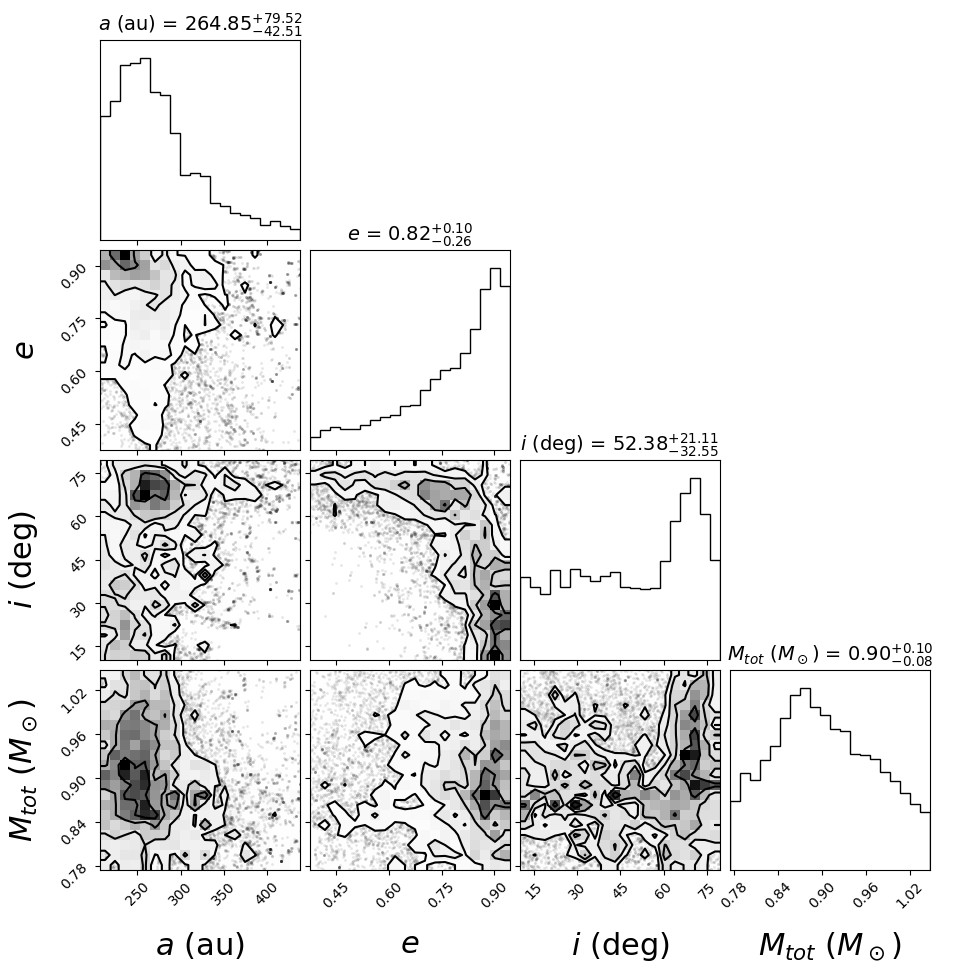}
  \caption{TYC~8047-232-1~B}
\end{subfigure}
\begin{subfigure}{0.44\textwidth}
  \centering
  \includegraphics[width=\linewidth]{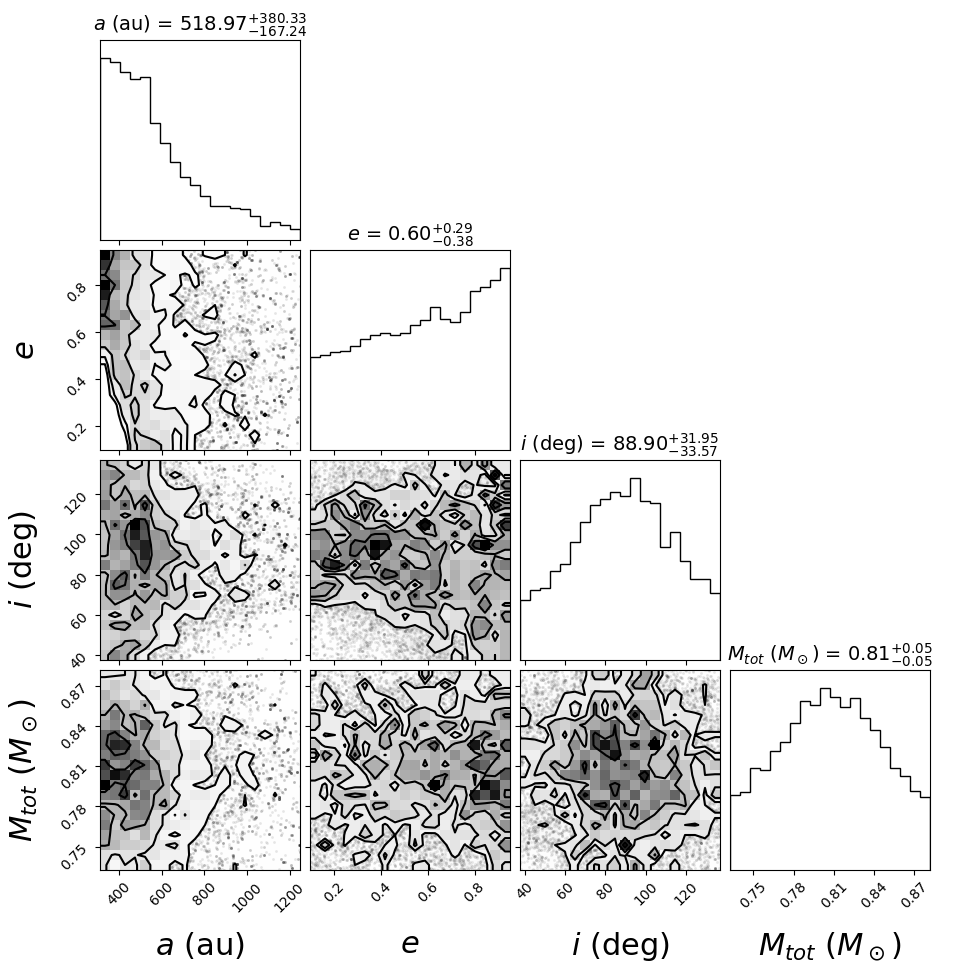}
  \caption{CT~Cha~b}
\end{subfigure}

\caption{Corner plots for AB~Pic~b (a), TYC~7084-794-1~B (b), TYC~8047-232-1~B (c), and CT~Cha~b (d).}
\label{corners1}
\end{figure*}

\begin{figure*}
\centering

\begin{subfigure}{0.44\textwidth}
  \centering
  \includegraphics[width=\linewidth]{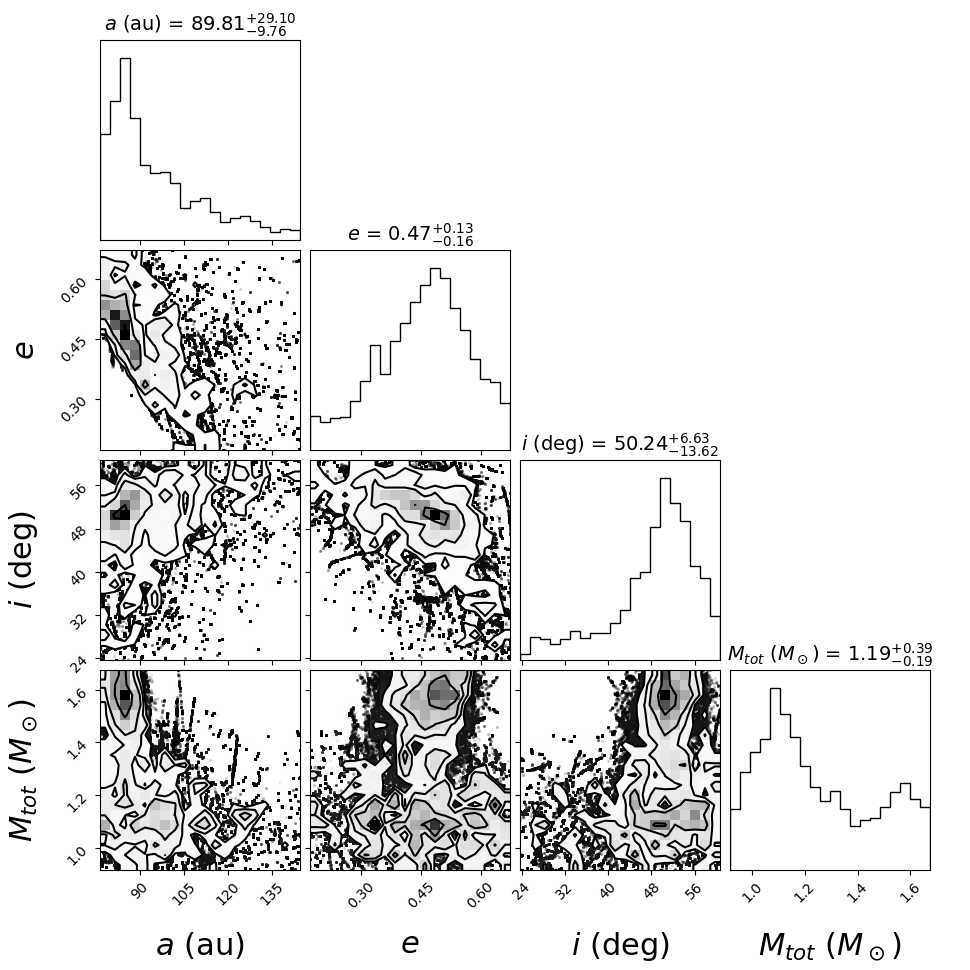}
  \caption{GQ~Lup~B}
\end{subfigure}
\begin{subfigure}{0.44\textwidth}
  \centering
  \includegraphics[width=\linewidth]{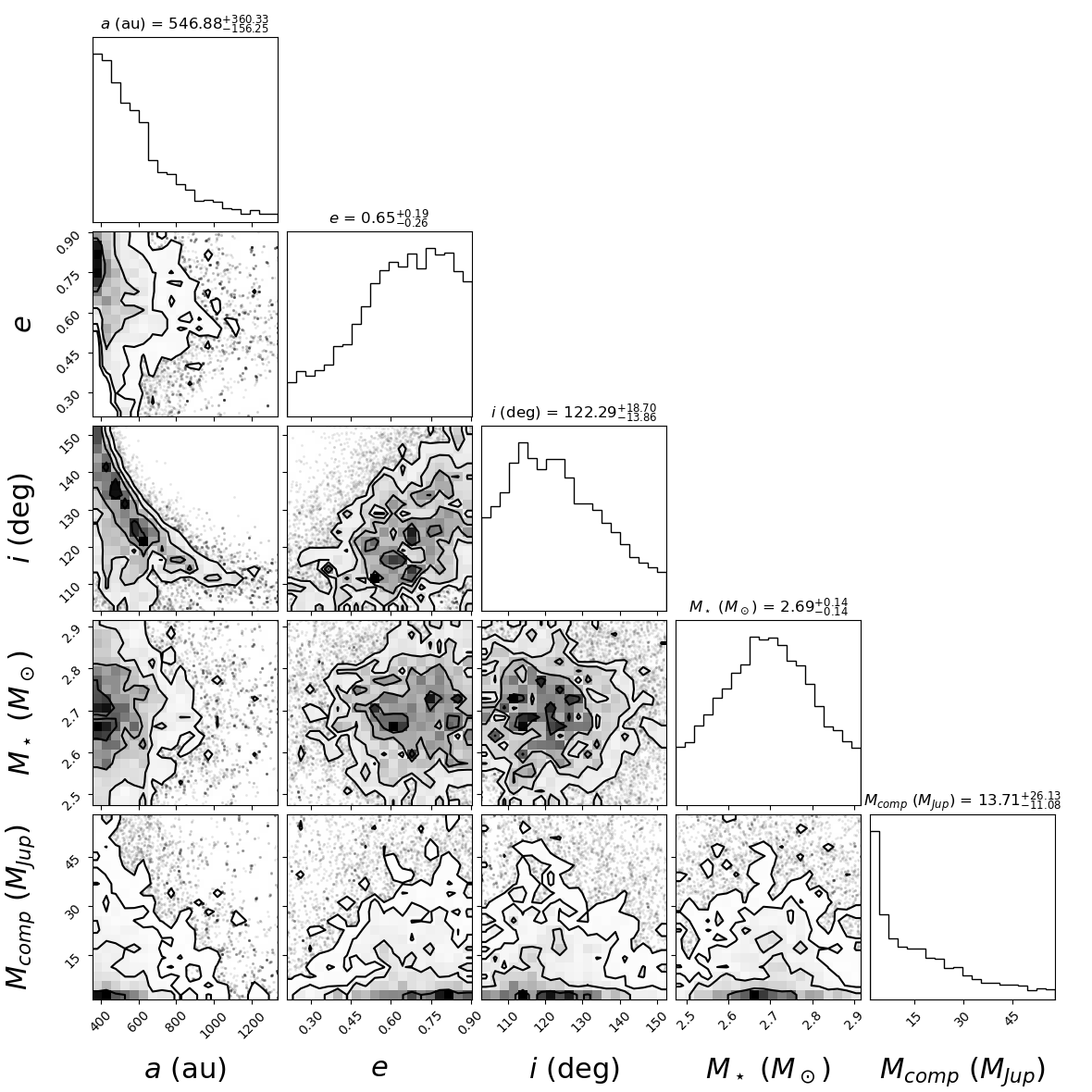}
  \caption{HIP~78530~B}
\end{subfigure}

\begin{subfigure}{0.44\textwidth}
  \centering
  \includegraphics[width=\linewidth]{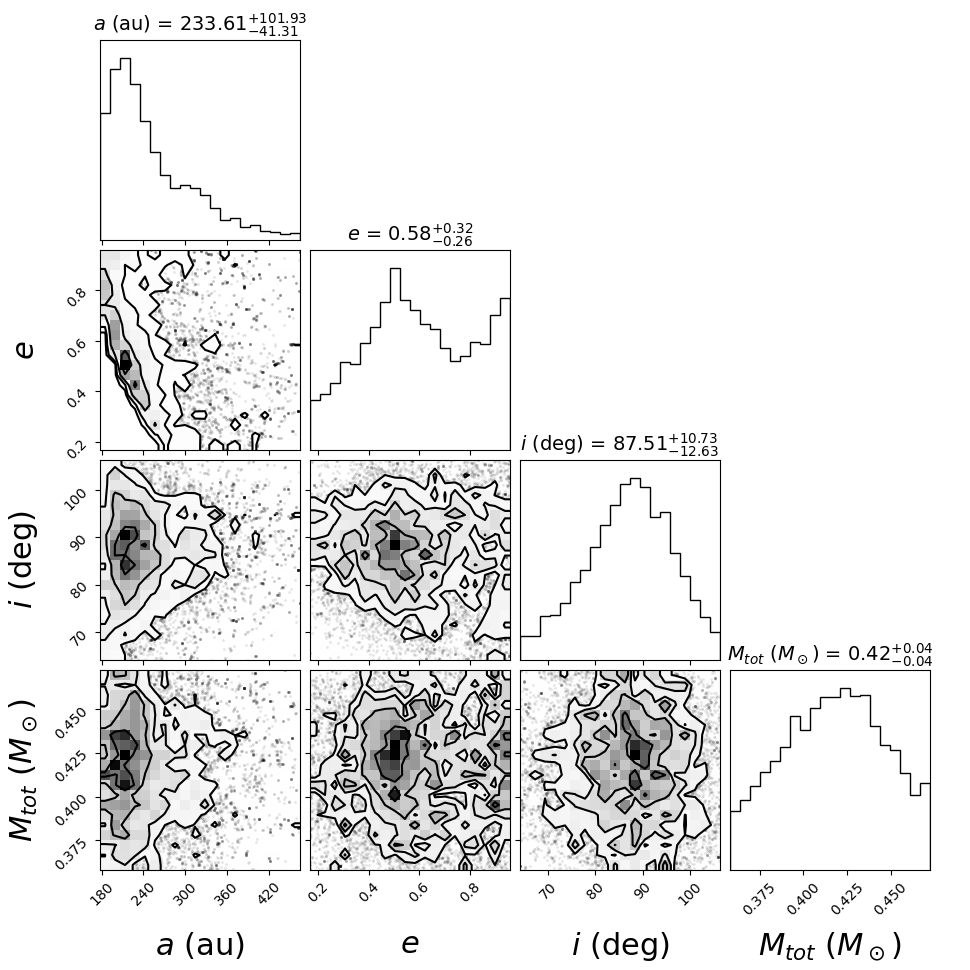}
  \caption{DH~Tau~b}
\end{subfigure}
\begin{subfigure}{0.44\textwidth}
  \centering
  \includegraphics[width=\linewidth]{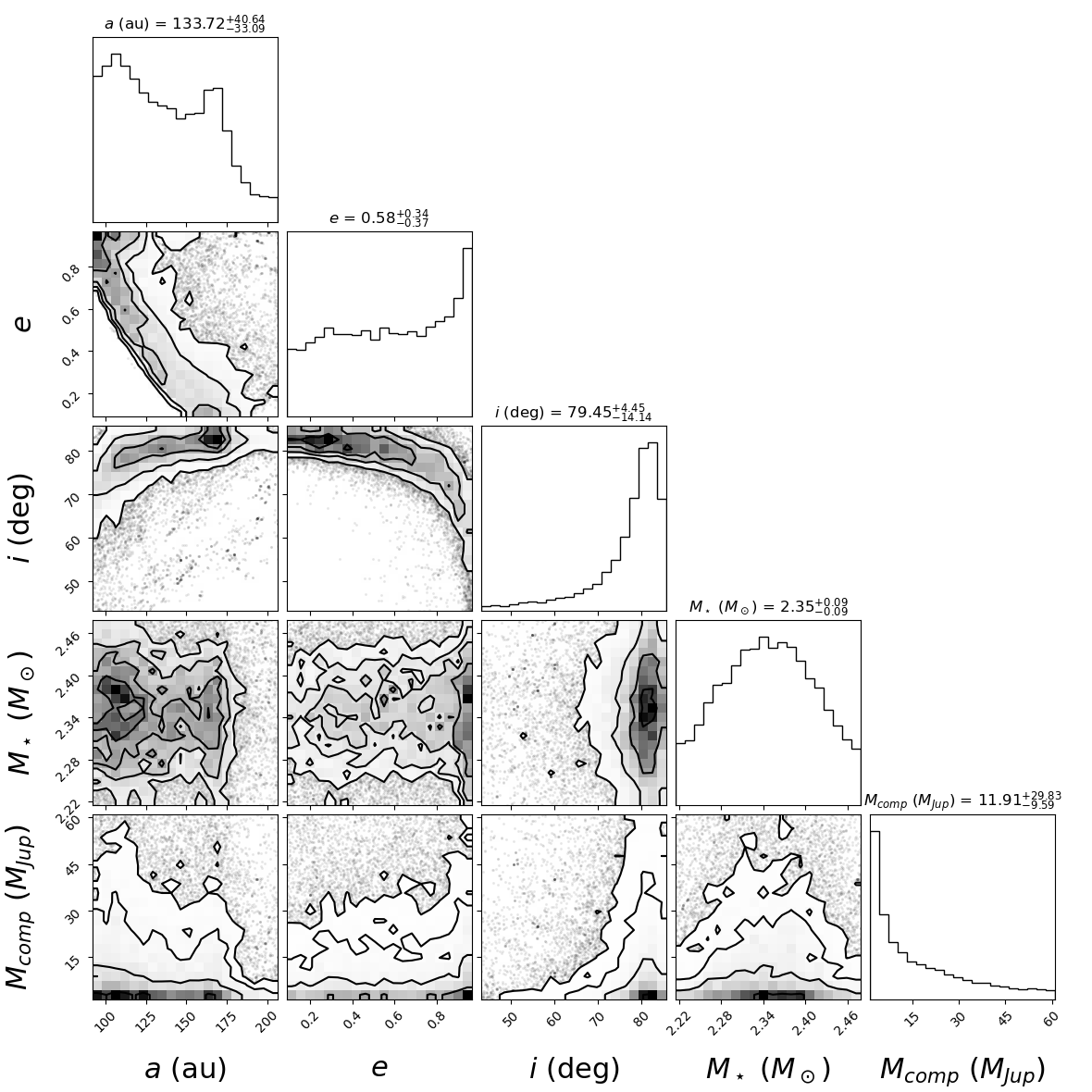}
  \caption{HIP~64892~B}
\end{subfigure}

\caption{Corner plots for GQ~Lup~B (a), HIP~78530~B (b), DH~Tau~b (c), and HIP~64892~B (d).}
\label{corners2}
\end{figure*}

\begin{figure*}
\centering

\begin{subfigure}{0.44\textwidth}
  \centering
  \includegraphics[width=\linewidth]{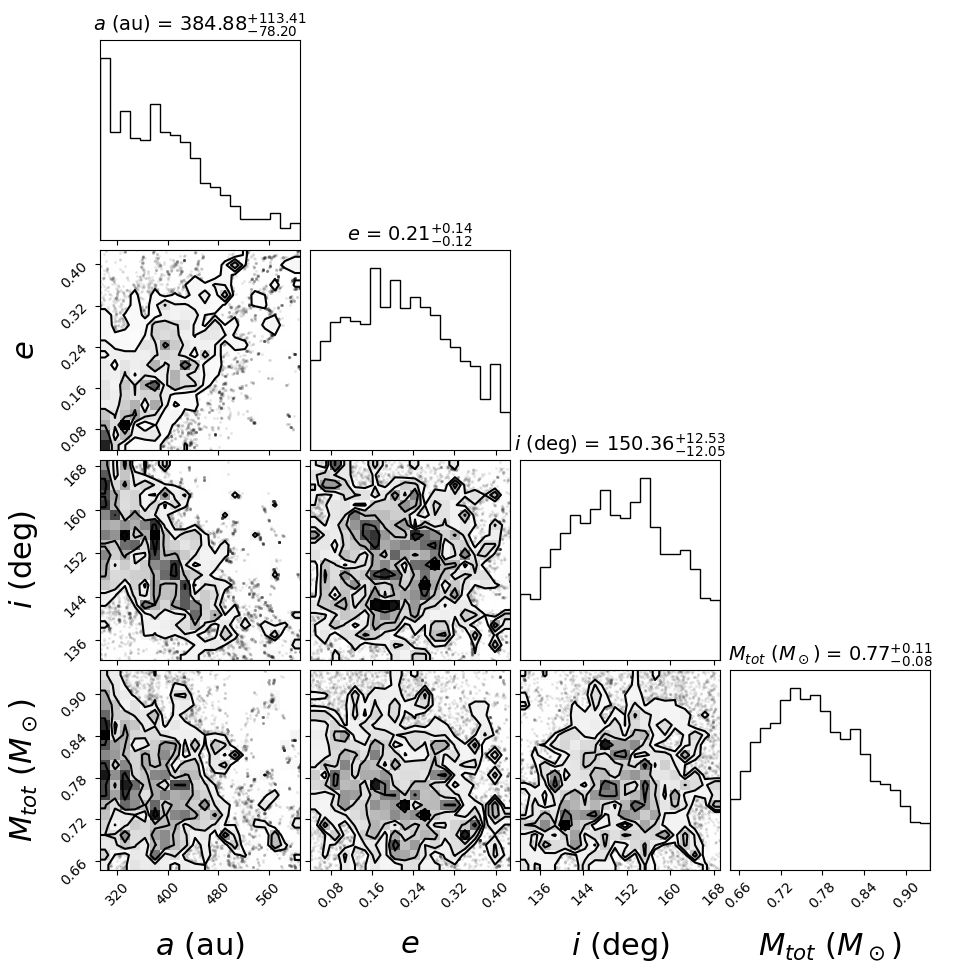}
  \caption{RX~J1609.5-2105~b}
\end{subfigure}
\begin{subfigure}{0.44\textwidth}
  \centering
  \includegraphics[width=\linewidth]{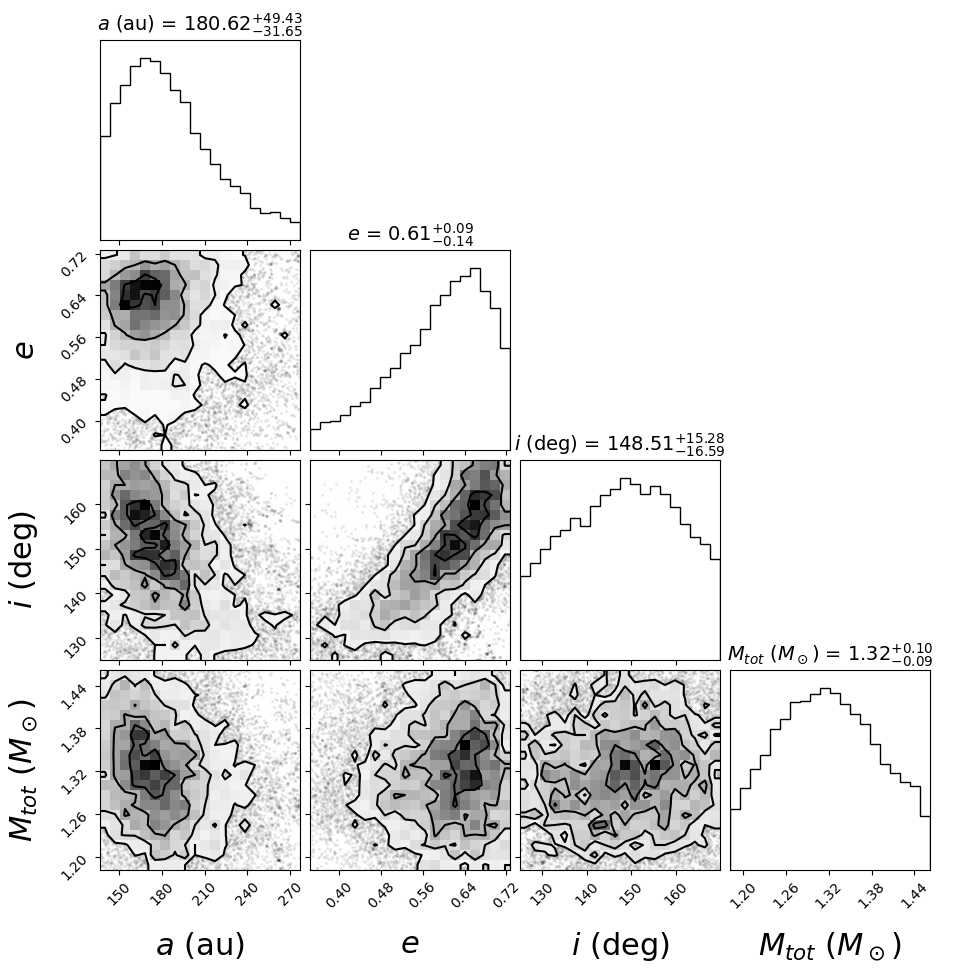}
  \caption{HII~1348~B}
\end{subfigure}

\begin{subfigure}{0.44\textwidth}
  \centering
  \includegraphics[width=\linewidth]{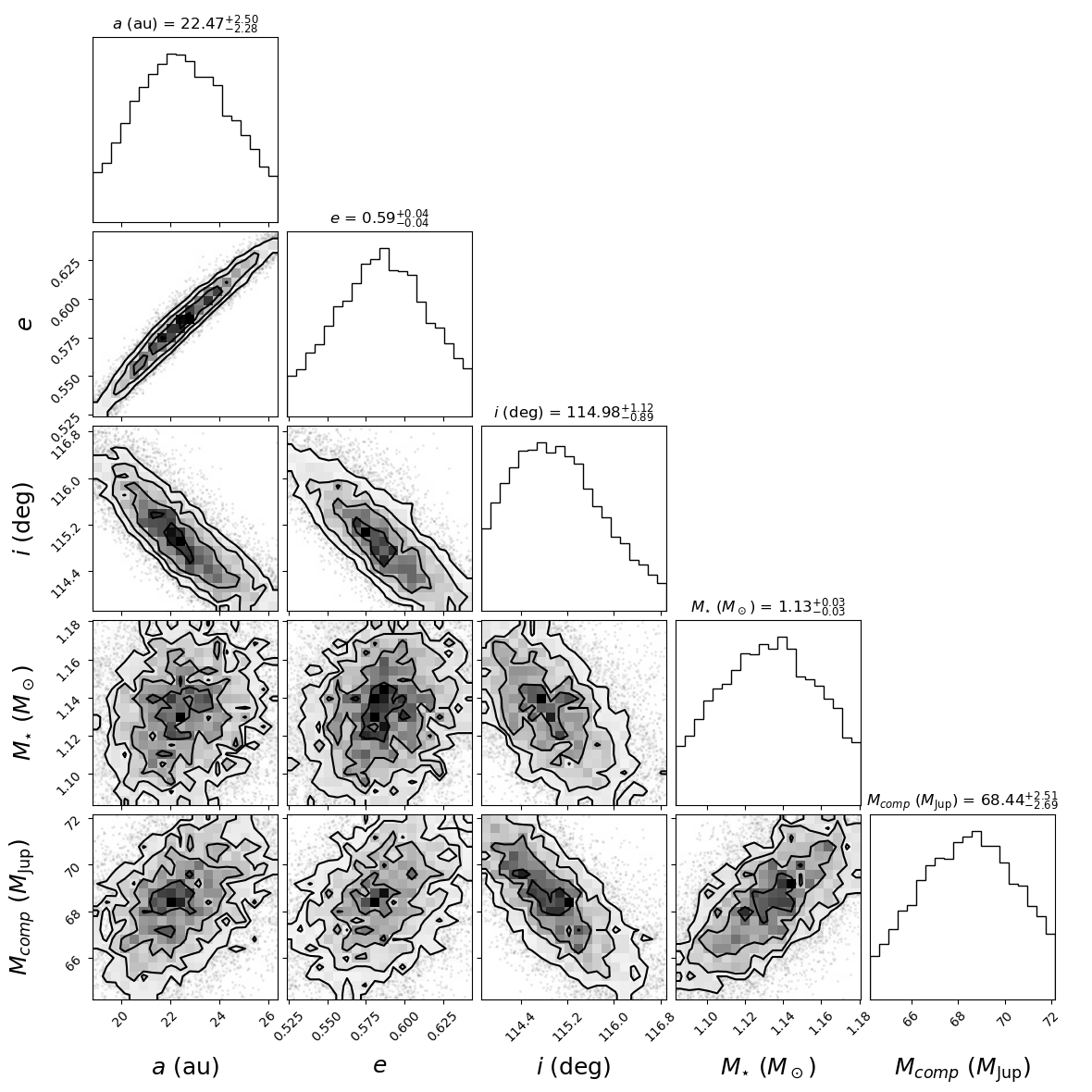}
  \caption{HD~984~B}
\end{subfigure}
\begin{subfigure}{0.44\textwidth}
  \centering
  \includegraphics[width=\linewidth]{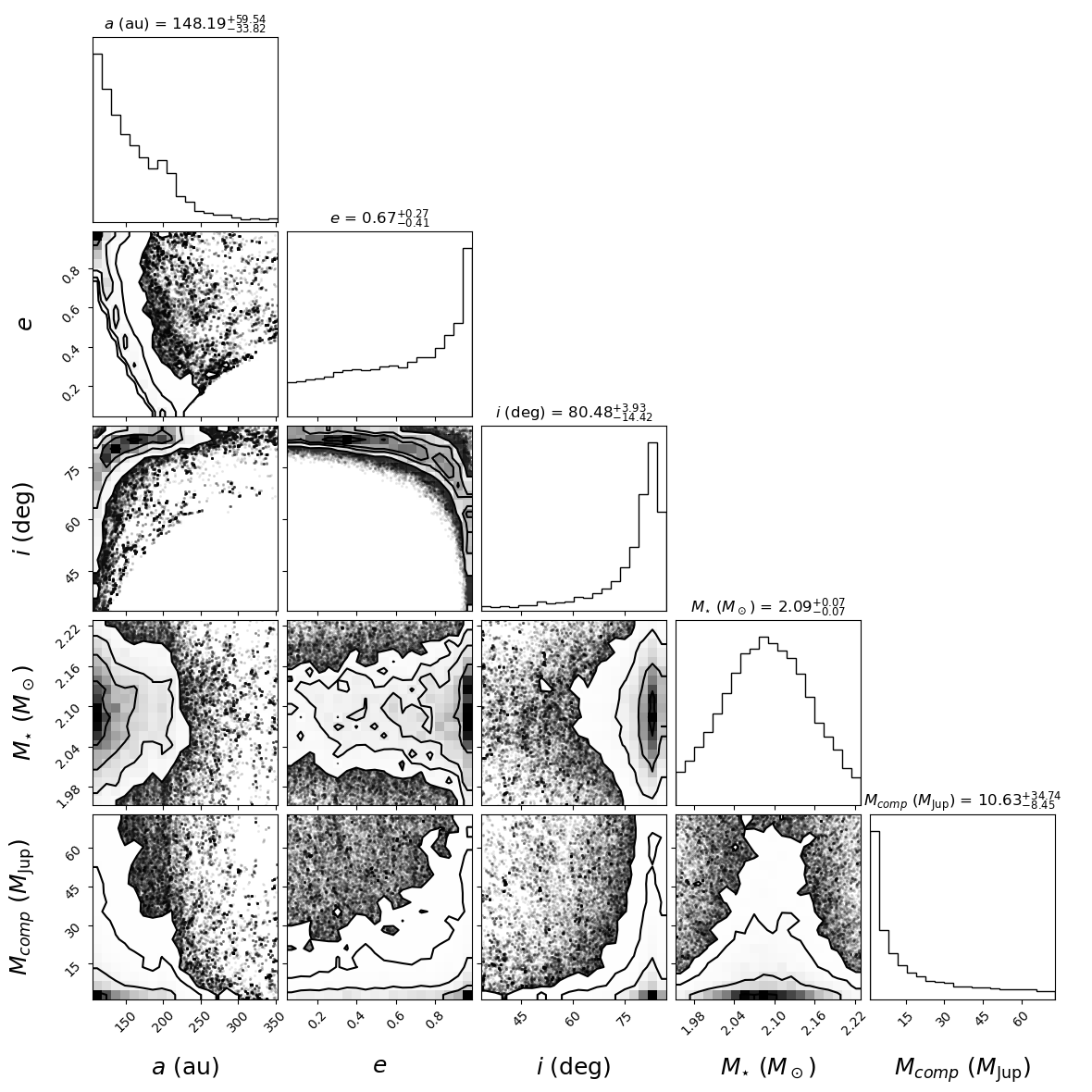}
  \caption{$\eta$~Tel~B}
\end{subfigure}

\caption{Corner plots for RX~J1609.5-2105~b (a), HII~1348~B (b), HD~984~B (c), and $\eta$~Tel~B (d).}
\label{corners3}
\end{figure*}
\FloatBarrier
\section{Astrometric measurements used for the orbital fitting}
\onecolumn
\begin{ThreePartTable}
\begin{longtable}{lcccl}
\caption{Astrometric measurements used for the orbital fitting.} \\
\hline\hline
Name & Epoch (UT) & $\rho$ (mas) & $\theta\,$ (deg) & Reference \\
\hline
\endfirsthead
\caption{continued} \\
\hline\hline
Name & Epoch (UT) & $\rho$ (mas) & $\theta\,$ (deg) & Reference \\
\hline
\endhead
\hline
\endfoot
\endlastfoot

AB~Pic~b &  $2003.213$ & $5460\pm10$ & $175.33\pm0.18$ & \citet{Chauvin2005a} \\
 & $2004.180$ & $5450\pm20$ & $175.13\pm0.21$ & \citet{Chauvin2005a} \\
 & $2004.735$ & $5450\pm10$ & $175.39\pm0.20$ & \citet{Chauvin2005a} \\
 & $2005.712$ & $5520\pm90$ & $175.40\pm0.30$ & \citet{Martinez2022} \\
 & $2009.901$ & $5420.0\pm9.0$ & $175.20\pm0.70$ & \citet{Rameau2013} \\
 & $2015.099$ & $5398.7\pm4.5$ & $175.26\pm0.13$ & \citet{PalmaBifani2023} \\
 & $2023.816$ & $5379.1\pm5.6$ & $175.17\pm0.06$ & \citet{Lazzoni2026} \\
 \hline
TYC~7084-794-1~B  & $2009.041$ & $3172.0\pm5.0$ & $244.13\pm0.25$ & \citet{Wahhaj2011} \\
 & $2010.025$ & $3137.0\pm5.0$ & $243.10\pm0.25$ & \citet{Wahhaj2011} \\
 & $2015.910$ & $2987.5\pm3.5$ & $241.88\pm0.06$ & \citet{Langlois2021} \\
 & $2016.049$ & $2986.6\pm4.7$ & $241.74\pm0.10$ & \citet{Langlois2021} \\
 & $2018.080$ & $2925.0\pm2.0$ & $241.07\pm0.10$ & \citet{Bowler2020} \\
 & $2023.912$ & $2777.1\pm4.0$ & $239.02\pm0.08$ & \citet{Lazzoni2026} \\
 \hline
TYC~8047-232-1~B  & $2001.504$ & $3240\pm20$ & $357.65\pm0.18$ & \citet{Neuhauser2003} \\
 & $2001.828$ & $3200\pm100$ & $359.0\pm2.0$ & \citet{Chauvin2003} \\
 & $2002.896$ & $3270\pm10$ & $358.85\pm0.23$ & \citet{Chauvin2005b} \\
 & $2003.686$ & $3266\pm10$ & $358.89\pm0.23$ & \citet{Chauvin2005b} \\
 & $2009.702$ & $3220.0\pm5.0$ & $358.67\pm0.13$ & \citet{Ginski2013} \\
 & $2012.437$ & $3209.0\pm6.0$ & $358.75\pm0.17$ & \citet{Ginski2013} \\
 & $2012.923$ & $3211.0\pm7.0$ & $358.67\pm0.13$ & \citet{Ginski2013} \\
 & $2015.732$ & $3207.1\pm2.7$ & $358.8\pm1.0$ & \citet{Langlois2021} \\
 & $2016.044$ & $3208.5\pm9.1$ & $358.5\pm1.1$ & \citet{Langlois2021} \\
 & $2023.748$ & $3165.5\pm4.0$ & $358.91\pm0.07$ & This work \\
 & $2024.653$ & $3168.3\pm4.0$ & $358.95\pm0.07$ & \citet{Lazzoni2026} \\
 \hline
CT~Cha~b  & $2006.129$ & $2670\pm36$ & $300.7\pm1.2$  & \citet{Schmidt2008} \\
 & $2007.165$ & $2670\pm38$ & $300.7\pm1.3$  & \citet{Schmidt2008} \\
 & $2008.135$ & $2674\pm40$ & $300.5\pm1.4$  & \citet{Schmidt2009} \\
 & $2013.261$ & $2687\pm11$ & $299.70\pm0.40$  & \citet{Wu2015} \\ 
 & $2017.211$ & $2681.5\pm4.0$ & $300.12\pm0.08$ & This work \\
 & $2025.068$ & $2688.4\pm4.0$ & $300.10\pm0.08$ & \citet{Lazzoni2026} \\
 \hline
GQ~Lup~B  & $1994.249$ & $714\pm36$ & $275.5\pm1.1$ & \citet{Janson2006} \\
 & $1999.271$ & $739\pm11$ & $275.62\pm0.86$ & \citet{Neuhauser2005}\\
 & $2004.480$ & $734.7\pm3.1$ & $275.48\pm0.25$ & \citet{Neuhauser2008} \\
 & $2005.400$ & $735.1\pm3.3$ & $276.00\pm0.34$ & \citet{Neuhauser2008} \\
 & $2005.482$ & $730\pm10$ & $276.20\pm0.30$ & \citet{McElwain2007} \\
 & $2005.600$ & $733.3\pm3.9$ & $275.87\pm0.37$ & \citet{Neuhauser2008} \\
 & $2006.140$ & $729.8\pm3.3$ & $276.14\pm0.35$ & \citet{Neuhauser2008}\\
 & $2006.380$ & $731.4\pm3.5$ & $276.06\pm0.38$ & \citet{Neuhauser2008} \\
 & $2006.540$ & $733.2\pm5.0$ & $276.26\pm0.68$ & \citet{Neuhauser2008}\\
 & $2007.130$ & $730.0\pm6.4$ & $276.04\pm0.63$ & \citet{Neuhauser2008}\\
 & $2008.451$ & $725.5\pm5.0$ & $276.66\pm0.50$ & \citet{Ginski2014}\\
 & $2009.490$ & $726.4\pm1.6$ & $276.54\pm0.17$ & \citet{Ginski2014}\\
 & $2010.340$ & $725.6\pm1.4$ & $276.86\pm0.18$ & \citet{Ginski2014}\\
 & $2011.425$ & $724.0\pm2.0$ & $276.94\pm0.23$ & \citet{Ginski2014}\\
 & $2012.169$ & $724.0\pm2.0$ & $277.04\pm0.24$ & \citet{Ginski2014}\\
 & $2013.373$ & $723\pm12$ & $277.4\pm1.4$ & \citet{Uyama2017}\\
 & $2013.378$ & $723\pm34$ & $275.7\pm1.5$ & \citet{Uyama2017}\\
 & $2015.288$ & $721.0\pm3.0$  & $277.60\pm0.40$ & \citet{Wu2017} \\
 & $2015.337$ & $701.1\pm3.0$ & $278.25\pm0.20$ & This work \\
 & $2016.484$ & $718.6\pm3.0$ & $277.82\pm0.20$ & This work \\
 & $2018.619$ & $711.6\pm2.4$ & $278.27\pm0.24$ & \citet{Stolker2021} \\
 & $2019.184$ & $720.3\pm3.4$ & $277.90\pm0.18$ & \citet{Stolker2021} \\
 & $2019.201$ & $715.0\pm2.8$ & $277.83\pm0.17$ & \citet{Stolker2021}\\
 & $2019.290$ & $701\pm20$ & $278.3\pm1.2$ & \citet{Stolker2021}\\
 & $2021.655$ & $710.690\pm0.026$ & $278.9336\pm0.0046$ & \citet{Venkatesan2025} \\
 & $2022.617$ & $708.143\pm0.031$ & $279.3715\pm0.0070$ & \citet{Venkatesan2025} \\
 & $2022.680$ & $708.166\pm0.036$ & $279.3552\pm0.0090$ & \citet{Venkatesan2025} \\
 & $2023.212$ & $708.214\pm0.076$ & $279.268\pm0.012$ & \citet{Venkatesan2025} \\
 & $2025.279$ & $705.7\pm3.0$ & $279.58\pm0.20$ & \citet{Lazzoni2026} \\
 \hline
HIP~78530~B  & $2008.394$ & $4529.0\pm6.0$ & $140.32\pm0.10$ & \citet{Lafreniere2011} \\
 & $2009.500$ & $4533.0\pm6.0$ & $140.35\pm0.10$ & \citet{Lafreniere2011} \\
 & $2010.661$ & $4536.0\pm6.0$ & $140.27\pm0.10$ & \citet{Lafreniere2011} \\
 & $2011.241$ & $4527.0\pm3.0$ & $140.30\pm0.10$ & \citet{Lachapelle2015} \\
 & $2011.397$* & $4540\pm90$ & $140.7\pm1.0$ & \citet{Bailey2013} \\
 & $2015.337$ & $4534.5\pm2.6$ & $139.98\pm0.04$ & \citet{Langlois2021} \\
 & $2025.279$ & $4525.5\pm4.7$ & $140.08\pm0.07$ & \citet{Lazzoni2026} \\
 \hline
DH~Tau~b  & $1999.044$* & $2351.0\pm1.0$ & $139.36\pm0.10$ & \citet{Itoh2005} \\
 & $1999.044$ & $2332.0\pm9.9$ & $138.68\pm0.19$  & \citet{Ginski2014} \\
 & $2002.893$* & $2340.0\pm6.0$ & $139.56\pm0.17$ & \citet{Itoh2005}\\
 & $2004.019$* & $2344.0\pm3.0$ & $139.83\pm0.06$ & \citet{Itoh2005} \\
 & $2009.748$ & $2339.3\pm4.1$ & $138.63\pm0.14$ & \citet{Ginski2014} \\
 & $2012.057$ & $2332.3\pm6.1$ & $138.76\pm0.16$ & \citet{Ginski2014} \\
 & $2012.926$ & $2342.7\pm5.7$ & $138.61\pm0.15$ & \citet{Ginski2014} \\
 & $2014.934$* & $2343.0\pm1.0$ & $140.25\pm0.02$ & \citet{Bryan2016} \\
 & $2015.816$ & $2350.8\pm2.8$ & $138.80\pm0.09$ & This work \\
 & $2015.844$* & $2339.0\pm1.0$ & $139.94\pm0.02$ & \citet{Bryan2016} \\
 & $2015.964$ & $2346.2\pm2.7$ & $138.80\pm0.09$ & This work \\
 & $2018.079$ & $2354.0\pm2.0$ & $138.46\pm0.10$ & \citet{Bowler2020} \\
 & $2018.956$ & $2346.1\pm3.0$ & $138.70\pm0.09$ & This work \\
 & $2020.011$ & $2345.7\pm2.7$ & $138.73\pm0.09$ & This work \\
 & $2023.921$ & $2348.4\pm2.7$ & $138.69\pm0.09$ & This work \\
 & $2024.762$ & $2347.6\pm2.7$ & $138.76\pm0.09$ & \citet{Lazzoni2026} \\
 & $2025.003$ & $2348.4\pm2.7$ & $138.74\pm0.09$ & \citet{Lazzoni2026} \\
 \hline
HIP~64892~B  & $2011.433$ & $1272\pm29$ & $310.0\pm1.3$ & \citet{Cheetham2018} \\
 & $2016.249$ & $1269.3\pm2.1$ & $311.70\pm0.07$ & \citet{Langlois2021} \\
 & $2017.104$ & $1271.3\pm1.7$ & $311.85\pm0.07$ & \citet{Langlois2021} \\
 & $2018.282$ & $1281.9\pm2.3$ & $311.94\pm0.09$ & \citet{Langlois2021} \\
 & $2025.148$ & $1298.3\pm3.0$ & $312.04\pm0.10$ & \citet{Lazzoni2026} \\
 \hline
RX~J1609.5-2105~b  & $2006.230$ & $2140\pm110$ & $27.00\pm0.60$ & \citet{Martinez2022} \\
 & $2008.321$ & $2215.0\pm6.0$ & $27.75\pm0.10$ & \citet{Lafreniere2010} \\
 & $2008.460$ & $2221.0\pm6.0$ & $27.76\pm0.10$ & \citet{Lafreniere2010} \\
 & $2008.462$ & $2210.1\pm1.0$ & $27.62\pm0.04$ & \citet{Ireland2011} \\
 & $2009.262$ & $2222.0\pm6.0$ & $27.65\pm0.10$ & \citet{Lafreniere2010} \\
 & $2009.408$ & $2211.3\pm0.9$ & $27.61\pm0.05$ & \citet{Ireland2011} \\
 & $2009.497$ & $2219.0\pm6.0$ & $27.74\pm0.10$ & \citet{Lafreniere2010} \\
 & $2009.619$ & $2198.9\pm3.5$ & $27.08\pm0.13$ & \citet{Ginski2014} \\
 & $2011.290$ & $2150\pm30$ & $28.30\pm0.40$ & \citet{Bailey2013} \\
 & $2013.260$ & $2210\pm10$ & $27.10\pm0.30$ & \citet{Wu2015b} \\
 & $2025.351$ & $2219.5\pm4.0$ & $26.39\pm0.10$ & \citet{Lazzoni2026} \\
 \hline
HII~1348~B  & $1996.732$ & $1090\pm20$ & $347.90\pm0.70$ & \citet{Bouvier1997} \\
 & $2004.754$ & $1097.0\pm5.0$ & $346.80\pm0.20$ & \citet{Geissler2012} \\
 & $2005.888$ & $1120\pm20$ & $346.80\pm0.60$ & \citet{Geissler2012} \\
 & $2011.975$ & $1120\pm30$ & $346.10\pm0.90$ & \citet{Yamamoto2013} \\
 & $2019.712$ & $1140.0\pm5.0$ & $345.20\pm0.30$ & \citet{Weible2025} \\
 & $2023.934$ & $1153.5\pm3.0$ & $343.88\pm0.10$ & \citet{Lazzoni2026} \\
 \hline
PZ~Tel~B  & $2007.446$ & $255.6\pm2.5$ & $61.68\pm0.60$ & \citet{Mugrauer2012} \\
 & $2009.739$ & $336.6\pm1.2$ & $60.52\pm0.22$ & \citet{Mugrauer2012} \\
 & $2010.274$ & $330\pm10$ & $59.0\pm1.0$ & \citet{Biller2010} \\
 & $2010.340$ & $356.4\pm1.1$ & $60.32\pm0.20$ & \citet{Mugrauer2012} \\
 & $2010.345$ & $354.7\pm1.2$ & $60.34\pm0.21$ & \citet{Mugrauer2012} \\
 & $2010.350$ & $360.0\pm3.0$ & $59.40\pm0.50$ & \citet{Biller2010} \\
 & $2010.734$ & $365.0\pm8.0$ & $59.20\pm0.80$ & \citet{Beust2016} \\
 & $2010.821$ & $369.3\pm1.1$ & $59.91\pm0.18$ & \citet{Mugrauer2012} \\
 & $2011.227$ & $382.2\pm1.0$ & $59.84\pm0.19$ & \citet{Mugrauer2012} \\
 & $2011.334$ & $394.0\pm2.0$ & $60.40\pm0.20$ & \citet{Beust2016} \\
 & $2011.419$ & $387.8\pm1.2$ & $59.66\pm0.23$ & \citet{Mugrauer2012} \\
 & $2011.422$ & $388.5\pm0.8$ & $59.66\pm0.16$ & \citet{Mugrauer2012} \\
 & $2011.424$ & $387.1\pm1.4$ & $59.67\pm0.25$ & \citet{Mugrauer2012} \\
 & $2011.427$ & $389.0\pm1.0$ & $59.70\pm0.20$ & \citet{Mugrauer2012} \\
 & $2011.430$ & $390.0\pm5.0$ & $60.00\pm0.60$ & \citet{Beust2016} \\
 & $2012.435$ & $419.5\pm1.4$ & $59.58\pm0.22$ & \citet{Ginski2014} \\ 
 & $2014.530$ & $477.4\pm1.6$ & $59.80\pm0.40$ &  \citet{Maire2016} \\ 
 & $2014.600$ & $479.6\pm0.4$ & $59.80\pm0.40$ & \citet{Maire2016} \\ 
 & $2014.780$ & $483.5\pm0.5$ & $59.48\pm0.16$ & \citet{Maire2016} \\ 
 & $2015.575$ & $501.0\pm2.0$ & $59.50\pm0.20$ & \citet{Squicciarini2025} \\
 & $2016.295$ & $519.9\pm1.3$ & $59.32\pm0.12$ &  \citet{Langlois2021} \\
 & $2016.710$ & $530.1\pm1.4$ & $59.14\pm0.13$ & \citet{Langlois2021} \\
 & $2017.375$ & $543.7\pm1.5$ & $59.13\pm0.13$ & \citet{Langlois2021} \\
 & $2018.408$ & $566.8\pm3.6$ & $59.17\pm0.50$ & \citet{MussoBarcucci2019} \\
 & $2018.433$ & $565.5\pm1.8$ & $58.74\pm0.43$ & \citet{Stolker2020} \\
 & $2022.710$ & $636.0\pm5.0$ & $59.30\pm0.50$ & \citet{Franson2023} \\
 & $2025.468$ & $682.7\pm2.5$ & $58.79\pm0.20$ & This work \\ 
 \hline
HD~984~B  & $2012.545$ & $190\pm20$ & $108.8\pm3.0$ & \citet{Meshkat2015} \\
 & $2012.550$ & $208\pm23$ & $108.9\pm3.1$ & \citet{Meshkat2015} \\
 & $2014.691$ & $201.6\pm0.4$ & $92.20\pm0.50$ & \citet{Meshkat2015} \\
 & $2015.661$ & $217.1\pm0.9$ & $83.50\pm0.30$ & \citet{JohnsonGroh2017} \\ 
 & $2019.514$ & $233.8\pm1.8$ & $57.64\pm0.29$ & \citet{Franson2022} \\
 & $2020.578$ & $242.9\pm1.7$ & $51.61\pm0.27$ & \citet{Franson2022} \\
 & $2024.781$ & $288.5\pm2.5$ & $31.37\pm0.50$ & This work \\
 \hline
$\eta$~Tel~B  & $1998.492$ & $4170\pm33$ & $166.95\pm0.36$ & \citet{Neuhauser2011} \\
 & $2000.307$ & $4107\pm57$ & $166.90\pm0.42$ & \citet{Neuhauser2011} \\
 & $2000.378$ & $4097\pm48$ & $166.90\pm0.42$ & \citet{Guenther2001} \\
 & $2004.329$ & $4196\pm23$ & $167.04\pm0.22$ & \citet{Neuhauser2011} \\
 & $2006.431$ & $4170\pm110$ & $167.0\pm1.4$ & \citet{Geissler2008} \\
 & $2007.753$ & $4212\pm33$ & $167.42\pm0.35$ & \citet{Neuhauser2011} \\
 & $2008.312$ & $4214\pm17$ & $166.81\pm0.22$ & \citet{Neuhauser2011} \\
 & $2008.599$ & $4195\pm17$ & $166.54\pm0.29$ & \citet{Neuhauser2011} \\
 & $2009.27$ & $4175\pm9$ & $167.60\pm0.20$ & \citet{Nielsen2013} \\
 & $2009.351$ & $4240\pm100$ & $168.50\pm1.3$ & \citet{Neuhauser2011} \\
 & $2009.496$ & $4199\pm31$ & $166.99\pm0.30$ & \citet{Neuhauser2011} \\
 & $2011.576$ & $4170.0\pm9.0$ & $167.43\pm0.70$ & \citet{Rameau2013} \\
 & $2015.341$ & $4215.0\pm4.0$ & $167.33\pm0.20$ & \citet{Nogueira2024} \\
 & $2016.454$ & $4218.0\pm4.0$ & $167.26\pm0.13$ & \citet{Nogueira2024} \\
 & $2017.452$ & $4218.0\pm4.0$ & $167.35\pm0.14$ & \citet{Nogueira2024} \\
 & $2023.362$ & $4199\pm15$ & $167.49\pm0.18$ & \citet{Chai2024} \\
& $2025.554$ & $4222.6\pm4.9$ & $167.43\pm0.07$ & \citet{Lazzoni2026} \\
\hline
\end{longtable}
\begin{TableNotes}
\footnotesize
\item Data points marked with an asterisk were not included in the orbital fits.
\end{TableNotes}
\end{ThreePartTable}
\label{astrometry}
\twocolumn

\section{Radial velocity measurements used for the orbital fitting}
\onecolumn
\begin{longtable}{lccl}
\caption{Radial velocity measurements used for the orbital fitting.} \\

\hline\hline
Name & Epoch (UT) &  RV (m s$^{-1}$) & Reference \\
\hline
\endfirsthead
\caption{(Continued)} \\
\hline\hline
Name & Epoch (UT) &  RV (m s$^{-1}$) & Reference \\
\hline
\endhead

\hline
\endlastfoot

PZ~Tel & $2009.3020$ & $-126.8\pm4.2$ & \citet{Trifonov2020} \\
 & $2009.3021$ & $-278.9\pm3.7$ & \citet{Trifonov2020} \\
 & $2009.3023$ & $-117.9\pm3.3$ & \citet{Trifonov2020} \\
 & $2009.3047$ & $-271.3\pm4.1$ & \citet{Trifonov2020} \\
 & $2009.3049$ & $19.4\pm3.6$ & \citet{Trifonov2020} \\
 & $2009.3051$ & $299.7\pm3.4$ & \citet{Trifonov2020} \\
 & $2009.3101$ & $139.7\pm3.9$ & \citet{Trifonov2020} \\
 & $2009.3106$ & $-169.3\pm3.3$ & \citet{Trifonov2020} \\
 & $2009.3108$ & $-128.7\pm3.2$ & \citet{Trifonov2020} \\
 & $2009.3487$ & $-486.5\pm5.1$ & \citet{Trifonov2020} \\
 & $2009.3488$ & $-19.4\pm6.7$ & \citet{Trifonov2020} \\
 & $2009.3491$ & $-359.6\pm6.6$ & \citet{Trifonov2020} \\
 & $2009.4250$ & $1504\pm27$ & \citet{Trifonov2020} \\
 & $2009.5482$ & $259.2\pm4.4$ & \citet{Trifonov2020} \\
 & $2009.5483$ & $143.8\pm4.8$ & \citet{Trifonov2020} \\
 & $2009.5565$ & $792.4\pm6.2$ & \citet{Trifonov2020} \\
 & $2009.5566$ & $669.2\pm7.4$ & \citet{Trifonov2020} \\
 & $2009.5569$ & $423.3\pm6.4$ & \citet{Trifonov2020} \\
 & $2009.5570$ & $542.3\pm6.8$ & \citet{Trifonov2020} \\
 & $2014.3132$ & $649.6\pm4.5$ & \citet{Trifonov2020} \\
 & $2014.3133$ & $796.9\pm4.6$ & \citet{Trifonov2020} \\
 & $2014.3789$ & $-705.8\pm5.6$ & \citet{Trifonov2020} \\
 & $2014.3790$ & $-546.0\pm6.3$ & \citet{Trifonov2020} \\
 & $2014.5130$ & $-695.0\pm6.6$ & \citet{Trifonov2020} \\
 & $2014.5131$ & $-776.1\pm6.5$ & \citet{Trifonov2020} \\
 & $2016.5523$ & $-222.9\pm2.9$ & \citet{Trifonov2020} \\
 & $2016.5524$ & $-303.5\pm2.7$ & \citet{Trifonov2020} \\
 & $2016.5603$ & $-695.8\pm3.5$ & \citet{Trifonov2020} \\
 & $2016.5604$ & $-794.6\pm3.5$ & \citet{Trifonov2020} \\
 & $2017.4195$ & $-26.6\pm4.3$ & \citet{Trifonov2020} \\
 & $2017.4196$ & $-114.6\pm4.0$ & \citet{Trifonov2020} \\
 & $2017.4500$ & $449.4\pm8.4$ & \citet{Trifonov2020} \\
 & $2017.4501$ & $594.3\pm7.9$ & \citet{Trifonov2020} \\
 & $2017.7725$ & $723.3\pm4.2$ & \citet{Trifonov2020} \\
 & $2017.7726$ & $498.1\pm3.9$ & \citet{Trifonov2020} \\
 & $2017.7727$ & $696.7\pm4.2$ & \citet{Trifonov2020} \\
 & $2017.7728$ & $238.5\pm4.0$ & \citet{Trifonov2020} \\
 & $2017.7729$ & $466.1\pm3.7$ & \citet{Trifonov2020} \\
 & $2017.7730$ & $316.6\pm3.8$ & \citet{Trifonov2020} \\
 & $2017.7808$ & $175.23\pm4.2$ & \citet{Trifonov2020} \\
 & $2017.7809$ & $-115.5\pm3.8$ & \citet{Trifonov2020} \\
 \hline 
 HD~984 & $2014.5107$ & $264.3\pm3.6$ & \citet{Grandjean2020} \\
 & $2014.5108$ & $144.9\pm3.7$ & \citet{Grandjean2020} \\
 & $2014.5135$ & $-14.3\pm4.1$ & \citet{Grandjean2020} \\
 & $2014.5136$ & $94.5\pm3.8$ & \citet{Grandjean2020} \\
 & $2014.8823$ & $86.4\pm2.0$ & \citet{Grandjean2020} \\
 & $2014.8824$ & $86.7\pm2.0$ & \citet{Grandjean2020} \\
 & $2014.8880$ & $-124.6\pm2.6$ & \citet{Grandjean2020} \\
 & $2014.8881$ & $-52.5\pm2.6$ & \citet{Grandjean2020} \\
 & $2014.9015$ & $-68.0\pm2.3$ & \citet{Grandjean2020} \\
 & $2014.9015$ & $-59.1\pm2.4$ & \citet{Grandjean2020} \\
 & $2015.5489$ & $93.8\pm4.5$ & \citet{Grandjean2020} \\
 & $2015.5490$ & $151.7\pm4.5$ & \citet{Grandjean2020} \\
 & $2015.5515$ & $48.5\pm3.0$ & \citet{Grandjean2020} \\
 & $2015.5516$ & $-48.1\pm3.0$ & \citet{Grandjean2020} \\
 & $2015.8498$ & $-26.3\pm2.3$ & \citet{Grandjean2020} \\
 & $2015.8499$ & $-3.7\pm2.4$ & \citet{Grandjean2020} \\
 & $2015.8551$ & $-69.4\pm5.7$ & \citet{Grandjean2020} \\
 & $2015.8552$ & $3.7\pm3.3$ & \citet{Grandjean2020} \\
 & $2016.5666$ & $13.5\pm2.9$ & \citet{Grandjean2020} \\
 & $2016.5667$ & $213.9\pm2.7$ & \citet{Grandjean2020} \\
 & $2016.8826$ & $-86.8\pm3.3$ & \citet{Grandjean2020} \\
 & $2016.8827$ & $-23.2\pm3.5$ & \citet{Grandjean2020} \\
 & $2019.6941$ & $-30\pm110$ & \citet{Franson2022} \\
 & $2019.7733$ & $-10\pm90$ & \citet{Franson2022} \\
 & $2019.7870$ & $-30\pm50$ & \citet{Franson2022} \\
 & $2019.9181$ & $-40\pm90$ & \citet{Franson2022} \\
 & $2019.9208$ & $10\pm3$ & \citet{Franson2022} \\
 & $2019.9236$ & $-10\pm80$ & \citet{Franson2022} \\
 & $2019.9727$ & $10\pm60$ & \citet{Franson2022} \\
 & $2020.5859$ & $140\pm90$ & \citet{Franson2022} \\
 & $2020.6104$ & $20\pm60$ & \citet{Franson2022} \\
 & $2020.6458$ & $-45\pm11$ & \citet{Franson2022} \\
 & $2020.7657$ & $-40\pm40$ & \citet{Franson2022} \\
 & $2020.8094$ & $10\pm80$ & \citet{Franson2022} \\
 & $2020.8474$ & $20\pm100$ & \citet{Franson2022} \\

\end{longtable}
\label{rvs}
\twocolumn

\end{appendix}

\end{document}